\documentclass{acmtrans2m}
\usepackage{latexsym}
\usepackage{proof}

\newenvironment{definition}{\begin{define} \rm}{\end{define}}
\newenvironment{example}{\begin{exa} \rm}{\end{exa}}
\newtheorem{theorem}{Theorem}
\newtheorem{corollary}[theorem]{Corollary}
\newtheorem{define}[theorem]{Definition}
\newtheorem{exa}[theorem]{Example}
\newtheorem{lemma}[theorem]{Lemma}
\newtheorem{proposition}[theorem]{Proposition}

\newcommand{\ie}{{\em i.e.}}
\newcommand{\eg}{{\em e.g.}}

\def \mTrue {\hbox{true}}
\def \mFalse {\hbox{false}}
\def \mAnd {\&}
\def \mOr {\hat \lor}

\newcommand \mBox[1] {[#1]}
\newcommand \mDia[1] {\langle #1 \rangle}
\newcommand \stf[2] {#1 \models #2}
\newcommand \matchBox[3] {[#1 \dot = #2] #3}
\newcommand \matchDia[3] {\langle #1 \dot = #2\rangle #3}
\newcommand \actBox[2] {[#1] #2}
\newcommand \actDia[2] {\langle #1 \rangle #2}
\newcommand \outBox[2] {[\outact\! #1] #2}
\newcommand \outDia[2] {\langle \outact\! #1 \rangle #2}
\newcommand \inBox[2] {[\inact\! #1] #2}
\newcommand \inDia[2] {\langle \inact\! #1 \rangle #2}
\newcommand \inBoxL[2] {[\inact\! #1]^l #2}
\newcommand \inDiaL[2] {\langle \inact\! #1 \rangle^l #2}

\newcommand \inBoxE[2] {[\inact\! #1]^e #2}
\newcommand \inDiaE[2] {\langle \inact\! #1\rangle^e #2}

\def\Ascr{{\mathcal A}}
\def\Bscr{{\mathcal B}}
\def\Cscr{{\mathcal C}}
\def\Dscr{{\mathcal D}}
\def\Escr{{\mathcal E}}

\def\Hscr{{\mathcal H}}

\def\Lscr{{\mathcal L}}
\def\Mscr{{\mathcal M}}

\def\Qscr{{\mathcal Q}}
\def\Rscr{{\mathcal R}}
\def\Sscr{{\mathcal S}}

\def\Xscr{{\mathcal X}}

\def\Api{{\mathtt A}}
\def\Bpi{{\mathtt B}}
\def\Cpi{{\mathtt C}}
\def\Dpi{{\mathtt D}}

\def\Ppi{{\mathtt P}}
\def\Qpi{{\mathtt Q}}
\def\Rpi{{\mathtt R}}
\def\Spi{{\mathtt S}}
\def\Tpi{{\mathtt T}}

\def\relbar{\mathrel{\smash-}}
\def\joinrelm{\mathrel{\mkern-3.2mu}}
\def\tailpiece{\kern 1pt\vrule height 1ex width 0.3ex depth -.25ex}
\def\seqsym{\mathrel{\tailpiece\joinrelm\relbar}}

\newcommand{\sep}{\;\mid\;}
\newcommand{\FOL   }{FO\lambda}
\newcommand{\FOLD  }{\FOL^{\Delta}}

\newcommand{\FOLDNb}{\FOL^{\Delta\nabla}}

\newcommand{\Linc}{{\rm Linc}}
\newcommand{\Judg}[2]{#1\triangleright#2}

\newcommand{\NSeq}[3]{#1\,;\,#2 \seqsym #3}

\newcommand{\Seq}[2]{#1\seqsym #2}

\newcommand{\TSeq}[3]{#1 \vdash #2 : #3}

\newcommand{\action}{\hbox{\sl a}}

\newcommand{\llbra}{[\![}
\newcommand{\rrbra}{]\!]}

\newcommand{\ebisim}[2]{\hbox{\sl ebisim}~#1~#2}
\newcommand{\lbisim}[2]{\hbox{\sl lbisim}~#1~#2}

\newcommand{\botL}{\bot{\cal L}}

\newcommand{\cL}{\hbox{\sl c}{\cal L}}

\newcommand{\cut}{\hbox{\sl cut}}
\newcommand{\defL}{\hbox{\sl def}{\cal L}}
\newcommand{\defR}{\hbox{\sl def\/{}}{\cal R}}

\newcommand{\defeq}{\mathrel{\stackrel{{\scriptscriptstyle\triangle}}{=}}}
\newcommand{\defmu}{\mathrel{\stackrel{\mu}{=}}}
\newcommand{\defnu}{\mathrel{\stackrel{\nu}{=}}}

\newcommand{\dom}[1]{{\rm dom}(#1)}

\newcommand{\existsL}{\exists{\cal L}}
\newcommand{\existsR}{\exists{\cal R}}

\newcommand{\forallL}{\forall{\cal L}}
\newcommand{\forallR}{\forall{\cal R}}

\newcommand{\fv}[1]{{\rm fv}(#1)}

\newcommand{\hc }{\hbox{\sf hc}}

\newcommand{\hcun}{\hc^{\forall\nabla} }
\newcommand{\iand}{\wedge}
\newcommand{\iexists}{\exists}

\newcommand{\iforall}{\forall}
\newcommand{\iimp}{\supset}

\newcommand{\inact}{\mathop{\downarrow}}
\newcommand{\indR}{\mu {\cal R}}
\newcommand{\indL}{\mu {\cal L}}

\newcommand{\coindR}{\nu {\cal R}}
\newcommand{\coindL}{\nu {\cal L}}

\newcommand{\init}{\hbox{\sl init}}
\newcommand{\inpi}[2]{\hbox{\sl in}~#1~#2}

\newcommand{\landL}{\land{\cal L}}
\newcommand{\landR}{\land{\cal R}}
\newcommand{\level}[1]{{\rm lvl}(#1)}

\newcommand{\lorL}{\lor{\cal L}}
\newcommand{\lorR}{\lor{\cal R}}

\newcommand{\matchpi}[3]{{match}~#1~#2~#3}

\newcommand{\nablaL}{\nabla{\cal L}}
\newcommand{\nablaR}{\nabla{\cal R}}
\newcommand{\name}{\hbox{\sl n}}

\newcommand{\nupi}[2]{\nu #1.#2}

\newcommand{\oimpL}{\oimp{\cal L}}
\newcommand{\oimpR}{\oimp{\cal R}}
\newcommand{\oimp}{\supset}
\newcommand{\one  }[3]{#1\stackrel{#2}{-\!\!-\!\!\!\rightarrow    } #3}

\newcommand{\onep }[3]{#1\stackrel{#2}{-\!\!-\!\!\!\rightharpoonup} #3}

\newcommand{\outact}{\mathop{\uparrow}}
\newcommand{\outpi}[3]{\hbox{\sl out}~#1~#2~#3}
\newcommand{\parpi}[2]{#1\mathbin{|}#2}
\newcommand{\barpi}{\mathbin{|}}

\newcommand{\pluspi}[2]{#1 + #2}

\newcommand{\proc}{\hbox{\sl p}}

\newcommand{\ra}{\rightarrow}

\newcommand{\taupi}[1]{\tau~#1}

\newcommand{\topR}{\top{\cal R}}
\newcommand{\trans}[1]{[\![ #1 ]\!]}

\newcommand{\wL}{\hbox{\sl w}{\cal L}}

\newcommand{\fn}[1]{{\rm fn}(#1)}
\newcommand{\bn}[1]{{\rm bn}(#1)}
\newcommand{\n}[1]{{\rm n}(#1)}

\title{Proof Search Specifications of Bisimulation and\\
       Modal Logics for the $\pi$-calculus}
\author{Alwen Tiu \\ Logic and Computation Group\\ 
College of Engineering and Computer Science\\
The Australian National University
\and 
Dale Miller\\ INRIA-Saclay \& LIX, \'Ecole polytechnique}

\begin{abstract}

We specify the operational semantics and bisimulation relations for
the finite $\pi$-calculus within a logic that contains the
$\nabla$ quantifier for encoding {\em generic judgments} and
definitions for encoding fixed points.
Since we restrict to the finite case, the ability of the logic to unfold fixed
points allows this logic to be complete for both the inductive nature
of operational semantics and the coinductive nature of bisimulation.
The $\nabla$ quantifier helps with
the delicate issues surrounding the scope of variables within
$\pi$-calculus expressions and their executions (proofs).  We 
illustrate several merits of the logical specifications permitted by
this logic: they
are natural and declarative; they contain no side-conditions
concerning names of variables while maintaining a completely formal
treatment of such variables; differences between late and open
bisimulation relations arise from familar logic distinctions; 
the interplay between the three quantifiers ($\forall$, $\exists$, and
$\nabla$) and their scopes can explain the differences between early
and late bisimulation and between various modal operators based on
bound input and output actions;
and proof search involving the application of inference rules, unification, and
backtracking can provide complete proof systems for one-step
transitions, bisimulation, and satisfaction in modal logic.
We also illustrate how one can encode the $\pi$-calculus with replications,
in an extended logic with induction and co-induction.
\end{abstract}



\begin{document}
\begin{bottomstuff}
Authors' addresses: A. Tiu, Logic and Computation Group, College of Engineering
and Computer Science, Building 115, The Australian National University,
Canberra, ACT 0200, Australia; D. Miller, Laboratoire d'Informatique
(LIX), \'Ecole Polytechnique, Rue de Saclay, 91128 Palaiseau Cedex,
France. 
\end{bottomstuff}
\maketitle

\section{Introduction}
\label{sec:intro}

We present formal specifications of various aspects of the
$\pi$-calculus, including its syntax, operational semantics,
bisimulation relations, and modal logics.  We shall do this by using
the $\FOLDNb$ logic \cite{miller05tocl}.  We provide a high-level
introduction to this logic here before presenting more technical
aspects of it in the next section.

Just as it is common to use meta-level application to represent
object-level application (for example, the encoding of $P+Q$ is via
the meta-level application of the encoding for plus to the encoding of
its two arguments), we shall use meta-level $\lambda$-abstractions to
encode object-level abstractions.  The term {\em higher-order abstract
  syntax} (HOAS) \cite{pfenning88pldi} is commonly used to describe
this approach to mapping object-level abstractions into some
meta-level abstractions.  Of course, the nature of the resulting
encodings varies as one varies the meta-level.  For example, if the
meta-level is a higher-order functional programming language or a
higher-order type theory, the usual abstraction available constructs
function spaces.  In this case, HOAS maps object-level abstractions to
semantically rich function spaces: determining whether or not two
syntactic objects are equal is then mapped to the question of
determining if two functions are equal (typically, an undecidable
judgment).  In such a setting, HOAS is less about syntax and more
about a particular mathematical denotation of the syntax.  In this
paper, we start with an intuitionistic subset of the Simple Theory of
Types \cite{church40} that does not contain the {\em mathematical}
axioms of extensionality, description, choice, and infinity.  In this
setting, $\lambda$-abstraction is not strong enough to denote general
computable functions and equality of $\lambda$-terms is decidable.  As
a result, this weaker logic provides term-level bindings that can be
used to encode syntax with bindings.  This style of describing syntax
via a meta-logic containing a weak form of $\lambda$-abstraction has
been called the {\em $\lambda$-tree syntax} \cite{miller00cl} approach
to HOAS in order to distinguish it from the approaches that use
function spaces.  The $\lambda$-tree syntax approach to encoding
expressions is an old one ({\em cf.}
\cite{huet78,miller86acl,miller87slp,paulson86jlp}) and is used in
specifications written in the logic programming languages
$\lambda$Prolog \cite{nadathur88iclp} and Twelf \cite{pfenning99cade}.

Following Church, we shall use $\lambda$-abstractions to encode
both {\em term-level abstractions} and {\em formula-level
abstractions} ({\em e.g.}, quantifiers).  The computational aspects of the
$\pi$-calculus are usually specified via structured operational
semantics \cite{plotkin81}: here, such specifications are encoded
directly as inference rules and proofs over primitive relational
judgments ({\em e.g.}, one-step transitions).  As a result, a formal account
of the interaction of binding in syntax and binding in computation
leads to notions of {\em proof-level abstractions}.  One such binding
is the familiar {\em eigenvariable} abstraction of
\cite{gentzen35} used to encode a universally quantified variable that has
scope over an entire sequent.  A second proof-level binding was
introduced in \cite{miller05tocl} to capture a notion of {\em generic
judgment}: this proof-level binding has a scope over individual
entries within a sequent and is closely associated with the
formula-level binding introduced by the $\nabla$-quantifier.  A major
goal of this paper is to illustrate how the $\nabla$-quantifier
and this second proof-level abstraction can be used to specify and
reason about computation: the $\pi$-calculus has been chosen, in part,
because it is a small calculus in which bindings play an important role
in computation.

A reading of the truth condition for $\nabla x_\gamma.Bx$ is something
like the following: this formula is true if $B x$ is true for the new
element $x$ of type $\gamma$.  In particular, the formula $\nabla
x_\gamma\nabla y_\gamma.x\neq y$ is a theorem regardless of the
intended interpretation of the domain $\gamma$ since the bindings for
$x$ and $y$ are distinct.  In contrast, the truth value of the
formula $\forall x_\gamma\forall y_\gamma.x\neq y$ is dependent on the
domain $\gamma$: this quantified inequality is true if and only if 
the interpretation of $\gamma$ is empty.

The $\FOLDNb$ logic is based on
intuitionistic logic, a weaker logic than classical logic.  One of the
principles missing from intuitionistic logic is that of the excluded
middle: that is, $A\lor \lnot A$ is not generally provable in
intuitionistic logic.  Consider, for example, the following formula
concerning the variable $w$:
$$\forall x_\gamma [x=w \lor x\not= w].
  \hbox to 0pt{\qquad\qquad\qquad\qquad$(*)$\hss}$$
In classical logic, this formula is a trivial theorem.  From a
constructive point-of-view, it might not be desirable to admit this
formula as a theorem in some cases. 
If the type of quantification $\gamma$ is
a conventional (closed) first-order datatype, then we might expect to have a
decision procedure for equality.  For example, if $\gamma$ is the type
for lists, then it is a simple matter to construct a procedure that
decides whether or not two members of $\gamma$ are equal by
considering the top constructor of the list and, in the event of
comparing two non-empty lists, making a recursive call (assuming a
decision procedure is available for the elements of the list).  In fact, it
is possible to prove in an intuitionistic logic augmented with
induction (see, for example, \cite{tiu04phd}) the formula $(*)$ for
closed, first-order datatypes.  

If the type $\gamma$ is not given inductively, as is the usual case
for names in intuitionistic formalizations of the $\pi$-calculus (see
\cite{despeyroux00ifiptcs} and below), then
the corresponding instance of $(*)$ is not provable.
Thus, whether or not
we allow instances of $(*)$ to be assumed can change the nature
of a specification.  In fact, we show in Section~\ref{sec:bisim},
that if we add to our specification of {\em open bisimulation}
\cite{sangiorgi96acta} assumptions corresponding to $(*)$, then we get
a specification of {\em late bisimulation}.  If we were working with a
classical logic, such a declarative presentation of these two
bisimulations would not be so easy to describe.  

The authors first presented the logic used in this paper in
\cite{miller03lics} and illustrated its usefulness with the
$\pi$-calculus: in particular, the specifications of one-step
transitions in Figure~\ref{late pi def} and of late bisimulation in
Figure~\ref{bisim} also appear in \cite{miller03lics} but without
proof.  In this paper, we state the formal properties of our specifications,
provide a specification of late bisimulation, and provide a novel
comparison between open and late bisimulation.
In particular, we show that the difference between open and late
bisimulation (apart from the difference that arises from the use of
types defined inductively or not) can be captured by the different
quantification of free names using $\forall$ and $\nabla$.
We show in Section~\ref{sec:bisim} that a natural class of name {\em
distinctions} can be captured by the alternation of $\forall$ and
$\nabla$ quantifiers and, in the case where we are interested only in
checking open bisimilarity modulo the empty distinction, the notion of
distinction that arises in the process of checking bisimilarity is
completely subsumed by quantifier alternation.
In Section~\ref{sec:modal} we show that ``modal logics
for mobility'' can easily be handled as well and present, for the
first time, a modal characterization of open bisimulation.  Since our
focus in this paper is on names, scoping of names, dependency of
names, and distinction of names, we have chosen to focus on the finite
$\pi$-calculus.  The treatment of the $\pi$-calculus with replication is
presented in Section~\ref{sec:pi rep} through an example.  In
Section~\ref{sec:auto} we outline the automation of proof search based
on these specifications: when such automation is applied to our
specification of open bisimulation, a symbolic bisimulation
procedures arises.  In Section~\ref{sec:related} we present some related and
future work and Section~\ref{sec:conc} concludes the paper. 
In order to improve the readability of the main part of the paper, 
numerous technical proofs have been moved to the appendices.

Parts of this paper, in their preliminary forms and without proofs,
have been presented in \cite{tiu04fguc,tiu05concur}: in particular,
the material on encoding bisimulations (Section~\ref{sec:bisim})
corresponds to \cite{tiu04fguc} and the material on encoding modal
logics for the $\pi$-calculus (Section~\ref{sec:modal}) corresponds to
\cite{tiu05concur}.

\section{Overview of the logic}
\label{sec:foldn}

This paper is about the use of a certain logic to specify and reason
about computation.  We shall assume that the reader is not interested
in an in-depth analysis of the logic but with its application.  We
state the most relevant results we shall need about this logic in
order to reason about our $\pi$-calculus specifications.  The reader
who is interested in more details about this logic is referred to
\cite{tiu04phd} and \cite{miller05tocl}.

At the core of the logic $\FOLDNb$ (pronounced ``fold-nabla'') is a
first-order logic for $\lambda$-terms (hence, the prefix $\FOL$) that
is the result of extending Gentzen's LJ sequent calculus for
first-order intuitionistic logic \cite{gentzen35} with simply typed
$\lambda$-terms and with quantifiers that range over non-predicate
types.
The full logic is the result of making two extensions to this core.
First, ``fixed points'' are added via the technical
device of ``definitions,'' presented below and marked with the symbol
$\defeq$.  Fixed points can capture important forms of ``must
behavior'' in the treatment of operational semantics
\cite{mcdowell00tcs,mcdowell03tcs}.  Fixed points also strengthen
negation to encompass ``negation-as-finite-failure.''  In the presence
of this stronger negation, the usual treatment of $\lambda$-tree
syntax via ``generic judgments'' encoded as universal quantifiers is inadequate:
a more intensional
treatment of such judgments is provided by the addition of the
$\nabla$-quantifier \cite{miller05tocl}.

A {\em sequent} is an expression of the form
$
\Seq{B_1, \dots, B_n}{B_0}
$
where $B_0,\ldots,B_n$ are formulas and the elongated turnstile
$\seqsym$ is the sequent arrow.
To the left of the turnstile is a multiset: thus
repeated occurrences of a formula are allowed.  If the formulas $B_0,
\dots, B_n$ contain free variables, they are considered universally
quantified outside the sequent, in the sense that if the above sequent
is provable then every instance of it is also provable.  In proof
theoretical terms, such free variables are called {\em eigenvariables}.

A first attempt at using sequent calculus to capture judgments about
the $\pi$-calculus could be to use eigenvariables to encode names in
the $\pi$-calculus, but this is certainly problematic.  For example, if we
have a proof of the sequent $\Seq{}{P x y}$, where $x$ and $y$ are
different eigenvariables, then logic dictates that the sequent
$\Seq{}{P z z}$ is also provable (given the universal quantifier
reading of 
eigenvariables).  If the judgment $P$ is about, say,
bisimulation, then it is not likely that a statement about
bisimulation involving two different names $x$ and $y$ remains true if
they are identified to the same name $z$.

To address this problem, the logic $\FOLDNb$ extends sequents with a
new notion of ``local scope'' for proof-level bound variables
(originally motivated in \cite{miller03lics} to encode ``generic
judgments'').  In particular, sequents in $\FOLDNb$ are of the form
$$
\NSeq{\Sigma}{\Judg{\sigma_1}{B_1},\dots, \Judg{\sigma_n}{B_n}}
   {\Judg{\sigma_0}{B_0}}
$$
where $\Sigma$ is a {\em global signature}, \ie, the set of eigenvariables 
whose scope is over the entire sequent, and $\sigma_i$ is a {\em local signature}, \ie, 
a list of variables scoped over $B_i$. 
We shall consider sequents to be binding structures in the sense 
that the signatures, both the global and local
ones, are abstractions over their respective scopes.
The variables in 
$\Sigma$ and $\sigma_i$ will admit $\alpha$-conversion by systematically
changing the names of variables in signatures as well as those in
their scope, following the usual convention of the $\lambda$-calculus.
The meaning of eigenvariables is as before except that 
now instantiation of eigenvariables has to be capture-avoiding with respect
to the local signatures.
The variables in local signatures act as locally scoped {\em generic constants}:
that is, they do not vary in proofs since they will not be instantiated.
The expression $\Judg{\sigma}{B}$ is called a {\em generic judgment}
or simply a {\em judgment}. 
We use script letters $\Ascr$, $\Bscr$, {\em etc} to denote judgments.
We write simply $B$ instead of $\Judg{\sigma}{B}$ if the
signature $\sigma$ is empty. We shall often write 
the list $\sigma$ as a string of variables: \eg,
a judgment $\Judg{(x_1,x_2,x_3)}{B}$ will be written as $\Judg{x_1x_2x_3}{B}$.
If the list $x_1,x_2,x_3$ is known from context we shall also abbreviate the
judgment as $\Judg{\bar{x}}{B}$.

Following Church \citeyear{church40}, the type $o$ is used to denote the
type of formulas.  The propositional constants of $\FOLDNb$ are
$\land$ (conjunction), $\lor$ (disjunction), $\oimp$ (implication),
$\top$ (true) and $\bot$ (false). We shall abbreviate $B \oimp \bot$
as $\neg B$ (intuitionistic negation).
Syntactically, logical constants can be seen as typed constants: for example, the binary connectives have
type $o\ra o\ra o$.  For each simple type $\gamma$ that does not
contain $o$, there are three quantifiers in $\FOLDNb$: namely,
$\forall_\gamma$ (universal quantifier), $\exists_\gamma$ (existential
quantifier), $\nabla_\gamma$ (nabla), each one of type $(\gamma\ra
o)\ra o$.  The subscript type $\gamma$ is often dropped when it can be
inferred from context or its value is not important.  Since we do not
allow quantification over predicates, this logic is
proof-theoretically similar to first-order logic.
The inference rules for $\FOLDNb$ that do not deal with
definitions are given in Figure~\ref{fig:core foldnb}.  

\begin{figure}
$$
\infer[\init]
      {\NSeq{\Sigma}{\Judg{\sigma}{B},\Gamma}{\Judg{\sigma}{B}}}{}
\qquad 
\infer[\cut]
	{\NSeq{\Sigma}{\Delta,\Gamma}{\Cscr}}
	{\NSeq{\Sigma}{\Delta}{\Bscr} \qquad
	 \NSeq{\Sigma}{\Bscr,\Gamma}{\Cscr}}
$$
$$
\infer[\landL]
      {\NSeq
	{\Sigma}
	{\Judg{\sigma}{B \land C},\Gamma}
	{\Dscr}
      }
      {\NSeq
	{\Sigma}
	{\Judg{\sigma}{B}, \Judg{\sigma}{C}, \Gamma}
	{\Dscr}
      }
\qquad
\infer[\landR]
      {\NSeq{\Sigma}{\Gamma}{\Judg{\sigma}{B \land C}}}
      {
	\NSeq{\Sigma}{\Gamma}{\Judg{\sigma}{B} }
	\qquad 
	\NSeq{\Sigma}{\Gamma}{\Judg{\sigma}{C}}
      }
$$
$$
\infer[\lorL]
      {\NSeq{\Sigma}
	{\Judg{\sigma}{B \lor C},\Gamma}{\Dscr}}
      {\NSeq{\Sigma}{\Judg{\sigma}{B},\Gamma}{\Dscr}
	\qquad
	\NSeq{\Sigma}{\Judg{\sigma}{C},\Gamma}{\Dscr}
      }
\qquad
\infer[\lorR]
      {\NSeq{\Sigma}{\Gamma}{\Judg{\sigma}{B \lor C}}}
      {\NSeq{\Sigma}{\Gamma}{\Judg{\sigma}{B}}}
$$
$$
\infer[\botL]
      {\NSeq{\Sigma}{\Judg{\sigma}{\bot},\Gamma}{\Bscr}}
      {}
\qquad 
\infer[\lorR]{\NSeq{\Sigma}{\Gamma}{\Judg{\sigma}{B \lor C}}}
	{\NSeq{\Sigma}{\Gamma}{\Judg{\sigma}{C}}}
$$
$$
\infer[\oimpL]{\NSeq{\Sigma}{\Judg{\sigma}{B \oimp C},\Gamma}{\Dscr}}
	{\NSeq{\Sigma}{\Gamma}{\Judg{\sigma}{B}}
	\qquad \NSeq{\Sigma}{\Judg{\sigma}{C},\Gamma}{\Dscr}}
\qquad 
\infer[\oimpR]{\NSeq{\Sigma}{\Gamma}{\Judg{\sigma}{B \oimp C}}}
	{\NSeq{\Sigma}{\Judg{\sigma}{B},\Gamma}{\Judg{\sigma}{C}}}
$$
$$
\infer[\forallL]
      {\NSeq
	{\Sigma}
	{\Judg{\sigma}{\forall_\gamma x.B},\Gamma}
	{\Cscr}
      }
      {
	\TSeq{\Sigma, \sigma}{t}{\gamma}
	\qquad
	\NSeq
	{\Sigma}
	{\Judg{\sigma}{B[t/x]},\Gamma}{\Cscr}
      }
\qquad
\infer[\forallR]
      {\NSeq{\Sigma}{\Gamma}{\Judg{\sigma}{\forall x.B}}}
      {\NSeq{\Sigma, h}{\Gamma}
	{\Judg{\sigma}{B[(h~\sigma)/x]}}}
$$
$$
\infer[\existsL]{\NSeq{\Sigma}{\Judg{\sigma}{\exists x.B},\Gamma}{\Cscr}}
	{\NSeq{\Sigma, h}{\Judg{\sigma}{B[(h~\sigma)/x]},\Gamma}{\Cscr}}
\qquad 
\infer[\existsR]
      {\NSeq{\Sigma}{\Gamma}{\Judg{\sigma}{\exists_\gamma x.B}}}
      {
	\TSeq{\Sigma, \sigma}{t}{\gamma} \qquad
	\NSeq{\Sigma}{\Gamma}{\Judg{\sigma}{B[t/x]}}}
$$
$$
\infer[\nablaL]
{\NSeq{\Sigma}{\Judg{\sigma}{\nabla x\ B},\Gamma}{\Cscr}}
{
\NSeq{\Sigma}{\Judg{(\sigma,y)}{B[y/x]}, \Gamma }{\Cscr}
}
\qquad
\infer[\nablaR]
{\NSeq{\Sigma}{\Gamma}{\Judg{\sigma}{\nabla x\ B}}}
{
\NSeq{\Sigma}{\Gamma}{\Judg{(\sigma,y)}{B[y/x]}}
}
$$
$$
\infer[\cL]{\NSeq{\Sigma}{\Bscr,\Gamma}{\Cscr}}
	{\NSeq{\Sigma}{\Bscr,\Bscr,\Gamma}{\Cscr}}
\qquad
\infer[\wL]
      {\NSeq{\Sigma}{\Bscr, \Gamma}{\Cscr}}
      {\NSeq{\Sigma}{\Gamma}{\Cscr}}
\qquad
\infer[\topR]{\NSeq{\Sigma}{\Gamma}{\Judg{\sigma}{\top}}}{}
$$
\caption{The inference rules of $\FOLDNb$ not dealing with definitions.}
\label{fig:core foldnb}
\end{figure}

During the search for proofs (reading rules bottom up), inference rules 
for $\forall$ and $\exists$ quantifier place new variables
(eigenvariables) into 
the global signature while the inference rules for $\nabla$ place new
variables into a local signature. 
In the $\forallR$ and $\existsL$ rules, {\em raising} \cite{miller92jsc} is
used when replacing the bound variable $x$ (which can be substituted for by
terms containing variables in both the global signature and the local signature
$\sigma$) with the variable $h$ (which can only be instantiated with
terms containing variables in
the global signature). In order not to miss substitution terms, the
variable $x$ is replaced by the term $(h\,x_1\dots x_n)$: the latter
expression is written simply as $(h\,\sigma)$ where $\sigma$ is the list
$x_1,\dots,x_n$.
As is usual, the eigenvariable $h$ must not be free in the lower
sequent of these rules. In $\forallL$ and $\existsR$, the term $t$ can have free
variables from both $\Sigma$ and $\sigma$, a fact that is given
by the typing judgment $\TSeq{\Sigma,\sigma}{t}{\tau}$.  
The $\nablaL$ and $\nablaR$ rules have the proviso that $y$ is 
not free in $\nabla x\ B$. The introduction rules for propositional connectives are 
the standard ones for intuitionistic logic. Reading the rules top down, 
the structural rule $\cL$ (contraction) allows removal of duplicate judgments from the sequent 
and the rule $\wL$ (weakening) allows introduction of a (possibly new) 
judgment into the sequent. Note that since the initial rule $\init$ has 
implicit weakening, the weakening rule $\wL$ can actually be shown admissible, hence
it is strictly speaking not necessary. It is, however, convenient for interactive proof
search, since it allows one to remove irrelevant formulae (reading the rule bottom up)
in a sequent.

While sequent calculus introduction rules generally only introduce
logical connectives, the full logic $\FOLDNb$ additionally allows introduction
of atomic judgments; that is, judgments which do not contain any
occurrences of logical constants.  To each atomic judgment, $\Ascr$, we
associate a defining judgment, $\Bscr$, the {\em definition} of
$\Ascr$.  The introduction rule for the judgment $\Ascr$ is in effect
done by replacing $\Ascr$ with $\Bscr$ during proof search.  This
notion of definitions is an extension of work by Schroeder-Heister
\citeyear{schroeder-heister93lics}, Eriksson \citeyear{eriksson91elp}, Girard
\citeyear{girard92mail}, St\"ark \citeyear{staerk92jfcs}, and McDowell and
Miller \citeyear{mcdowell00tcs}.  These inference rules for definitions
allow for modest reasoning about the fixed points of (recursive) definitions.

\begin{definition}
\label{def:def}
A {\em definition clause} is written 
$\forall \bar x [p \, \bar{t} \defeq B]$, 
where $p$ is a predicate constant, every free variable of
the formula $B$ is also free in at least one term in the list
$\bar{t}$ of terms, and all variables free in $p\,\bar{t}$ are
contained in the list $\bar{x}$ of variables.  The atomic formula $p
\, \bar{t}$ is called the {\em head} of the clause, and the formula
$B$ is called the {\em body}.  The symbol $\defeq$ is used simply to
indicate a definitional clause: it is not a logical connective.  

Let $\forall_{\tau_1} x_1\ldots \forall_{\tau_n} x_n.H \defeq B$ be a
definition clause. Let $y_1, \ldots, y_m$ be a list of variables of types
$\alpha_1,\ldots,\alpha_m$, respectively. 
The {\em raised definition clause} of $H$ with respect to the signature
$\{y_1:\alpha_1,\ldots,y_m:\alpha_m\}$ is defined as
$$
\forall h_1\ldots \forall h_n.\Judg{\bar{y}}{H\theta}
\defeq \Judg{\bar{y}}{B\theta}
$$
where $\theta$ is the substitution
$[(h_1\,\bar{y})/x_1,\ldots,(h_n\,\bar{y})/x_n]$ and 
$h_i$ is of type
$\alpha_1 \ra \ldots \ra \alpha_m \ra \tau_i$, for every
$i\in\{1,\ldots,n\}$. 
A {\em definition} is a set of definition clauses
together with their raised clauses.
\end{definition}

Recall that we use script letters, such as $\Bscr$, $\Hscr$, etc., 
to refer to generic judgments. In particular, in referring to a raised definition clause, e.g., 
$$
\forall h_1\ldots \forall h_n.\Judg{\bar{y}}{H\theta}
\defeq \Judg{\bar{y}}{B\theta}
$$
we shall sometimes simply write $\Hscr \defeq \Bscr$ when the 
local signatures can be inferred from context or are unimportant to the discussion.

To guarantee the consistency (and cut-elimination) of the logic
$\FOLDNb$, we need some kind of stratification of definition so as to 
avoid a situation where a definition of a predicate depends
negatively on itself. 
For this purpose, we associate to each predicate $p$ a 
natural number $\level{p}$, the {\em level} of $p$.  The notion of
level is generalized to formulas as follows.
\begin{definition}
\label{def:level}
Given a formula $B$, its {\em level} $\level{B}$ is defined as follows:
\begin{enumerate}
\item $\level{p \, \bar{t}} = \level{p}$
\item $\level{\bot} = \level{\top} = 0$
\item $\level{B \land C} = \level{B \lor C} = \max(\level{B},\level{C})$
\item $\level{B \oimp C} = \max(\level{B}+1,\level{C})$
\item $\level{\forall x.B} = \level{\nabla x.B} = \level{\exists x.B}
       = \level{B}$. 
\end{enumerate}
We shall require that for every definition clause $\forall\bar{x}[p
\, \bar{t} \defeq B]$, $\level{B} \leq \level{p}$.  
\end{definition}
Note that the stratification condition above implies that in
a stratified definition, say $\forall \bar x [p\,\bar t \defeq B]$,
the predicate $p$ can only occur strictly positively in $B$ (if it occurs at all).
All definitions considered in this paper can be easily stratified according
to the above definition and cut-elimination holds for the logic using them.
For the latter, we refer the reader to \cite{miller05tocl} for the full details.

The introduction rules for a defined judgment are as follows.
When applying the introduction rules, we shall omit the 
outer quantifiers in a definition clause 
and assume implicitly that the free variables in the definition
clause are distinct from other variables in the sequent.
$$
\infer[\defL]
      {\NSeq{\Sigma}{\Ascr,\Gamma}{\Cscr}}
      {\{\NSeq{\Sigma\theta}
              {\Bscr\theta,\Gamma\theta}{\Cscr\theta}
       \;\vert\; \theta \in CSU(\Ascr,\Hscr) \mbox{ for some raised clause 
       $\Hscr \defeq \Bscr$}\}}
$$
$$       
\infer[\defR, \mbox{\quad where\ } \Hscr \defeq \Bscr \mbox{ is a raised definition clause and }
\Hscr\theta = \Ascr]
      {\NSeq{\Sigma}{\Gamma}{\Ascr}}
      {\NSeq{\Sigma}{\Gamma}{\Bscr\theta}}       
$$
In the above rules, we apply substitution to judgments.
The result of applying a substitution $\theta$ to a generic judgment
$\Judg{x_1,\ldots,x_n}{B}$, written as
$(\Judg{x_1,\ldots,x_n}{B})\theta$, is
$\Judg{y_1,\ldots,y_n}{B'}$, if $(\lambda x_1\ldots\lambda
x_n.B)\theta$ is equal (modulo $\lambda$-conversion) to $\lambda
y_1\ldots\lambda y_n.B'$.  If $\Gamma$ is a multiset of generic
judgments, then $\Gamma\theta$ is the multiset $\{J\theta\mid
J\in\Gamma\}$.  
In the $\defL$ rule, we use the notion of {\em complete set of
unifiers} (CSU) \cite{huet75tcs}. We denote by $CSU(\Ascr, \Hscr)$ the
complete set of unifiers for the pair $(\Ascr, \Hscr)$: that is, for
any substitution $\theta$ such that $\Ascr\theta = \Hscr\theta$, there
is a substitution $\rho \in CSU(\Ascr,\Hscr)$ such that $\theta = \rho
\circ \theta'$ for some substitution $\theta'$.  In all the
applications of $\defL$ in this paper, the set $CSU(\Ascr, \Hscr)$ is
either empty (the two judgments are not unifiable) or contains a
single substitution denoting the most general unifier.  The signature
$\Sigma\theta$ in $\defL$ denotes a signature obtained from $\Sigma$
by removing the variables in the domain of $\theta$ and adding the
variables in the range of $\theta$.  In the $\defL$ rule, reading the
rule bottom-up, eigenvariables can be instantiated in the premise,
while in the $\defR$ rule, eigenvariables are not instantiated.  The
set that is the premise of the $\defL$ rule means that that rule
instance has a premise for every member of that set: if that set is
empty, then the premise is proved.

Equality for terms can be defined in $\FOLDNb$ using the single
definition clause $[\forall x.\;x = x \defeq \top]$.
Specializing the $\defL$ and $\defR$ rules to equality yields the
inference rules
$$
 \infer{\NSeq{\Sigma}{\Judg{\bar y}{s=t},\Gamma}{\Cscr}}
       {\{\NSeq{\Sigma\theta}{\Gamma\theta}{\Cscr\theta}
       \;\vert\; \theta \in CSU(\lambda\bar y.s, \lambda\bar y.t)\}}
  \qquad
  \infer{\NSeq{\Sigma}{\Gamma}{\Judg{\bar y}{t=t}}}{}
$$
Disequality $s\neq t$, the negation of equality, is an abbreviation
for $(s = t)\oimp \bot$.

One might find the following analogy with logic programming helpful:
if a definition is viewed as a logic program, then the $\defR$ rule
captures backchaining and the $\defL$ rule corresponds to {\em case
analysis} on all possible ways an atomic judgment could be proved.  In
the case where the program has only finitely many computation paths,
we can effectively encode {\em negation-as-failure} using $\defL$
\cite{hallnas91jlc}.

\section{Some meta-theory of the logic}
\label{sec:meta}

Once we have written a computational specification as logical
formulas, it is important that the underlying logic has 
formal properties that allow us to reason about that
specification.  In this section, we list a few formal properties of
$\FOLDNb$ that will be used later in this paper.

Cut-elimination for $\FOLDNb$ \cite{miller05tocl,tiu04phd} is probably
the single most important meta-theoretic property needed.  Beside
guaranteeing the consistency of the logic, it also provides
completeness for {\em cut-free} proofs: these proofs are 
used to help prove the {\em adequacy} of a logical specification.  For
example, the proof that a certain specification actually encodes the
one-step transition relation or the bisimulation relation starts by
examining the highly restricted structure of cut-free proofs.  Also,
cut-elimination allows use of modus ponens and substitutions into
cut-free proofs and to be assured that another cut-free proof arises
from that operation.

Another important structural property of provability is the {\em
invertibility} of inference rules.  An inference rule of logic is {\em
invertible} if the provability of the conclusion implies the
provability of the premise(s) of the rule.  The following rules in
$\FOLDNb$ are invertible: $\landR, \landL, \lorL, \oimpR, \forallR,
\existsL, \defL$ (see \cite{tiu04phd} for a proof).  Knowing the
invertibility of a rule can be useful in determining some structure of
a proof.  For example, if we know that a sequent $\Seq{A\lor B,
\Gamma}{C}$ is provable, then by the invertibility of $\lorL$, we know
that it must be the case that $\Seq{A,\Gamma}{C}$ and
$\Seq{B,\Gamma}{C}$ are provable.

We now present several meta-theoretic properties of provability that
are specifically targeted at the $\nabla$-quantifier.  These
properties are useful when proving the adequacy of our
specifications of bisimulation and modal logic in the following
sections.  These properties also provide some insights into
the differences between the universal and the $\nabla$ quantifiers.  The
proofs of the propositions listed in this section can be found in 
\cite{tiu04phd}.

Throughout the paper, we shall use the following notation for provability:
We shall write $\vdash \NSeq \Sigma \Gamma C$ 
to denote the fact the sequent $\NSeq \Sigma \Gamma C$ is provable, and 
$\vdash B$ to denote provability of the sequent $\NSeq . . B.$

The following proposition states that the global scope of an
eigenvariable can be weakened to be a locally scoped variable when
there are no assumptions.

\begin{proposition}
\label{prop:forall nabla}
If $\vdash\forall x B$  then $\vdash\nabla x B$.
\end{proposition}

Notice that the implication $\forall_\tau x B \oimp
\nabla_\tau x B$ does not necessarily hold.   For example, if the type $\tau$ is
empty, then $\forall_\tau x B$ may be true vacuously, independently of the
structure of $B$, whereas attempting to prove $\nabla x B$ reduces to
attempting to prove $B$ given the fresh element $x$ of type $\tau$. 

As we suggested in Section~\ref{sec:intro} with the formula $\nabla
x_\gamma\nabla y_\gamma.x\neq y$, the converse of
Proposition~\ref{prop:forall nabla} is not true in general.  That
converse does hold, however, if we use definitions and formulas that
do not contain implications and, consequently, do not contain
negations (since these are formally defined as implications).
Horn clauses provide an 
interesting fragment of logic that does not contain negations: in
that setting, the distinction between $\nabla$ and $\forall$ cannot be
observed using the proof system.  More precisely, let {\em
$\hcun$-formulas} (for Horn clauses formulas with $\forall$ and
$\nabla$) be a formulas that do not contain occurrences of the logical
constant $\oimp$ (implication).  A $\hcun$-definition is a definition
whose bodies are $\hcun$-formulas.  For example, the definition of the
one-step transition in Figure~\ref{late pi def} is an
$\hcun$-definition but the definition of bisimulation in
Figure~\ref{bisim} is not a $\hcun$-definition.

\begin{proposition}\label{prop:nabla forall}
Let $\Dscr$ be a $\hcun$-definition and $\forall x G$ be a $\hcun$-formula.
Then, assuming $\Dscr$ is the only definition used, $\forall x G$ is provable  
if and only if $\nabla x G$ is provable.
\end{proposition}

The above proposition highlights the fact that positive occurrences of
$\nabla$ are interchangeable with $\forall.$ The specification of the
operational semantics of the $\pi$-calculus in the next section uses
only positive occurences of $\nabla$, hence its specification can be
done also in a logic without $\nabla$.  However, our specifications of
bisimulation and modal logics in the subsequent sections make use of
implications in definitions and, as a result, $\nabla$ cannot be
replaced with $\forall$.  We shall come back to this discussion on the
distinction between $\nabla$ and $\forall$ when we present the
specification of bisimulation in Section~\ref{sec:bisim}.

Finally, we state a technical result about proofs in $\FOLDNb$ that
states that provability of a sequent is not affected by the
application of substitutions.

\begin{proposition}
\label{prop:subst}
Let $\Pi$ be a proof of $\NSeq{\Sigma}{\Gamma}{\Cscr}$. 
Then for any substitution $\theta$, there exists a proof $\Pi'$ of 
$\NSeq{\Sigma\theta}{\Gamma\theta}{\Cscr\theta}$ such that
the height of proof of $\Pi'$ is less than or equal to the height of $\Pi$.
\end{proposition}

\section{Logical specification of one-step transition}

The finite $\pi$-calculus is the fragment of the $\pi$-calculus
without recursion (or replication). 
In particular, process expressions are defined as
$$
\Ppi ::= 0 \sep \bar{x}y.\Ppi \sep x(y).\Ppi \sep \tau.\Ppi \sep
       (x)\Ppi \sep [x = y] \Ppi \sep \Ppi | \Ppi \sep \Ppi + \Ppi.
$$
We use the symbols $\Ppi$, $\Qpi$, $\Rpi$, $\Spi$, $\Tpi$ to
denote processes and  lower case letters, {\em e.g.}, $a, b, c, d, x, y, z$
to denote names.
The occurrence of $y$ in the processes $x(y).\Ppi$ and $(y)\Ppi$ is a binding 
occurrence with $\Ppi$ as its scope. 
The set of free names in $\Ppi$ is denoted by $\fn{\Ppi}$, the set of bound
names is denoted by $\bn{\Ppi}$. We write $\n{\Ppi}$ for the set
$\fn{\Ppi} \cup \bn{\Ppi}$. We consider processes to be equivalent 
if they are identical up to a renaming of bound variables.

The relation of one-step (late) transition \cite{milner92icII} for the
$\pi$-calculus is denoted by $\one{\Ppi}{\alpha}{\Qpi}$, 
where $\Ppi$ and $\Qpi$ are processes and $\alpha$ is an action. 
The kinds of actions are {\em the silent action} $\tau$, 
{\em the free  input action} $x y$, 
{\em the free output action} $\bar{x}y$, 
{\em the bound input action} $x(y)$, and
{\em the bound output action} $\bar{x}(y)$. The name $y$ in $x(y)$ and
$\bar{x}(y)$ is a binding occurrence. Just as we did with processes, we use
$\fn{\alpha}$, $\bn{\alpha}$ and $\n{\alpha}$ to denote free names,
bound names, and names in $\alpha$. An action without binding
occurrences of names is a {\em free action} (this includes the silent action); 
otherwise it is a {\em bound action}. 

Three primitive syntactic categories are used to encode the
$\pi$-calculus into $\lambda$-tree syntax: \name\ for names, \proc\
for processes, and $\action$ for actions.  We do not assume any
inhabitants of type $\name$: as a consequence, a free name is
translated to a variable of type $\name$ that is either universally or
$\nabla$-quantified, depending on whether we want to allow names to
be instantiated or not. For instance, when encoding late bisimulation,
free names correspond to $\nabla$-quantified variables, while when
encoding open bisimulation, free names correspond to universally
quantified variables (Section~\ref{sec:bisim}).  Since the rest of
this paper is about the $\pi$-calculus, the $\nabla$ quantifier will
from now on only be used at type $\name$.

There are three constructors for actions: $\tau : \action$ (for the
silent action) and the two constants $\inact$ and $\outact$, both of
type $\name \ra \name \ra \action$ (for building input and output
actions, respectively).  The free output action $\bar{x}y$, is encoded
as $\outact x y$ while the bound output action $\bar{x}(y)$ is encoded
as $\lambda y\ (\outact x y)$ (or the $\eta$-equivalent term $\outact
x$).  The free input action $x y$, is encoded as $\inact x y$ while
the bound input action $x(y)$ is encoded as $\lambda y\ (\inact x y)$
(or simply $\inact x$).  Notice that bound input and bound output actions
have type $\name\ra\action$ instead of $\action$.

The following are process constructors, where $+$ and $|$ are written
as infix: 
$$
\begin{array}{c}
0 : \proc \qquad
\tau : \proc \ra \proc \qquad
{out} : \name\ra\name\ra\proc\ra\proc \qquad 
{in} : \name\ra (\name \ra \proc)\ra \proc \\
+ : \proc\ra\proc\ra \proc \quad\ 
| : \proc\ra\proc\ra\proc \quad\ 
match : \name\ra\name\ra\proc\ra\proc \quad\hfil
\nu : (\name\ra\proc)\ra\proc
\end{array}
$$
Notice $\tau$ is overloaded by being used as a constructor
of actions and of processes.
The one-step transition relation is represented using two predicates:
The predicate $\one{\cdot^1}{\cdot^2}{\cdot^3}$ of type
$\proc\ra\action\ra\proc\ra o$, where the first argument (indicated with $\cdot^1$)
is of type $\proc$, the second argument is of type $\action$, and
the third argument is of type $\proc$, 
encodes transitions involving the free actions while the 
predicate $\onep{\cdot^1}{\cdot^2}{\cdot^3}$ of type
$\proc\ra(\name\ra\action)\ra(\name\ra\proc)\ra o$ encodes transitions
involving bound values.  The precise translation of the $\pi$-calculus
syntax into simply typed $\lambda$-terms is given in the following
definition. We assume that names in $\pi$-calculus processes are
translated to variables (of the same names) in the meta logic.

\begin{definition}
The following function $\trans{.}$ translates process expressions
to $\beta\eta$-long normal terms of type $\proc$. 
$$
\begin{array}{c}
\trans{0} = 0 \qquad
\trans{\Ppi + \Qpi} = \trans{\Ppi} + \trans{\Qpi}\qquad
\trans{\Ppi | \Qpi} = \trans{\Ppi}\, | \, \trans{\Qpi}\qquad
\trans{\tau.\Ppi}  =  \tau\,\trans{\Ppi}\\
\trans{[x = y]\Ppi} = \matchpi{x}{y}{\trans{\Ppi}}\qquad\qquad
\trans{\bar{x}y.\Ppi} = \outpi{x}{y}{\trans{\Ppi}}\\
\trans{x(y).\Ppi} = \inpi{x}{\lambda y.\trans{\Ppi}}\qquad\qquad
\trans{(x)\Ppi}  =  \nu \lambda x.\trans{\Ppi}\\
\end{array}
$$
We abbreviate $\nu\lambda x.P$ as simply $\nupi{x}{P}$.  
The one-step transition judgments are translated to atomic
formulas as follows (we overload the symbol $\trans{.}$).
$$
\begin{array}{rcl@{\qquad}rcl}
\trans{\one{\Ppi}{x y}{\Qpi}} & = & \one{\trans{\Ppi}}{\inact x y}{\trans{\Qpi}}&
\trans{\one{\Ppi}{x(y)}{\Qpi}}&=&\onep{\trans{\Ppi}}{\inact x}{\lambda y.\trans{\Qpi}}\\
\trans{\one{\Ppi}{\bar{x}y}{\Qpi}} &=& \one{\trans{\Ppi}}{\outact x y}{\trans{\Qpi}} &
\trans{\one{\Ppi}{\bar{x}(y)}{\Qpi}} &=& 
     \onep{\trans{\Ppi}}{\outact x}{\lambda y.\trans{\Qpi}} \\
\trans{\one{\Ppi}{\tau}{\Qpi}} &=& \one{\trans{\Ppi}}{\tau}{\trans{\Qpi}}
\end{array}
$$
\end{definition}

Notice that we mention encodings of free input actions and free input transition
judgments. Since we shall be concerned only with late transition systems, these 
will not be needed in subsequent specifications.  Giving these actions
and judgments explicit encodings, however, simplifies the argument for the adequacy of
representations of these syntactic judgments: that is, every
$\beta\eta$-normal term
of type $\action$ corresponds to an action in the $\pi$-calculus, and similarly, every
atomic formula encoding of a one-step transition judgment (in $\beta\eta$-normal form) 
corresponds to a one-step transition judgment in the $\pi$-calculus.

\begin{figure}
$$
\begin{array}{rrcl}
\hbox{\sc tau:}   & \one{\taupi{P}}{\tau}{P} &\defeq& \top\\
\hbox{\sc in:}    & \onep{\inpi{X}{M}}{\inact X}{M} &\defeq& \top\\
\hbox{\sc out:}   & \one{\outpi{x}{y}{P}}{\outact x y}{P} &\defeq& \top\\
\hbox{\sc match:} & \one{\matchpi{x}{x}{P}}{A}{Q}  &\defeq& \one{P}{A}{Q} \\ 
                  & \onep{\matchpi{x}{x}{P}}{A}{Q} &\defeq& \onep{P}{A}{Q}\\
\hbox{\sc sum:}   & \one{\pluspi{P}{Q}}{A}{R} &\defeq& \one{P}{A}{R} \\
                  & \one{\pluspi{P}{Q}}{A}{R} &\defeq& \one{Q}{A}{R}\\
                  & \onep{\pluspi{P}{Q}}{A}{R} &\defeq& \onep{P}{A}{R} \\
                  & \onep{\pluspi{P}{Q}}{A}{R} &\defeq&  \onep{Q}{A}{R}\\
\hbox{\sc par:}   & \one{\parpi{P}{Q}}{A}{\parpi{P'}{Q}} &\defeq& \one{P}{A}{P'} \\
                  & \one{\parpi{P}{Q}}{A}{\parpi{P}{Q'}} &\defeq& \one{Q}{A}{Q'}\\
                  & \onep{\parpi{P}{Q}}{A}{\lambda n(\parpi{M\,n}{Q})} &\defeq& 
                    \onep{P}{A}{M}\\
                  & \onep{\parpi{P}{Q}}{A}{\lambda n(\parpi{P}{N\,n})} &\defeq& 
                    \onep{Q}{A}{N}.\\
\hbox{\sc res:}   & \one{\nupi{n}{P n}}{A}{\nupi{n}{Q n}} &\defeq& 
                      \nabla n(\one {P n}{A}{Q n})\\
                  & \onep{\nupi{n}{P n}}{A}{\lambda m\ \nupi{n}{P' n m}} &\defeq& 
                       \nabla n(\onep {P n}{A}{P' n}) \\  
\hbox{\sc open:}  & \onep{\nupi n M n}{\outact X}{M'} &\defeq& 
                     \nabla n(\one{M n}{\outact X n}{M' n}) \\
\hbox{\sc close:} &  \one{\parpi P Q}{\tau}{\nupi{n}{(\parpi{M n}{N n})}}  &\defeq& 
                         \exists X. \onep{P}{\inact X}{M}\land\onep{Q}{\outact X}{N}\\
                  & \one{\parpi P Q}{\tau}{\nupi{n}{(\parpi{M n}{N n})}} &\defeq& 
                     \exists X. \onep{P}{\outact X}{M}\land\onep{Q}{\inact X}{N}\\
\hbox{\sc com:}   & \one{\parpi P Q}{\tau}{\parpi{M Y}{Q'}} &\defeq& 
                       \exists X. \onep{P}{\inact X}{M}\land\one{Q}{\outact X Y}{Q'}\\
                  & \one{\parpi P Q}{\tau}{\parpi{P'}{N Y}} &\defeq& 
                     \exists X. \one{P}{\outact X Y}{P'} \land \onep{Q}{\inact X}{N}
\end{array}
$$
\caption{Definition clauses for the late transition system.}
\label{late pi def}
\end{figure}

Figure~\ref{late pi def} contains a definition, called ${\bf D}_\pi$,
that encodes the operational semantics of the late transition system
for the finite $\pi$-calculus.
In this specification, free variables are schema
variables that are assumed to be universally scoped over the 
definition clause in which they appear.  These schema
variables have primitive types such as $\action$, $\name$, and $\proc$
as well as functional types such as $\name\ra\action$ and
$\name\ra\proc$.  

Notice that, as a consequence of using $\lambda$-tree syntax for this
specification, the usual side conditions in the original specifications of
the $\pi$-calculus \cite{milner92icII} are no longer present. 
For example, the side condition that $X \not = n$ in
the open rule is implicit, since $X$ is outside the scope of $n$ and, therefore,
cannot be instantiated with $n$ (substitutions into logical
expressions cannot capture bound variable names).
The adequacy of our encoding is stated in the following lemma and 
proposition (their proofs can be found in \cite{tiu04phd}).

\begin{lemma}
\label{lemma: adequacy}
The function $\trans{.}$ is a bijection between $\alpha$-equivalence
classes of process expressions and $\beta\eta$-equivalence classes of
terms of type $\proc$ whose free variables (if any) are of type $\name$.
\end{lemma}

\begin{proposition}
\label{prop:one step}
Let $\Ppi$ and $\Qpi$ be processes and $\alpha$ an action. Let $\bar{n}$ be
a list of free names containing the free names in $\Ppi$, $\Qpi$, and $\alpha$.
The transition $\one{\Ppi}{\alpha}{\Qpi}$ is derivable in the $\pi$-calculus if and only
if $\NSeq{.}{.}{\nabla \bar{n}.\trans{\one{\Ppi}{\alpha}{\Qpi}}}$ is provable 
in $\FOLDNb$ with the definition ${\bf D}_\pi$.
\end{proposition}

If our goal was only to correctly encode one-step transitions for the
$\pi$-calculus then we would need neither $\nabla$ nor definitions.
In particular, let ${\bf D}^\forall_\pi$ be the result of replacing
all $\nabla$ quantifiers in ${\bf D}_\pi$ with $\forall$ quantifiers.
A slight generalization of Proposition~\ref{prop:nabla forall} (see
\cite{miller05tocl,tiu04phd}) allows us to conclude that
$\NSeq{.}{.}{\nabla \bar{n}.\trans{\one{\Ppi}{\alpha}{\Qpi}}}$ is provable 
in $\FOLDNb$ with the definition ${\bf D}_\pi$ if and only if
$\NSeq{.}{.}{\forall \bar{n}.\trans{\one{\Ppi}{\alpha}{\Qpi}}}$ is provable 
in $\FOLDNb$ with the definition ${\bf D}^\forall_\pi$.  Furthermore, we
can also do with the simpler notions of {\em theory} or {\em
assumptions} and not {\em definition}.  In particular, let ${\bf
P}_\pi$ be the set of implications that result from changing all
definition clauses in ${\bf D}^\forall_\pi$ into reverse implications
(i.e., the head is implied by the body).  We can then conclude that
$\NSeq{.}{.}{\forall \bar{n}.\trans{\one{\Ppi}{\alpha}{\Qpi}}}$ is provable 
in $\FOLDNb$ with the definition ${\bf D}^\forall_\pi$ if and only if 
$\NSeq{.}{{\bf P}_\pi}{\forall \bar{n}.\trans{\one{\Ppi}{\alpha}{\Qpi}}}$ 
is provable in intuitionistic (and classical) logic.  In fact, such a specification 
of the one-step transitions in the $\pi$-calculus as a theory without $\nabla$ dates
back to at least Miller and Palamidessi \citeyear{miller99surveys}.

Definitions and $\nabla$ are needed, however, for proving non-Horn
properties (that is, properties requiring a strong notion of
negation).  The following proposition is a dual of
Proposition~\ref{prop:one step}. Its proof can be found in the appendix.

\begin{proposition}
\label{prop:neg one step}
Let $\Ppi$ and $\Qpi$ be processes and $\alpha$ an action.   Let $\bar{n}$ be
a list of free names containing the free names in $\Ppi$, $\Qpi$, and $\alpha$.
The transition $\one{\Ppi}{\alpha}{\Qpi}$ is not derivable in the $\pi$-calculus if and only
if $\NSeq{.}{.}{\lnot\nabla \bar{n}.\trans{\one{\Ppi}{\alpha}{\Qpi}}}$ is provable 
in $\FOLDNb$ with the definition ${\bf D}_\pi$.
\end{proposition}

The following example illustrates how a negation can be proved in
$\FOLDNb$.  When writing encoded process expressions, we shall use,
instead, the syntax of the $\pi$-calculus along with the usual
abbreviations: for example, when a name $z$ is used as a prefix, it
denotes the prefix $z(w)$ where $w$ is vacuous in its scope; when a
name $\bar{z}$ is used as a prefix it denotes the output prefix
$\bar{z}a$ for some fixed name $a$.  We also abbreviate
$(y)\bar{x}y.P$ as $\bar{x}(y).P$ and the process term $0$ is omitted
if it appears as the continuation of a prefix. We assume that the
operators $|$ and $+$ associate to the right, {\em e.g.}, we write $P + Q + R$
to denote $P + (Q + R)$.

\begin{example}
\label{ex:one step negative}
Consider the process $(y)([x=y]\bar{x}z)$, which could be the
continuation of some other process which inputs $x$ on some channel,
{\em e.g.}, $a(x).(y)[x = y] \bar{x} z$.  
Since the bound variable $y$ is different from
any name substituted for $x$, that process cannot make
a transition and the following formula should be provable.
$$
\forall x\forall z\forall Q \forall \alpha.[(\one{(y)[x = y]
    (\bar{x}z )}{\alpha}{Q}) \oimp \bot]
$$ 
Since $y$ is bound inside the
scope of $x$,  no instantiation for $x$ can be equal to $y$.
The formal derivation of the above formula is (ignoring the initial 
uses of $\oimpR$ and $\forallR$):
$$
\infer[\defL]
{\NSeq{\{x,z,Q,\alpha \}}{\Judg{.}{(\one{(y)[x = y] (\bar{x}z.0)}{\alpha}{Q})}}{\bot}}
{
\infer[\nablaL]
{\NSeq{\{x,z,Q',\alpha \}}{\Judg{.}{\nabla y.(\one{[x = y] (\bar{x}z.0)}{\alpha}{Q' y})}}{\bot}}
{
\infer[\defL]
{\NSeq{\{x,z,Q',\alpha \}}{\Judg{y}{(\one{[x = y](\bar{x}z.0)}{\alpha}{Q' y})}}{\bot}}
{}
}
}
$$
The success of the topmost instance of $\defL$ depends on the failure of 
the unification problem
$
\lambda y.x = \lambda y.y.
$
Notice that the scoping of term-level variables is maintained at the proof-level
by the separation of (global) eigenvariables and (locally bound) generic variables.
The ``newness'' of $y$ is internalized as a $\lambda$-abstraction and, hence,
it is not subject to instantiation. 
\end{example}

The ability to prove a negation is implied by any proof system that
can also prove bisimulation for the $\pi$-calculus (at least for
the finite fragment): for
example, the negation above holds because the process
$(y)([x=y]\bar{x}z)$ is bisimilar to $0$ (see the next
section).

\section{Logical specifications of strong bisimilarity}
\label{sec:bisim}

We consider specifying three notions of bisimilarity tied to the late
transition system: the strong early bisimilarity, the strong late bisimilarity 
and the strong open bisimilarity.  As it turns out, the definition clauses
corresponding to strong late and strong open bisimilarity coincide.
Their essential differences are in the quantification of free names
and in the presence (or the absence) of the axiom of excluded middle
on the equality of names.
The difference between early and late bisimulation is tied to the scope
of the quantification of names in the case involving bound input (see the
definitions below).
The original definitions of early, late, and open bisimilarity are given in 
\cite{milner92icII,sangiorgi01}.  Here we choose to make the
side conditions explicit, instead of adopting the bound variable
convention in \cite{sangiorgi01}.

Given a relation on processes $\Rscr$, we write $P~\Rscr ~ Q$
to denote $(P,Q) \in \Rscr.$

\begin{definition} 
\label{def:lbisim}
A process relation $\Rscr$ is a {\em strong late bisimulation}
if $\Rscr$ is symmetric and whenever $\Ppi ~ {\cal R} ~ \Qpi$,
\begin{enumerate}
\item if $\one{\Ppi}{\alpha}{\Ppi'}$ and $\alpha$ is a free action,
  then there is $\Qpi'$ such that
  $\one{\Qpi}{\alpha}{\Qpi'}$ and $\Ppi'  ~ \Rscr ~ \Qpi'$;
\item if $\one{\Ppi}{x(z)}{\Ppi'}$ and $z \not \in \n{\Ppi,\Qpi}$ 
  then there is $\Qpi'$ such that
  $\one{\Qpi}{x(z)}{\Qpi'}$ and, for every name $y$, 
  $\Ppi'[y/z] ~\Rscr ~ \Qpi'[y/z]$; and 
\item if $\one{\Ppi}{\bar{x}(z)}{\Ppi'}$ and $z \not \in \n{\Ppi,\Qpi}$ 
  then there is $\Qpi'$
  such that $\one{\Qpi}{\bar{x}(z)}{\Qpi'}$ and $\Ppi' ~\Rscr ~ \Qpi'$.
\end{enumerate}
The processes $\Ppi$ and $\Qpi$ are {\em strong late bisimilar},
written $\Ppi \sim_l \Qpi$, if there is a strong late bisimulation
$\Rscr$ such that $\Ppi ~ \Rscr ~ \Qpi.$
\end{definition}

\begin{definition} 
\label{def:ebisim}
A process relation $\Rscr$ is a {\em strong early bisimulation}
if $\Rscr$ is symmetric and whenever $\Ppi ~ {\cal R} ~ \Qpi$,
\begin{enumerate}
\item if $\one{\Ppi}{\alpha}{\Ppi'}$ and $\alpha$ is a free action,
  then there is $\Qpi'$ such that
  $\one{\Qpi}{\alpha}{\Qpi'}$ and $\Ppi' ~\Rscr ~ \Qpi'$,
\item if $\one{\Ppi}{x(z)}{\Ppi'}$ and $z \not \in \n{\Ppi,\Qpi}$ 
  then for every name $y$, there is $\Qpi'$ such that
  $\one{\Qpi}{x(z)}{\Qpi'}$ and $\Ppi'[y/z] ~\Rscr ~ \Qpi'[y/z]$,
\item if $\one{\Ppi}{\bar{x}(z)}{\Ppi'}$ and $z \not \in \n{\Ppi,\Qpi}$ 
  then there is $\Qpi'$
  such that $\one{\Qpi}{\bar{x}(z)}{\Qpi'}$ and $\Ppi' ~\Rscr ~ \Qpi'$.
\end{enumerate}
The processes $\Ppi$ and $\Qpi$ are {\em strong early bisimilar}, written 
$\Ppi \sim_e \Qpi$, if there is a strong early bisimulation $\Rscr$ such that
$\Ppi ~ \Rscr ~ \Qpi.$
\end{definition}

\begin{definition} 
A {\em distinction} $D$ is a finite symmetric and irreflexive relation
on names.  A substitution $\theta$ {\em respects} a distinction $D$ if
$(x,y) \in D$ implies $x\theta \not= y\theta$. We refer to the
substitution $\theta$ as a {\em $D$-substitution}.  Given a distinction
$D$ and a $D$-substitution $\theta$, the result of applying $\theta$
to all variables in $D$, written $D\theta$, is another distinction.
We denote by $\fn{D}$ the set of names occurring in $D$.
\end{definition}

Since distinctions are symmetric by definition, when we enumerate a
distinction, we often omit the symmetric part of the distinction.  For
instance, we shall write $\{(a,b)\}$ to mean the distinction $\{(a,b),
(b,a)\},$ and we shall also write $D\cup(S \times T)$, for some
distinction $D$ and finite sets of names $S$ and $T$, to mean the
distinction $D\cup (S\times T) \cup (T\times S)$.
 
Following Sangiorgi~\cite{sangiorgi96acta}, we use a set of relations,
each indexed by a distinction, to define open bisimulation.

\begin{definition} 
\label{def:obisim}
The indexed set $\Sscr = \{\Sscr_D \}_D$ of process relations is an
{\em indexed
open bisimulation} if for every distinction $D$, the relation
$\Sscr_D$ is symmetric and for every 
$\theta$ that respects $D$, if $\Ppi ~ \Sscr_D ~ \Qpi$ then:
\begin{enumerate}
\item if $\one{\Ppi\theta}{\alpha}{\Ppi'}$ and $\alpha$ is a free action,
  then there is $\Qpi'$ such that
  $\one{\Qpi\theta}{\alpha}{\Qpi'}$ and $\Ppi' \Sscr_{D\theta} \Qpi'$,
\item if $\one{\Ppi\theta}{x(z)}{\Ppi'}$ and $z \not \in \n{\Ppi\theta,\Qpi\theta}$ 
  then there is $\Qpi'$ such that
  $\one{\Qpi\theta}{x(z)}{\Qpi'}$ and $\Ppi' ~\Sscr_{D\theta} ~ \Qpi'$,
\item \label{def:obisim3} if $\one{\Ppi\theta}{\bar{x}(z)}{\Ppi'}$ and $z \not \in \n{\Ppi\theta,\Qpi\theta}$ 
  then there is $\Qpi'$
  such that $\one{\Qpi\theta}{\bar{x}(z)}{\Qpi'}$ and 
  $\Ppi' ~ \Sscr_{D'} ~ \Qpi'$ where $D' = D\theta \cup 
  (\{z\} \times \fn{\Ppi\theta,\Qpi\theta, D\theta})$.
\end{enumerate}
The processes $\Ppi$ and $\Qpi$ are {\em strong open $D$-bisimilar},
written $\Ppi \sim^D_o \Qpi$, if there is an indexed open bisimulation
$\Sscr$ such that $P~\Sscr_D ~\Qpi$. The processes $\Ppi$ and $\Qpi$ are
{\em strong open bisimilar} if $\Ppi \sim_o^\emptyset  \Qpi.$
\end{definition}

Note that we strengthen a bit the condition 3 in
Definition~\ref{def:obisim} to include the distinction $(\{z\} \times
\fn{D\theta})$.  Strengthening the distinction this way does not change the
open bisimilarity, as noted in \cite{sangiorgi01}, but in our encoding
of open bisimulation, the distinction $D$ is part of the specification
and the modified definition above helps us account for names better.

\begin{figure}
{\small
$$
\ebisim{P}{Q}\defeq
\begin{array}[t]{l}
 \iforall A\iforall P'
    \begin{array}[t]{l}
      [\one{P}{A}{P'}\iimp\iexists Q'.\;\one{Q}{A}{Q'}
       \iand \ebisim{P'}{Q'}]\iand\null
    \end{array} \\
 \iforall A \iforall Q'
    \begin{array}[t]{l}
      [\one{Q}{A}{Q'}\iimp\iexists P'.\;\one{P}{A}{P'}
         \null\iand \ebisim{Q'}{P'}]\iand\null
    \end{array} \\    
 \iforall X\iforall P'
        \begin{array}[t]{l}
        [\onep{P}{\inact X}{P'}\iimp
         \forall w \iexists Q'.\;\onep{Q}{\inact X}{Q'}
          \iand \ebisim{(P'w)}{(Q' w)}]\iand \null          
    \end{array} \\    
 \iforall X\iforall Q'
        \begin{array}[t]{l}
        [\onep{Q}{\inact X}{Q'}\iimp
         \forall w  \iexists P'.\;\onep{P}{\inact X}{P'}
         \null\iand \ebisim{(Q' w)}{(P' w)}]\iand \null          
    \end{array} \\    
 \iforall X \iforall P'
    \begin{array}[t]{l}
      [\onep{P}{\outact X}{P'}\iimp\iexists Q'.\;\onep{Q}{\outact X}{Q'}
         \null\iand\nabla w. \ebisim{(P' w)}{(Q' w)}]\iand\null
    \end{array}\\
 \iforall X \iforall Q'
    \begin{array}[t]{l}
      [\onep{Q}{\outact X}{Q'}\iimp\iexists P'.\;\onep{P}{\outact X}{P'}
         \null\iand\nabla w. \ebisim{(Q' w)}{(P' w)}]\quad
    \end{array}    
\end{array}
$$
$$
\lbisim{P}{Q}\defeq
\begin{array}[t]{l}
 \iforall A\iforall P'
    \begin{array}[t]{l}
      [\one{P}{A}{P'}\iimp\iexists Q'.\;\one{Q}{A}{Q'}
         \null\iand \lbisim{P'}{Q'}]\iand\null
    \end{array} \\
 \iforall A\iforall Q'
    \begin{array}[t]{l}
      [\one{Q}{A}{Q'}\iimp\iexists P'.\;\one{P}{A}{P'}
         \iand \lbisim{Q'}{P'}]\iand\null
    \end{array} \\    
 \iforall X\iforall P'
    \begin{array}[t]{l}
      [\onep{P}{\inact X}{P'}\iimp\iexists Q'.\;\onep{Q}{\inact X}{Q'}
          \iand\iforall w. \lbisim{(P' w)}{(Q' w)}]\iand\null
    \end{array} \\
 \iforall X\iforall Q'
    \begin{array}[t]{l}
      [\onep{Q}{\inact X}{Q'}\iimp\iexists P'.\;\onep{P}{\inact X}{P'}
          \iand\iforall w.\; \lbisim{(Q' w)}{(P' w)}]\iand\null
    \end{array} \\    
 \iforall X \iforall P'
    \begin{array}[t]{l}
      [\onep{P}{\outact X}{P'}\iimp\iexists Q'.\;\onep{Q}{\outact X}{Q'}
         \iand\nabla w.\; \lbisim{(P' w)}{(Q' w)}]\iand\null
    \end{array}\\
 \iforall X \iforall Q'
    \begin{array}[t]{l}
      [\onep{Q}{\outact X}{Q'}\iimp\iexists P'.\;\onep{P}{\outact X}{P'}
         \iand\nabla w.\; \lbisim{(Q' w)}{(P' w)}]\quad
    \end{array}    
\end{array}
$$
}
\caption{Specification of strong early, {\sl ebisim}, and late, {\sl
    lbisim}, bisimulations.}
\label{bisim}
\end{figure}

Early and late bisimulation can be specified in $\FOLDNb$ using the
definition clauses in Figure~\ref{bisim}.  The definition clause for
open bisimulation is the same as the one for late bisimulation.  The
exact relationship between these definitions and the bisimulation
relations repeated above will be stated later in this section.

In reasoning about the specifications of early/late bisimulation, we encode
free names as $\nabla$-quantified variables whereas in the
specification of open bisimulation we encode free names as
$\forall$-quantified variables.  For example, the processes $P x y =
(x | \bar{y})$ and $Q x y = (x.\bar{y} + \bar{y}.x)$ are late
bisimilar. The corresponding encoding in $\FOLDNb$ would be $\nabla x
\nabla y. \lbisim{(Pxy)}{(Qxy)}$.  The free names $x$ and $y$ should
not be $\forall$-quantified for the following, simple reason: in logic
we have the implication $\forall x \forall y\ \lbisim{(P x y)}{(Q x
y)} \oimp \forall z\ \lbisim{(P z z)}{(Q z z)}$.  That is, either
$\forall x\forall y\ \lbisim{(Pxy)}{(Qxy)}$ is not provable, or it is
provable and we have a proof of $\forall z\ \lbisim{(Pzz)}{(Qzz)}$. In
either case we lose the adequacy of the encoding.

The definition clauses shown in Figure~\ref{bisim} 
do not fully capture early and late bisimulations, since there is an
implicit assumption in the definition of these bisimulations that
name equality is decidable. This basic assumption on the ability to 
decide the equality among names is one of the differences between 
open and late bisimulation.  Consider,
for example,  the processes (taken from \cite{sangiorgi96acta})
$$P = x(u).(\tau.\tau + \tau) \qquad \hbox{ and } \qquad
Q = x(u).(\tau.\tau + \tau + \tau.[u = z]\tau).$$
As shown in \cite{sangiorgi96acta} $P$ and $Q$ are late bisimilar but
not open bisimilar: establishing late bisimulation makes use 
of a case analysis that depends on
whether the input name $u$ is equal to $z$ or not.  
Decidability of name equality, in the case of early and late bisimulation,
is encoded as an additional axiom of excluded middle on names, i.e., 
the formula $\forall x\forall y (x = y \lor x \not = y)$.
Note that since we allow dynamic creation of scoped names (via
$\nabla$), we must also state this axiom for arbitrary extensions of
local signatures.  The following set collects together such 
generalized excluded middle formulas:
$$
\Escr = \{\nabla n_1\cdots\nabla n_k\forall x\forall y (x = y\lor x\not = y)\mid k\geq 0\}.
$$
We shall write $\Xscr \subseteq_f \Escr$ to indicate that $\Xscr$ is a
finite subset of $\Escr$.

The following theorem states the soundness and completeness of the
$\hbox{\sl ebisim}$ and $\hbox{\sl lbisim}$ specifications with respect to the notions of early and 
late bisimilarity in the $\pi$-calculus. By soundness we mean that, given a
pair of processes $\Ppi$ and $\Qpi$, if the encoding of the late (early)
bisimilarity is provable in $\FOLDNb$ then the processes $\Ppi$ and $\Qpi$ are late (early) bisimilar.
Completeness is the converse.  The soundness and completeness of the
open bisimilarity encoding is presented at the end of this section, where we
consider the encoding of the notion of distinction in the $\pi$-calculus.

\begin{theorem}
\label{thm:lbisim}
Let $\Ppi$ and $\Qpi$ be two processes and let $\bar{n}$ be
the free names in $\Ppi$ and $\Qpi$. 
Then $\Ppi \sim_l \Qpi$ if and only if the sequent
$\NSeq{.}{\Xscr}{\nabla \bar{n}. \lbisim \Ppi \Qpi}$
is provable for some $\Xscr \subseteq_f \Escr$.
\end{theorem}

\begin{theorem}
\label{thm:ebisim}
Let $\Ppi$ and $\Qpi$ be two processes and let $\bar{n}$ be
the free names in $\Ppi$ and $\Qpi$. 
Then $\Ppi \sim_e \Qpi$ if and only if the sequent
$\NSeq{.}{\Xscr}{\nabla \bar{n}. \ebisim \Ppi \Qpi}$
is provable for some $\Xscr \subseteq_f \Escr$.
\end{theorem}

It is well-known that the late bisimulation relation is
not a congruence since it is not preserved by the input prefix.
Part of the reason why the congruence property fails 
is that in the late bisimilarity there is no syntactic distinction
made between names which can be instantiated and names which cannot be
instantiated. 
Addressing this difference between names
is one of the motivations behind the introduction 
of distinctions and open bisimulation. 
There is another important
difference between open and late bisimulation; in open bisimulation
names are instantiated {\em lazily}, i.e., only when needed. 
The lazy instantiation of
names is intrinsic in $\FOLDNb$; eigenvariables are instantiated
only when applying the $\defL$-rule. 
The syntactic distinction between names that can be instantiated and 
those that cannot be instantiated are reflected in $\FOLDNb$ by the difference between 
the quantifier $\forall$ and $\nabla$. 
The alternation of quantifiers in $\FOLDNb$ gives rise to
a particular kind of distinction, the precise definition of which
is given below.

\begin{definition}
A {\em quantifier prefix} 
is a list $\Qscr_1 x_1 \Qscr_2 x_2\, \dots \, \Qscr_n x_n$ for some $n \geq 0$,
where $\Qscr_i$ is either $\nabla$ or $\forall$.
If $\Qscr\bar{x}$ is the above quantifier prefix, then the
{\em $\Qscr\bar{x}$-distinction} is the distinction
$$
\{ (x_i, x_j), (x_j, x_i) \mid i \not = j \mbox{ and } \Qscr_i = \Qscr_j = \nabla, \mbox{ or }
    i < j \mbox{ and } \Qscr_i = \forall \mbox{ and } \Qscr_j = \nabla \}.
$$
\end{definition}
Notice that if $\Qscr\bar{x}$ consists only of universal quantifiers
then the $\Qscr\bar{x}$-distinction is empty.  Obviously, the
alternation of quantifiers does not capture all possible distinction,
{\em e.g.}, the distinction 
$$\{(x,y), (y,x), (x,z), (z,x), (u,z),(z,u) \}$$
does not correspond to any quantifier prefix.  However, we can encode
the full notion of distinction by an explicit encoding of the unequal
pairs, as shown later.

It is interesting to see the effect of  substitutions on $D$
when $D$ corresponds to a prefix $\Qscr\bar{x}$. 
Suppose $\Qscr\bar{x}$ is the prefix 
$\Qscr_1\bar{u}\forall x\Qscr_2\bar{v}\forall y\Qscr_3\bar{w}.$
Since any two $\forall$-quantified
variables are not made distinct in the definition of $\Qscr\bar{x}$ prefix,
there is a $\theta$ which respects $D$ and which can identify $x$ and $y$.
Applying $\theta$ to $D$ changes 
$D$ to some $D'$ which corresponds to the prefix 
$\Qscr_1\bar{u}\forall z\Qscr_2\bar{v}\Qscr_3\bar{w}$. 
Interestingly, these two prefixes are related by logical implication:
$$
\Qscr_1\bar{u}\forall x\Qscr_2\bar{v}\forall y\Qscr_3\bar{w}.P
\oimp
\Qscr_1\bar{u}\forall z\Qscr_2\bar{v}\Qscr_3\bar{w}.P[z/x, z/y]
$$
for any formula $P$. This observation suggests the following lemma.

\begin{lemma}
\label{lm:prefix}
Let $D$ be a $\Qscr\bar{x}$-distinction and let $\theta$
be a $D$-substitution. Then the distinction $D\theta$
corresponds to some prefix $\Qscr'\bar{y}$ such that
$\Qscr\bar{x}.P \oimp \Qscr'\bar{y}.P\theta$ for any 
formula $P$ such that $\fv{P} \subseteq \{ \bar x\}$.
\end{lemma}

\begin{definition}
Let $D = \{ (x_1,y_1), \dots, (x_n, y_n) \}$ be a distinction.  The
distinction $D$ is translated as the formula $\llbra D\rrbra = x_1
\not = y_1 \land \dots \land x_n \not = y_n$.  If $n = 0$ then $\llbra
D \rrbra$ is the logical constant $\top$ (the empty conjunction).
\end{definition}

\begin{theorem}
\label{thm:open bisim sound}
Let $\Ppi$ and $\Qpi$ be two processes, let $D$ be a distinction and
let $\Qscr \bar{x}$ be a quantifier prefix, where $\bar{x}$ contains
the free names in $\Ppi, \Qpi$ and $D$.
If the formula 
$
\Qscr \bar{x}. (\llbra D \rrbra \oimp \lbisim \Ppi \Qpi)
$
is provable then $\Ppi \sim^{D'}_o \Qpi$, where $D'$ is the union of
$D$ and the $\Qscr\bar{x}$-distinction.
\end{theorem}

\begin{theorem}
\label{thm:open bisim complete}
If $\Ppi \sim^D_o \Qpi$ then the formula
$\forall \bar{x}. \llbra D \rrbra \oimp \lbisim \Ppi \Qpi$ is provable, where $\bar{x}$
are the free names in $\Ppi, \Qpi$ and $D$.
\end{theorem}

If a distinction $D$ corresponds to a quantifier prefix $\Qscr \vec x$, 
then it is easy to show that $\Qscr \vec x.\llbra D \rrbra$ is derivable in $\FOLDNb.$
Therefore, we can state more concisely the adequacy result for the class of 
$D$-open bisimulations in which $D$ corresponds to a quantifier
prefix.  The following
corollary follows from Theorem~\ref{thm:open bisim sound}, 
Theorem~\ref{thm:open bisim complete} and Proposition~\ref{prop:forall nabla}. 

\begin{corollary}
\label{cor:open bisim prefix}
Let $D$ be a distinction, let $\Ppi$ and $\Qpi$ be two processes and let $\Qscr \vec x$
be a quantifier prefix such that $\vec x$ contains the free names of $D$, $\Ppi$ and $\Qpi$,
and $D$ corresponds to the $\Qscr \vec x$-distinction. Then
$\Ppi \sim^D_o \Qpi$ if and only if $\vdash \Qscr \vec x. \lbisim \Ppi \Qpi.$
\end{corollary}

Note that, by Lemma~\ref{lm:prefix}, the property of being a quantifier-prefix distinction
is closed under $D$-substitution. Note also that in Definition~\ref{def:obisim}(\ref{def:obisim3}),
if $D\theta$ is a quantifier-prefix distinction then so is 
$$
D' = D\theta \cup 
  (\{z\} \times \fn{\Ppi\theta,\Qpi\theta, D\theta}).
$$
That is, if $D\theta$ corresponds to a quantifier prefix $\Qscr \vec x$,
then $D'$ corresponds to the quantifier prefix $\Qscr \vec x \nabla z.$
Taken together, these facts imply that one can define 
an open bisimulation relation which is indexed only by 
quantifier-prefix distinctions. That is, the family of relations $\{\Sscr_D \}_D$,
where each $D$ is a quantifier-prefixed distinction and each $\Sscr_D$ is defined as
$$
\Sscr_D = \{(\Ppi,\Qpi) \mid \Ppi \sim^D_o \Qpi \},
$$
is an indexed open bisimulation.

Notice the absence of the excluded middle assumption on names in the
specification of open bisimulation. Since $\FOLDNb$ is
intuitionistic, this difference between late and open bisimulation is
easily observed.  This would not be the case if the specification logic were
classical.  Since the axiom of excluded middle is present as well in
the specification of early bisimulation (Theorem~\ref{thm:ebisim}),
one might naturally wonder if there is a meaningful notion of
bisimulation obtained from removing the excluded middle in the
specification of early bisimulation and $\forall$-quantify the free names.
In other words, we would like to see if there is a
notion of ``open-early'' bisimulation. In fact, 
the resulting bisimulation relation is exactly the same as open ``late''
bisimulation.

\begin{theorem}
\label{thm:open-early-bisim}
Let $\Ppi$ and $\Qpi$ be two processes and let $\bar n$ be the 
free names in $\Ppi$ and $\Qpi$. Then 
$\forall \bar n. \lbisim {\Ppi}{\Qpi}$ is provable if and only if
$\forall \bar n. \ebisim{\Ppi}{\Qpi}$ is provable.
\end{theorem}

We note that while it is possible to prove the impossibility
of transitions (Proposition~\ref{prop:neg one step}) within $\FOLDNb$, 
it is in general not the case with non-bisimilarity (which is 
not even recursively enumerable in the infinite setting). 
If we have evidence that two processes are not
bisimilar, say, because one has a trace that the other does not have,
then this trace information can be used in the proof a
non-bisimulation.  Probably a good approach to this is to rely on
the modal logics developed later in the paper: if processes are not
bisimilar, there is an assertion formula that separates them.  We
have not planned to develop this particular theme since it seems to
us to not be the main thrust of this paper: describing proofs of
non-bisimilarity in the finite pi-calculus case is an interesting thing 
that could be developed on top of the foundation we provide.

To conclude this section, we should explicitly compare 
the two specifications of early bisimulation in Definition~\ref{def:ebisim}
and in Theorem~\ref{thm:ebisim},
the two specifications of late bisimulation in Definition~\ref{def:lbisim} and
in Theorem~\ref{thm:lbisim} and
the two specifications of open bisimulation in Definition~\ref{def:obisim}
and in Corollary~\ref{cor:open bisim prefix}.
Notice that those specifications that rely on logic are
written without the need for any explicit conditions on variable names
or any need to mention distinctions explicitly.  These various
conditions are, of course, present in the detailed description of the
proof theory of our logic, but it seems desirable to push
the details of variable names, substitutions, free and
bound-occurrence, and equalities into logic, where they have elegant
and standard solutions.

\section{Specification of modal logics}
\label{sec:modal}

\begin{figure}[t]
(a) Propositional connectives and {\em basic} modality:
$$
\begin{array}{ll@{\ \null\defeq\null\ }l}
(\mTrue:) & \stf P \mTrue  & \top. \\
(\hbox{and}:) & \stf P {A \mAnd B}  &  \stf P A \land \stf P B.\\
(\hbox{or}:) & \stf P {A \mOr B}  &  \stf P A \lor \stf P B.\\
(\hbox{match}:)&\stf P {\matchDia X X A} & \stf P A. \\
(\hbox{match}:)&\stf P {\matchBox X Y A} & (X = Y) \oimp \stf P A.\\
(\hbox{free}:) &\stf P {\actDia{X} A}&\exists P'(\one{P}{X}{P'}\land \stf{P'}{A}).\\
(\hbox{free}:) &\stf P {\actBox{X} A}&\forall P'(\one{P}{X}{P'}\oimp \stf{P'}{A}).\\
(\hbox{out}:)  &\stf P {\outDia X A} &\exists P'(\onep{P}{\outact X}{P'} \land 
                                      \nabla y.\stf {P'y} {A y}).\\
(\hbox{out}:) & \stf P {\outBox X A}  &  
       \forall P'(\onep{P}{\outact X}{P'} \oimp \nabla y.\stf {P' y}{A y}).\\
(\hbox{in}:) & \stf P {\inDia X A}  &  \exists P'(\onep{P}{\inact X}{P'} \land 
                 \exists y.\stf {P'y} {A y}).\\
(\hbox{in}:) & \stf P {\inBox X A} & \forall P'(\onep{P}{\inact X}{P'} \oimp 
                       \forall y. \stf {P' y}{A y}).
\end{array}
$$
(b) {\em Late} modality:
$
\begin{array}{l@{\ \null\defeq\null\ }l}
\stf P {\inDiaL X A} & \exists P'(\onep{P}{\inact X}{P'} \land \forall y. \stf {P'y} {A y}).\\
\stf P {\inBoxL X A} & \forall P'(\onep{P}{\inact X}{P'} \oimp \exists y.\stf {P' y}{A y}).
\end{array}
$

\vskip11pt
(c) {\em Early} modality:
$
\begin{array}{l@{\ \null\defeq\null\ }l}
\stf P {\inDiaE X A} & \forall y\exists P'(\onep{P}{\inact X}{P'} \land \stf {P'y} {A y}).\\
\stf P {\inBoxE X A} & \exists y\forall P'(\onep{P}{\inact X}{P'} \oimp \stf {P' y}{A y}).
\end{array}
$
\caption{Modal logics for the $\pi$-calculus in $\lambda$-tree syntax}
\label{fig:modal}
\end{figure}

We now present the modal logics for the $\pi$-calculus that were introduced
in \cite{milner93tcs}.  
In order not to confuse meta-level ($\FOLDNb$) formulas (or connectives) with
the formulas (connectives) of the modal logics under consideration, we shall
refer to the latter as object formulas (respectively, object connectives). 
We shall work only with positive object formulas, i.e., we do not permit
negations in those formulas. Note that since there are no atomic formulas
in these modal logics (in particular, $\mTrue$ or $\mFalse$ are not atomic),
de Morgan identities can be used to remove all occurrences of
negations from such formulas.
The syntax of the object formulas is as follows. 
$$
\begin{array}{ll}
\Api ::= & \mTrue ~\mid ~ \mFalse ~ \mid ~  \Api \land \Api ~ \mid ~ \Api \lor \Api 
 ~ \mid ~ [x = z] \Api
 ~ \mid ~ \langle x = z \rangle \Api\\
 & \mid ~ \actDia \alpha \Api ~ \mid ~ \actBox \alpha \Api 
 ~ \mid ~ \mDia {\bar x(y)} \Api
 ~ \mid ~ \mBox {\bar x(y)} \Api 
 ~ \mid ~ \mDia {x(y)} \Api ~ \mid ~ \mBox {x(y)} \Api \\
 & \mid ~ \mDia {x(y)}^L \Api ~ \mid ~ \mBox {x(y)}^L \Api
 ~ \mid ~ \mDia {x(y)}^E \Api ~ \mid ~ \mBox {x(y)}^E \Api
\end{array}
$$
The symbol $\alpha$ denotes a free action, i.e., a free input, a free
output, or the silent action.
In each of the formulas 
$\mDia{\bar x(y)}\Api$, $\mDia{x(y)}\Api$, $\mDia{x(y)}^L \Api$ and
$\mDia{x(y)}^E \Api$ (and their dual `boxed'-formulas), the occurrence of $y$ in parentheses is
a binding occurrence whose scope is $\Api$.
We use $\Api$, $\Bpi$, $\Cpi$, $\Dpi$ %, possibly with subscripts or primes,
to range over object formulas.
Note that we consider only finite conjunctions since the transition system
we are considering is finitely branching, and, therefore, an infinite
conjunction is not needed (as noted in \cite{milner93tcs}).
We consider object formulas equivalent up to renaming of bound variables. 

To encode object formulas we introduce the 
type $o'$ to denote such formulas and introduce the
following constants for encoding the object connectives:
$\mTrue$ and $\mFalse$ of type $o'$; 
$\mAnd$ and $\mOr$ of type $o' \ra o' \ra o'$; 
$\matchDia{\cdot^1}{\cdot^2}{\cdot^3}$ and $\matchBox{\cdot^1}{\cdot^2}{\cdot^3}$
of type $\name \ra \name \ra o' \ra o'$; 
$\actDia{\cdot^1}{\cdot^2}$ and $\actBox{\cdot^1}{\cdot^2}$ of type $\action
\ra o' \ra o'$; and
$\outDia{\cdot^1}{\cdot^2}$,
$\outBox{\cdot^1}{\cdot^2}$,
$\inDia{\cdot^1}{\cdot^2}$, 
$\inBox{\cdot^1}{\cdot^2}$, 
$\inDiaL{\cdot^1}{\cdot^2}$, 
$\inBoxL{\cdot^1}{\cdot^2}$, 
$\inDiaE{\cdot^1}{\cdot^2}$, and
$\inBoxE{\cdot^1}{\cdot^2}$ of type $\name\ra(\name\ra o')\ra o'$.
The translation of object formulas to $\lambda$-tree syntax is
given in the following definition.

\begin{definition}
The following function $\trans{.}$ translates object formulas 
to $\beta\eta$-long normal terms of type $o'$. 
$$
\begin{array}{l@{\qquad}l}
\trans{\mTrue} = \mTrue & \trans{\mFalse} = \mFalse\\
\trans{\Api \land \Bpi} = \trans{\Api} \mAnd \trans{\Bpi} & 
\trans{\Api \lor \Bpi} = \trans{\Api} \mOr \trans{\Bpi}\\
\trans{[x = y] \Api} = \matchBox x y {\trans{\Api}} &
\trans{\mDia{x = y} \Api} = \matchDia x  y {\trans \Api}\\
\trans{\actDia{\alpha} \Api} = \actDia \alpha {\trans \Api} &
\trans{\actBox{\alpha} \Api} = \actBox \alpha {\trans \Api}\\
\trans{\mDia{\bar x(y)}{\Api}} = \outDia x {(\lambda y\trans \Api)} &
\trans{\mBox{\bar x(y)}{\Api}} = \outBox x {(\lambda y\trans \Api)} \\
\trans{\mDia{x(y)}{\Api}} = \inDia x {(\lambda y\trans \Api)} &
\trans{\mBox{x(y)}{\Api}} = \inBox x {(\lambda y\trans \Api)}\\
\trans{\mDia{x(y)}^L {\Api}} = \inDiaL x {(\lambda y\trans \Api)} &
\trans{\mBox{x(y)}^L {\Api}} = \inBoxL x {(\lambda y\trans \Api)}\\
\trans{\mDia{x(y)}^E {\Api}} = \inDiaE x {(\lambda y\trans \Api)} &
\trans{\mBox{x(y)}^E {\Api}} = \inBoxE x {(\lambda y\trans \Api)}
\end{array}
$$
\end{definition}

In specifying the satisfaction relation $\models$ between processes and
formulas, we restrict to the class of formulas which do not contain
occurrences of the free input modality. This is because we consider only the late transition system
and the semantics of the free input modality is defined with respect to the early
transition system.
But we note that adding this input modality and the early transition system 
does not pose any difficulty. Following Milner et. al., we shall identify
an object logic with the set of formulas it allows. We shall refer
to the object logic without the free input modalities as $\Ascr^-$. 

The satisfaction relation $\models$ is encoded using the same symbol, 
which is given the type $\proc \ra o' \ra o$. This satisfaction
relation is defined by the clauses in Figure~\ref{fig:modal}.
This definition, called $\Dscr\Ascr^-$, 
corresponds to the modal logic $\Ascr$ defined in \cite{milner93tcs}, minus
the clauses for the free input modality.  
Notice that $\Dscr\Ascr^-$ interprets object-level disjunction and conjunction
with, respectively, meta-level disjunction and conjunction. Since the modal logic $\Ascr^-$
is classical and the meta-logic $\FOLDNb$ is intuitionistic, 
one may wonder whether such an encoding is complete. But since 
we consider only negation-free object formulas and since 
there are no atomic formulas, classical and intuitionistic provability
coincide for the non-modal fragment of $\Ascr^-$.
The definition $\Dscr\Ascr^-$ is, however, incomplete for the full logic $\Ascr^-$, 
in the sense that there are true assertions of modal logics that are not provable 
using this definition alone. Using the `box' modality, one can still encode 
some limited forms of negation, e.g., inequality of names. For instance, the modal judgment 
$$
\stf {x(y).x(z).0} {\mDia{x(y)} \mDia{x(z)} (\actDia{x = z} \mTrue ~ \mOr ~ \actBox{x = z} \mFalse)},
$$
which essentially asserts that any two names are equal or unequal, is valid in $\Ascr$, 
but its encoding in $\FOLDNb$ is not provable without additional assumptions. 
It turns out that, as in the case with the specification of late bisimulation,
the only assumption we need to assure completeness is the 
axiom of excluded middle on the equality of names: 
$
\forall x\forall y. x = y \lor x \not = y. 
$
Again, as in the specification of late bisimulation, we must also
state this axiom for arbitrary extensions of local signatures. 
The adequacy of the specification of modal logics is 
stated in the following theorem.

\begin{theorem}
\label{thm:modal adequacy}
Let $\Ppi$ be a process, let $\Api$ be an object formula of the modal logic $\Ascr^-$. 
Then $\Ppi \models \Api$ if and only if for some
list $\bar n$ such that $\fn{\Ppi, \Api} \subseteq \{\bar n \} $
and some  $\Xscr \subseteq_f \Escr$,  
the sequent 
$\Seq{\Xscr}{\nabla \bar n.(\trans{\Ppi} \models \trans{\Api})}$
is provable in $\FOLDNb$ with definition $\Dscr\Ascr^-$.
\end{theorem}

The adequacy result stated in Theorem~\ref{thm:modal adequacy}
subsumes the adequacy for the specifications of the sublogics of $\Ascr^-$.
Note that we quantify free names in the process-formula pair
in the above theorem since we do not assume any constants
of type $\name$. Of course, such constants can be introduced
without affecting the provability of the satisfaction 
judgments, but for simplicity, we repeat our treatment of names in the
late bisimulation setting here as well.

Notice that the list of names $\bar n$ in Theorem~\ref{thm:modal adequacy}
can contain more than just the free names of $\Ppi$ and $\Api.$
This is important for the adequacy of the specification, since in the modal
logics for the $\pi$-calculus, we can specify a modal formula $\Api$ and a process 
$\Ppi$ such that the assertion $\Ppi \models \Api$ is true only if 
there exists a new name which is not among the free names of both $\Ppi$
and $\Api$. Consider, for example, the assertion
$$
a(x).0 \models \mBox{a(x)}^L \mBox{x = a} \mFalse
$$
and its encoding in $\FOLDNb$ as the formula
$$
\inpi{a}{(\lambda x.0)} \models \inBoxL a {(\lambda x. \matchBox x a \mFalse)}.
$$
If we do not allow extra new names in the quantifier prefix
in Theorem~\ref{thm:modal adequacy}, then we would have to
prove the formula
$$
\nabla a.\big(\inpi{a}{(\lambda x.0)}\models\inBoxL a{(\lambda x.\matchBox x a\mFalse)}\big).
$$
It is easy to see that provability of this formula reduces to
provability of
$$
\nabla a \exists x.\big(0 \models \matchBox x a \mFalse\big).
$$
Since we do not assume any constants of type \name, the only way to prove
this would be to instantiate $x$ with $a$, hence, 
$$
\nabla a. (0 \models \matchBox a a \mFalse)\hbox{\quad and\quad}
\nabla a. (a = a) \oimp 0 \models \mFalse.
$$
must be provable.  This is, in turn, equivalent to $\nabla a. 0 \models
\mFalse$ which should not be 
provable for the adequacy result to hold. 
The key step here is the instantiation of $\exists x$. For the original formula
to be provable, $x$ has to be instantiated with a name that is
distinct from $a$.  This can be done only if we allow extra names
in the quantifier prefix: for example, the following formula is provable.
$$
\nabla a\nabla b.
  \big(\inpi{a}{(\lambda x.0)} \models \inBoxL a {(\lambda x. \matchBox x a \mFalse)}
  \big)
$$

Note that in the statement of Theorem~\ref{thm:modal adequacy}, the list of names $\bar n$
is existentially quantified. If one is to implement model checking for $\Ascr^-$ using
the specification in Figure~\ref{fig:modal}, the issue of how these names are chosen needs to
be addressed. Obviously, the free names of $\fn{\Ppi,\Api}$ needs to be among $\bar n$.
It remains to calculate how many new names need to be added.
An inspection on the definition in Figure~\ref{fig:modal} shows that such 
new names may be needed only when bound input modalities are present in the modal formula.
More specifically, when instantiating the name quantification ($\forall$ or $\exists$) in 
a definition clause for a bound input modality, such as in the definition clause
$$
\stf P {\inDiaL X A} \defeq \exists P'(\onep{P}{\inact X}{P'} \land \forall y. \stf {P'y} {A y}),
$$
we need to consider only cases where $y$ is instantiated to a free name in $\stf P {\inDiaL X A}$,
and where $y$ is instantiated to a new name. For the latter, the particular choice of the new name 
is unimportant, since the satisfiability relation for $\Ascr^-$ is closed under substitution 
with new names (cf. Lemma 3.4. in \cite{milner93tcs}). 
One can thus calculate the number of new names needed based on the number of bound 
input modalities in $\Api.$

In \cite{milner93tcs}, late bisimulation was characterized by the
sublogic $\Lscr\Mscr$ of $\Ascr^-$ that arises from restricting the
formulas to contain only the propositional connectives and the
following modalities: $\langle \tau \rangle$, $\langle \bar x y
\rangle$, $\langle \bar x(y) \rangle$, $[x = y]$, $\langle
x(y)\rangle^L$, and their duals.  We shall now show a similar 
characterization for open bisimulation.

The following theorem states that by dropping the excluded middle and
changing the quantification of free names from $\nabla$ to $\forall$,
we get exactly a characterization of open bisimulation by the encoding
of the sublogic $\Lscr\Mscr$.

\begin{theorem}
\label{thm:ch-open}
Let $\Ppi$ and $\Qpi$ be two processes.
Then $\Ppi \sim_o^{\emptyset} \Qpi$ if and only if
for every $\Lscr \Mscr$-formula $\Api$,
it holds that $\vdash \forall \bar n(\trans{\Ppi} \models \trans{\Api})$
if and only if 
$\vdash \forall \bar n(\trans{\Qpi} \models \trans{\Api})$,
where $\bar n$ is the list of free names in $\Ppi$, $\Qpi$ and $\Api$.
\end{theorem}

\section{Allowing replication in process expressions}
\label{sec:pi rep}

We now consider an extension to the finite $\pi$-calculus which will allow
us to represent non-terminating processes. 
There are at least two ways to encode non-terminating processes in the
$\pi$-calculus; {\em e.g.}, via recursive definitions or
replications \cite{sangiorgi01}. We consider here the latter approach since it leads to
a simpler presentation of the operational semantics. 
To the syntax of the finite $\pi$-calculus we add the process expression
$!P$. The process $!P$ can be understood as the infinite
parallel composition of $P$, i.e., $P | P | \cdots | P | \cdots$.
Thus it is possible to have a process 
that retains a copy of itself after making a transition;
{\em e.g.}, $\one{!P}{\alpha}{P' | !P}$. The operational semantics for one-step
transitions of the $\pi$-calculus with replication 
is given as the definition clauses Figure~\ref{fig:rep-pi def}, 
adapted to the $\lambda$-tree syntax from the original presentation in \cite{sangiorgi01}. 
We use the same symbol to encode replication in $\lambda$-tree syntax,
{\em i.e.}, $! : \proc \ra \proc$.

\begin{figure}
$$
\begin{array}{l}
\one{!P}{A}{P' \barpi  !P} \defeq \one{P}{A}{P'} \\
\onep{!P}{X}{\lambda y(M y \barpi  !P)} \defeq \onep{P}{X}{M} \\
\one{!P}{\tau}{(P' \barpi  M~Y) \barpi !P} \defeq \exists X. \one{P}{\outact X Y}{P'} \land 
     \onep{P}{\inact X}{M} \\
\one{!P}{\tau}{\nupi{z}{(Mz \barpi  N z)} \barpi  !P} \defeq
\exists X. \onep{P}{\outact X}{M} \land \onep{P}{\inact X}{N}
\end{array}
$$
\caption{Definition clauses for the $\pi$-calculus with replication}
\label{fig:rep-pi def}
\end{figure}

In order to reason about bisimulation of processes involving $!$, we need to move
to a stronger logic which incorporates both induction and co-induction
proof rules. We consider the logic $\Linc$ \cite{tiu04phd}, which is an extension of
$\FOLDNb$ with induction and co-induction proof rules. 
We first need to extend the notion of definitions to include inductive
and co-inductive definitions. 

\begin{definition}
An inductive definition clause is written
$$
\forall \bar x. p \bar x \defmu B~ p~\bar x
$$
where $B$ is a closed term. The symbol $\defmu$ is used to indicate
that the definition is inductive. 
Similarly, a co-inductive definition clause is written
$$
\forall \bar x. p \bar x \defnu B~ p~\bar x.
$$
The notion of definition given in Definition~\ref{def:def}
shall be referred to as {\em basic definition}.
An {\em extended definition} is a collection of 
basic, inductive, or co-inductive definition clauses.
\end{definition}
A definition clause can be seen as a fixed point 
equation: in fact, Baldle \& Miller \citeyear{baelde07lpar} provide an
alternative approach to inductive and co-inductive definitions similar
to what is available in the $\mu$-calculus.
When definitions are seen as fixed points, provability of
$p\,\bar t$, depending on whether $p$ is basic, inductive or co-inductive, 
means that $\bar t$ is, respectively, in a fixed point, the least fixed point, and 
the greatest fixed point of the underlying fixed point equation defining $p$.

Notice that the head of the (co-)inductive definition clauses contains
a predicate with arguments that are only variables and not more
general terms: this restriction simplifies the presentation
of the induction and co-induction inference rules.  Arguments that are more
general terms can
be encoded as explicit equalities in the body of the clause.
We also adopt a higher-order notation in describing the body
of clauses, {\em i.e.}, we use $B~p~\bar x$ to mean that $B$ is a
top-level abstraction that has
no free occurrences of the predicate symbol $p$ and the variables $\bar x$.
This notation simplifies the presentation of the (co-)induction
rules: in particular, it simplifies the presentation of predicate
substitutions.

There must be some stratification on the extended definition
so as not to introduce inconsistency into the logic.
For the details of such stratification we refer the interested 
readers to \cite{tiu04phd}. For our current purpose, it should be
sufficient to understand that mutual recursive (co-)inductive definitions are
not allowed, and dependencies through negation are forbidden
as it already is in basic definitions.

Let $p\bar x \defmu B~p~\bar x$ be an inductive definition. Its left and
right introduction rules are
$$
\infer[\indL]
{\NSeq{\Sigma}{\Judg{\bar z}{p~\bar t}, \Gamma}{\Cscr}}
{
\NSeq{\bar x}{B~S~\bar x}{S~\bar x}
&
\NSeq{\Sigma}{\Judg{\bar z}{S~\bar t}, \Gamma}{\Cscr}
}
\qquad
\infer[\indR]
{\NSeq{\Sigma}{\Gamma}{\Judg{\bar z}{p~\bar t}}}
{\NSeq{\Sigma}{\Gamma}{\Judg{\bar z}{B~p~\bar t}}}
$$
where $S$ is the {\em induction invariant}, and it is a closed term 
of the same type as $p$. 
The introduction rules for co-inductively defined predicates are dual
to the inductive ones. In this case, we suppose that $p$ is 
defined by the co-inductive clause $p \bar x \defnu B~p~\bar x$.
$$
\infer[\coindL]
{\NSeq{\Sigma}{\Judg{\bar z}{p~\bar t},\Gamma}{\Cscr}}
{
\NSeq{\Sigma}{\Judg{\bar z}{B~p~\bar t}, \Gamma}{\Cscr}
}
\qquad
\infer[\coindR]
{\NSeq{\Sigma}{\Gamma}{\Judg{\bar z}{p~\bar t}}}
{
\NSeq{\Sigma}{\Gamma}{\Judg{\bar z}{S~\bar t}}
&
\NSeq{\bar x}{S~\bar x}{B~S~\bar x}
}
$$
Here $S$ is a closed term denoting the co-induction invariant
or {\em simulation}. 
Induction rules cannot be applied to co-inductive predicates and
vice versa. The $\defR$ and $\defL$ rules, strictly speaking,
are applicable only to basic definitions. But as it is shown
in \cite{tiu04phd}, these rules are derivable for (co-)inductive
definitions: that is, for these definitions, $\defR$ can be shown to be 
a special case of $\coindR$ and $\defL$ a special case of $\indL$.

The definitions in $\FOLDNb$ we have seen so far can be carried over to
$\Linc$ with some minor bureaucratic changes: {\em e.g.}, in the case of bisimulations, 
we now need to indicate explicitly that it is a co-inductive definition.
For instance, the definition of $\hbox{lbisim}$ should now be indicated as 
a co-inductive definition by changing the symbol $\defeq$ with $\defnu$.
We shall now present an example of proving bisimulation using explicit
induction and co-induction rules. We shall not go into details of the technical 
theorems of the adequacy results: these can be found in \cite{tiu04phd}.

\begin{example}
Let $\Ppi = !(z)(\bar{z}a \barpi  z(y).\bar{x}y)$ and $\Qpi = !\tau.\bar{x}a$.
The only action $\Ppi$ can make is the silent action $\tau$ since
the channel $z$ is restricted internally within the process.
It is easy to see that $\one{\Ppi}{\tau}{(z)(0 \barpi \bar{x}a) \barpi  \Ppi}$.
That is, the continuation of $P$ is capable of outputting a free name $a$
or making a silent transition.
Obviously $\Qpi$ can make the same $\tau$ action and results in 
a bisimilar continuation.
Let us try to prove $\lbisim \Ppi \Qpi$.
The simple proof strategy of unfolding the
$\hbox{lbisim}$ clause via $\defR$ will not work here since
after the first $\defR$ on $\hbox{lbisim}$ 
(but before the second $\defR$ on $\hbox{lbisim}$) 
we arrive at the sequent 
$
\lbisim{((z)(0\barpi \bar{x}a) \barpi  \Ppi)}{(\bar{x}a \barpi  \Qpi).}
$
Since $\Ppi$ and $\Qpi$ still occur in the continuation pair,
it is obvious that this strategy is non terminating. 
We need to use the co-induction proof rules instead.

An informal proof starts by 
finding a bisimulation (a set of pairs of processes) $\Sscr$ such that
$(\Ppi,\Qpi) \in \Sscr$. 
Let 
$$
\begin{array}{ll}
\Sscr' = \{(\Rpi_1\barpi \cdots\barpi \Rpi_n\barpi \Ppi, ~ \Tpi_1\barpi \cdots\barpi \Tpi_n\barpi \Qpi) 
    \mid & n \geq 0, \hbox{ $\Rpi_i$ is $(z)(0 \barpi  \bar{x}a)$ or $(z)(0 \barpi  0)$ } \\
 & \hbox{ and $\Tpi_i$ is either $\bar{x}a$ or $0$} \}.
\end{array}
$$
Define $\Sscr$ to be the symmetric closure of $\Sscr'$. 
It can be verified that $\Sscr$ is a bisimulation set by showing the set is
closed with respect to one-step transitions. To prove this formally
in $\Linc$ we need to represent the set $\Sscr$.
We code the set $\Sscr$ as the following inductive definition
(we allow ourselves to put general terms in the head of this
definition and to have more than one clause: it is straightforward to
translate this definition to the restricted one give above).
\newcommand\invar[2]{{\hbox{\sl inv}~#1~#2}}
$$
\begin{array}{rcl}
\multicolumn{3}{c}{\invar \Ppi \Qpi  \defmu  \top. \quad \invar \Qpi \Ppi \defmu \top.}\\
\invar{((z)(0\barpi 0)\barpi M)}{(0\barpi N)} & \defmu & \invar M N. \\
\invar{(0\barpi N)}{((z)(0\barpi 0)\barpi M)} & \defmu & \invar N M.\\ 
\invar{((z)(0\barpi \bar{x}a) \barpi  M)}{(\bar{x}a \barpi  N)} & \defmu & \invar M N. \\
\invar{(\bar{x}a \barpi  N)}{((z)(0\barpi \bar{x}a) \barpi  M)} & \defmu & \invar N M. 
\end{array}
$$
Note that for simplicity of presentation, we assume that we have two
constants of type $\name$, namely, $x$ and $a$, in the logic (but we note that
this assumption is not necessary). 
The set of pairs encoded by {\sl inv} can be shown to be symmetric,
i.e., the formula $\forall R\forall T.\invar{R}{T} \oimp \invar{T}{R}$ is provable
inductively (using the same formula as the induction invariant).

To now prove the sequent $\Seq{}{\lbisim{\Ppi}{\Qpi}}$, we can use 
the $\coindR$ rule with the predicate {\sl inv} as the invariant.  The
premises of the $\coindR$ rule are the two sequents $\Seq{}{\invar \Ppi \Qpi}$ and 
$\NSeq{R,T}{\invar R T}	{B\,R\,T}$, where $B\,R\,T$ is the following
large conjunction
$$
	 \begin{array}[t]{l}
	 \iforall A\iforall R'
	    \begin{array}[t]{l}
	      [(\one{R}{A}{R'})\iimp\iexists T'.(\one{T}{A}{T'})
	         \null\iand \invar{R'}{T'}]\iand\null
	    \end{array} \\
	 \iforall A\iforall T'
	    \begin{array}[t]{l}
	      [(\one{T}{A}{T'})\iimp\iexists R'.(\one{R}{A}{R'})
	         \null\iand \invar{T'}{R'}]\iand\null
	    \end{array} \\    
	 \iforall X\iforall R'
	    \begin{array}[t]{l}
	      [(\onep{R}{\inact X}{R'})\iimp\iexists T'.(\onep{T}{\inact X}{T'})
	          \null\iand\iforall w.\invar{(R' w)}{(T' w)}]\iand\null
	    \end{array} \\
	 \iforall X\iforall T'
	    \begin{array}[t]{l}
	      [(\onep{T}{\inact X}{T'})\iimp\iexists R'.(\onep{R}{\inact X}{R'})
	          \null\iand\iforall w.\invar{(T' w)}{(R' w)}]\iand\null
	    \end{array} \\    
	 \iforall X \iforall R'
	    \begin{array}[t]{l}
	      [(\onep{R}{\outact X}{R'})\iimp\iexists T'.(\onep{T}{\outact X}{T'})
	         \null\iand\nabla w. \invar{(R' w)}{(T' w)}]\iand\null
	    \end{array}\\
	 \iforall X \iforall T'
	    \begin{array}[t]{l}
	      [(\onep{T}{\outact X}{T'})\iimp\iexists R'.(\onep{R}{\outact X}{R'})
	         \null\iand\nabla w. \invar{(T' w)}{(R' w)}].\quad
	    \end{array}    
	  \end{array}
$$
The sequent reads, intuitively, that the set defined by {\sl inv} 
is closed under one-step transitions. This is proved by induction on
{\sl inv}. Formally, this is done by applying $\indL$ to 
$\invar R T$, using the invariant 
$$\lambda R\lambda T.\invar{R}{T} \oimp B\,R\,T.$$ 
The sequents corresponding to the base cases of the induction are
$$
\Seq{\invar{\Ppi}{\Qpi}}{B\,\Ppi\,\Qpi} \quad \mbox{ and } \quad 
\Seq{\invar{\Qpi}{\Ppi}}{B\,\Qpi\,\Ppi}
$$
and the inductive cases are given by 
$$
\begin{array}{c}
\Seq{\invar{R}{T} \oimp B\,R\,T}
    {\invar{((z)(0\barpi 0)\barpi R)}{(0\barpi T)} \oimp B((z)(0\barpi 0)\barpi R)(0\barpi T),}\\
\Seq{\invar{R}{T} \oimp B\,R\,T}
    {\invar{((z)(0\barpi \bar{x}a)\barpi R)}{(\bar{x}a\barpi T)} 
    \oimp B((z)(0\barpi \bar{x}a)\barpi R)(\bar{x}a\barpi T)}
\end{array}
$$
and their symmetric variants.
The full proof involves a number of cases of which we show one here:
the other cases can be proved similarly.

We consider a case for free output, where we have the sequent (after applying
some right-introduction rules)
\begin{equation}
\Seq{
	\left\{
	\begin{array}{r}
	\invar{R}{T}\oimp B\,R\,T \\
	\invar{((z)(0\barpi \bar{x}a)\barpi R)}{(\bar{x}a\barpi T)}\\
	\one{((z)(0\barpi \bar{x}a)\barpi R)}{A}{R'}
	\end{array} \right\}
	}
	{
	\exists T'.\one{(\bar{x}a\barpi T)}{A}{T'} \land 
		\invar{R'}{T'}
	}
\label{eq:pi rep}		       
\end{equation}
to prove. 
Its symmetric case can be proved analogously. 
The sequent (\ref{eq:pi rep}) can be simplified by applying
$\defL$ to the {\sl inv} predicate, 
followed by an instance of $\oimpL$. The resulting sequent is
\begin{equation}
\Seq{
	\left\{
	\begin{array}{r}
	B\,R\,T, ~ \invar{R}{T}\\	
	\one{((z)(0\barpi \bar{x}a)\barpi R)}{A}{R'}
	\end{array} \right\}
	}
	{
	\exists T'.\one{(\bar{x}a\barpi T)}{A}{T'} \land 
		\invar{R'}{T'}
	}
\label{eq:pi rep2}
\end{equation}

\begin{figure}
$$
\infer[\existsR]
{
	\Seq{B\,R\,T, \invar{R}{T}}{	
	\exists T'.\one{(\bar{x}a\barpi T)}{\bar{x}a}{T'} \land 
	   \invar{((z)(0\barpi 0)\barpi R)}{T'}
	}
}
{
\infer[\landR]
{
\Seq{B\,R\,T, \invar{R}{T}}{	
	\one{(\bar{x}a\barpi T)}{\bar{x}a}{(0\barpi T)} \land 
	   \invar{((z)(0\barpi 0)\barpi R)}{(0\barpi T)}
	}
}
{
	\infer[\defR]
	{\Seq{\cdots}{\one{(\bar{x}a\barpi T)}{\bar{x}a}{(0\barpi T)}}}
	{
	\infer[\topR]
	{\Seq{\cdots}{\top}}
	{}	
	}
	&
	\infer[\defR]
	{\Seq{\cdots, \invar{R}{T}}{\invar{((z)(0\barpi 0)\barpi R)}{(0\barpi T)}}}
	{
	\infer[\init]
	{\Seq{\cdots, \invar{R}{T}}{\invar{R}{T}}}
	{}
	}
}
}
$$
\caption{A derivation in Linc}
\label{fig:pi rep ex1}
$$
\infer[\landL]
{
	\Seq{B\,R\,T, \one{R}{A}{R''}}
	{
	\exists T'.\one{(\bar{x}a\barpi T)}{A}{T'} \land 
	\invar{(z)(0\barpi \bar{x}a)\barpi R'')}{T'}
	}		
}
{
	\infer[\forallL;\forallL]
	{
		\Seq{\forall U\forall A'\ \one{R}{A}{U} \oimp 
			\exists V.\one{T}{A'}{V} \land \invar{U}{V}, 
		   \one{R}{A}{R''}}
		{
		\cdots
		}	
	}
	{
	\infer[\oimpL]
	{
		\Seq{\one{R}{A}{R''} \oimp 
		\exists V.\one{T}{A'}{V} \land \invar{R''}{V}, 
		   \one{R}{A}{R''}}
		{
			\cdots
		}			
	}
	{
	\infer[\init]
	{\Seq{\one{R}{A}{R''}}{\one{R}{A}{R''}} }
	{}
	&
	\deduce
	{
		\Seq{\exists V.\one{T}{A}{V} \land 
			\invar{R''}{V} }
		{
		\cdots
		}				
	}
	{\Pi}
	}
	}
}
$$
where $\Pi$ is
$$
\infer[\existsL;\landL]
{\Seq{\exists V.\one{T}{A}{V} \land \invar{R''}{V}}
	{	\exists T'.\one{(\bar{x}a\barpi T)}{A}{T'} \land 
	\invar{(z)(0\barpi \bar{x}a)\barpi R'')}{T'}
	}
}
{
\infer[\existsR]
{
\Seq{\one{T}{A}{V}, \invar{R''}{V}}
	{	
	 \exists T'.\one{(\bar{x}a\barpi T)}{A}{T'} \land 
	 \invar{(z)(0\barpi \bar{x}a)\barpi R'')}{T'}
	}
}
{
\infer[\landR]
{
\Seq{\one{T}{A}{V}, \invar{R''}{V}}
	{	
	 \one{(\bar{x}a\barpi T)}{A}{(\bar{x}a\barpi V)} \land 
	 \invar{(z)(0\barpi \bar{x}a)\barpi R'')}{(\bar{x}a\barpi V)}
	}
}
{
\infer[\defR]
{\Seq{\one{T}{A}{V}}{\one{(\bar{x}a\barpi T)}{A}{(\bar{x}a\barpi V)}}}
{
\infer[\init]
{\Seq{\one{T}{A}{V}}
	{\one{T}{A}{V}}}
{}
}
&
\infer[\defR]
{\Seq{\invar{R''}{V}}{\invar{((z)(0\barpi \bar{x}a)\barpi R'')}{(\bar{x}a\barpi V)}}}
{
\infer[\init]
{\Seq{\invar{R''}{V}}{\invar{R''}{V}}}
{}
}
}
}
}
$$
\caption{A derivation in $\Linc$ given in two parts}
\label{fig:pi rep ex2}
\end{figure}

There are three ways in which the one-step transition
in the left-hand side of the sequent (\ref{eq:pi rep}) can be inferred 
(via $\defL$), 
{\em i.e.}, either $A$ is $\bar{x}a$ and $R'$ is $((z)(0\barpi 0)\barpi R)$, or
$\one{R}{A}{R''}$ and $R'$ is $(z)(0\barpi \bar{x}a)\barpi R'')$, or $A$ is $\tau$ and 
$\onep{R}{\inact X}{M}$, $R'$ is $((z)(0\barpi 0) | Ma)$
for some $X$ and $M$. These three cases correspond to the following sequents.
$$
\begin{array}{l}
\Seq{B\,R\,T, \invar{R}{T}}{	
\exists T'.\one{(\bar{x}a\barpi T)}{\bar{x}a}{T'} \land 
   \invar{((z)(0\barpi 0)\barpi R)}{T'}
}\\
\Seq{B\,R\,T, \invar{R}{T}, \one{R}{A}{R''}}
	{
	\exists T'.\one{(\bar{x}a\barpi T)}{A}{T'} \land 
	\invar{(z)(0\barpi \bar{x}a)\barpi R'')}{T'}
	}	\\
\Seq{B\,R\,T, \invar{R}{T}, \one{R}{\inact X}{M}}
	{
	\exists T'.\one{(\bar{x}a\barpi T)}{\tau}{T'}
	\land \invar{((z)(0\barpi 0) | Ma)}{T'}
	}	
\end{array}
$$
The proof of the first sequent is given in Figure~\ref{fig:pi rep ex1} and of the
second sequent is given in Figure~\ref{fig:pi rep ex2}. 
The proof for the third sequent is not given but it is easy to see that
it has a similar structure to the proof of the second one.
\end{example}

\section{Automation of proof search}
\label{sec:auto}

The above specifications for one-step transitions, for late, early,
and open bisimulation, and for modal logics are not only declarative
and natural, they can also, in many cases, be turned into effective and
{\em symbolic} implementations by using techniques from the proof
search literature.  In this section we outline high-level aspects of
the proof theory of $\FOLDNb$ that can be directly exploited to
provide implementations of significant parts of this logic: we also
describe how such general aspects can be applied to some of our
$\pi$-calculus examples.

\subsection{Focused proof search}

Since the cut-elimination theorem holds for $\FOLDNb$, the search for
a proof can be restricted to {\em cut-free proofs}.  It is possible to
significantly constrain cut-free proofs to {\em focused proofs} while
still preserving completeness.  The search for focused proofs has a
simple structure that is organized into two phases.  The {\em
  asynchronous} phase applies only invertible inference rules in any
order and until no additional invertible rules can be applied.  The
{\em synchronous} phase involves the selection of (possibly)
non-invertible inference rule and the hereditary (focused) application
of such inference rules until invertible rules are possible again.
Andreoli \cite{andreoli92jlc} provided such a focused proof system for
linear logic and proved its completeness.  Subsequently, many 
focusing systems for intuitionistic and classical logic have been
developed, {\em cf.} \cite{liang07csl} for a description of several of
them. 
Baelde and Miller \citeyear{baelde07lpar} present a 
focusing proof system for the multiplicative and additive linear
logic (MALL) extended with fixed points and show that that proof
system provides a focusing proof system for a large
subset of $\FOLD$.
Focused proof systems are
generally the basis for the automation of logic programming languages
and they generalize the notion of {\em uniform proofs}
\cite{miller91apal}.  

\subsection{Unification}
Unification can be used in the implementation of $\FOLDNb$ proof
search in two different ways.  
One way involves the implementation of the $\defL$ inference rule
and the other way involves the determination of appropriate terms for
instantiating the 
$\exists$ quantifier in the $\existsR$ inference rule and the
$\forall$ quantifier in the $\forallL$ inference rule.  In the
specifications presented here, unification only requires the decidable
and determinate subset of higher-order unification called {\em
higher-order pattern} (or $L_\lambda$) unification \cite{miller91jlc}.
This style of unification, which can be described as first-order
unification extended to allow for bound variables and their mobility
within terms, formulas, and proofs, is known to have efficient and practical
unification algorithms that compute most general unifiers whenever
unifiers exist \cite{nipkow93lics,nadathur05iclp}.  The Teyjus implementation
\cite{nadathur99cade,nadathur05tplp} of $\lambda$Prolog provides an
effective implementation of such unification, as does Isabelle
\cite{paulson90abs} and Twelf \cite{pfenning99cade}.

\subsection{Proof search for one-step transitions.}
Computing one-step transitions can be done entirely using a
conventional, higher-order logic programming language, such as
$\lambda$Prolog: since the definition ${\bf D}_{\pi}$ for one-step
transitions is Horn, we can use Proposition~\ref{prop:nabla forall} to show
that for the purposes of computing one-step transitions, all
occurrences of $\nabla$ in ${\bf D}_{\pi}$ can be changed to
$\forall$.  The resulting definition is then a $\lambda$Prolog logic
program for which Teyjus provides an effective implementation.  In particular,
after loading that definition, we would simply ask the query
$\one{P}{A}{P'}$, where $P$ is the encoding of a particular
$\pi$-calculus expression and $A$ and $P'$ are free
variables.  Standard logic programming interpreters would then
systematically bind 
these two variables to the actions and continuations that $P$ can
make.  Similarly, if the query was $\onep{P}{A}{P'}$, logic
programming search would systematically return all bound actions
(here, $A$ has type $\name\to \action$) and corresponding bound
continuations (here, $P'$ has type $\name\to \proc$).

\subsection{Proof search for open bisimulation.}

Theorem proving establishing a bisimulation goal is not done via a
conventional logic programming system like $\lambda$Prolog since such
systems do not implement the $\nabla$-quantifier and the case analysis
and unification of eigenvariables that is required for the $\defL$
inference rule.  None-the-less, the implementation of proof search for
open bisimulation is easy to specify using the following key steps.
(Sequents missing from this outline are trivial to address.)  In the
following, we use the quantifier prefix $\Qscr$ to denote either
$\forall x$ or $\nabla x$ or the empty quantifier prefix.
\begin{enumerate}

\item When searching for a proof of
$\NSeq{\Sigma}{}{\Judg{\sigma}{\Qscr.\lbisim{P}{Q}}}$ apply
right-introduction rules: {\em i.e.}, simply introduce the quantifier
$\Qscr$ (if it is non-empty) and then open the definition of
{\sl lbisim}.

\item If the sequent has a formula on its left-hand side, then that
formula is $\Judg{\sigma}{\one{P}{A}{P'}}$, where $P$ denotes a
particular term where all its non-ground subterms are of type $\name$, 
and $A$ and $P'$ are terms, possibly containing eigenvariables.  In this
case, select the $\defL$ inference rule: the premises of this
inference rule will then be either $(i)$ the empty-set of premises
(which represents the only way that proof search terminates), or
$(ii)$ a set of premises that are all again of the form of one-step
judgments, or $(iii)$
the premise contains $\top$ instead of an atom on the left, in which
case, we must consider the remaining case that follows (after using
the weakening $\wL$ inference rule).

\item If the sequent has the form
$\NSeq{\Sigma}{}{\Judg{\sigma}{\exists Q' [\one{Q}{A}{Q'}\land
B(P',Q')]}}$, where $B(P',Q')$ involves a recursive call to {\sl
lbisim} and where $P'$ is a closed term, then we must instantiate the
existential quantifier with an appropriate substitution.  Standard
logic programming techniques can be
used to find a substitution for $Q'$ such that $\one{Q}{A}{Q'}$ is
provable (during this search, eigenvariables and locally scoped
variables are treated as constants and $P$ and $A$ denote particular
closed terms).  There might be several ways to prove such a formula
and, as a result, there might be several different substitutions for
$Q'$.  If one chooses the term $T$ to instantiate $Q'$, then one proceeds
to prove the sequent
$\NSeq{\Sigma}{}{\Judg{\sigma}{\Qscr.\lbisim{P'}{T}}}$.  
If the sequent has instead the form $\NSeq{\Sigma}{}{\Judg{\sigma}{\exists Q'
[\onep{Q}{A}{Q'}\land B(P',Q')]}}$, then one proceeds in an analogous manner.
\end{enumerate}
Proof search for the first two cases is invertible (no backtracking is
needed for those cases).  On the other hand, the third
case is not invertible and backtracking on possibly all choices of
substitution term $T$ might be necessary to ensure completeness.

\subsection{The Bedwyr model checker}

The various implementation techniques mentioned above---unification of
$\lambda$-terms, backtracking focused proof search, unfolding
definitions---have all been implemented within the Bedwyr model
checking system ~\cite{baelde07cade}, which implements proof search for a
simple fragment \cite{tiu05eshol} of $\FOLDNb$.  The definitions of
one-step transitions and of bisimulation are in this fragment and the
Bedwyr system is a complete implementation of open bisimulation for
the finite $\pi$-calculus: in particular, it provides a decision
procedure for open-bisimulation.  Bedwyr also implements limited forms
of the modal logic described in Section~\ref{sec:modal}.  It is also
possible to use Bedwyr to explore why two $\pi$-calculus
processes might not be bisimilar: for example, it easy to define
traces for such processes and then to search for a trace that holds of
one process but not of the other.

Since Bedwyr is limited to intuitionistic reasoning, it does not
fully implement late bisimulation.  We now speculate briefly on how
one might extend a system like Bedwyr to treat late bisimulation.

\subsection{Proof search for late bisimulation.}
The main difference between doing proof search for open bisimulation
and late bisimulation is that in the latter we need to select and
instantiate formulas from the set $\Escr$ and explore the cases
generated by the resulting $\lorL$ rule. 
For example, consider a sequent of the form $\NSeq{\Sigma,
  x}{\Escr,\Gamma_x}{C_x}$, where $\Gamma_x\cup\{C_x\}$ is a set of
formulas 
which may have $x$ free.  One way to proceed with the search for a proof
would be to instantiate $\forall z(x=z\lor x\not=z)$ twice with the
constants $a$ and $b$.  We would then need to consider proofs of the
sequent  
$\NSeq{\Sigma, x}{x=a\lor x\not=a,x=b\lor x\not=b,\Gamma_x}{C_x}$.
Using the $\lorL$ rule twice, we are left with four sequents to prove:
\begin{enumerate}
\item $\NSeq{\Sigma, x}{x=a,x=b,\Gamma_x}{C_x}$ which is 
  proved trivially since the equalities are contradictory; 
\item $\NSeq{\Sigma, x}{x=a,x\not=b,\Gamma_x}{C_x}$, which is
  equivalent to $\NSeq{\Sigma}{\Gamma_a}{C_a}$;
\item $\NSeq{\Sigma, x}{x\not=a,x=b,\Gamma_x}{C_x}$, which is
  equivalent to $\NSeq{\Sigma}{\Gamma_b}{C_b}$; and 
\item $\NSeq{\Sigma, x}{x\not=a,x\not=b,\Gamma_x}{C_x}$.
\end{enumerate}
In this way, the excluded middle can be used with a set of $n$ items
to produce $n+1$ sequents: one for each member of the set and
one extra sequent to handle all other cases (if there are any).

The main issue for implementing proof search with this specification
of late bisimulation is to determine what instances of the excluded
middle are needed: answering this question would then reduce proof
search to one similar to open bisimulation.  There seems to be two
extreme approaches to take. At one extreme, we can take instances for
all possible names that are present in our process expressions:
determining such instances is simple but might lead to many more cases
to consider than is necessary.  The other extreme would be more lazy:
an instance of the excluded middle is suggested only when there seems
to be a need to consider that instance.  The failure of a $\defR$ rule
because of a mismatch between an eigenvariable and
a constant would, for example, suggest that excluded middle should be
invoked for that eigenvariable and that constant.  The exact details
of such schemes and their completeness are left for future work.

\section{Related and future work}
\label{sec:related}

There are many papers on topics related to the encoding of the
operational semantics of the $\pi$-calculus into formal systems.  An
encoding of one-step transitions for the $\pi$-calculus using Coq was
presented in \cite{despeyroux00ifiptcs} but the problem of computing
bisimulation was not considered.  Honsell, Miculan, and Scagnetto
\cite{honsell01tcs} give a more involved encoding of the
$\pi$-calculus in Coq and assume that there are an infinite number of
global names.  They then build formal mechanisms to support notions
such as ``freshness'' within a scope, substitution of names,
occurrences of names in expressions, etc.  Gabbay
\cite{gabbay03automath} does something similar but uses the
set theory developed in \cite{gabbay01fac} to help develop his formal
mechanisms.  This formalism is later given a first-order axiomatization by
Pitts~\cite{pitts03ic}, resulting in an extension of first-order logic called 
{\em nominal logic}.  Aspects of nominal reasoning have been incorporated
into the proof assistant Isabelle~\cite{urban05cade} and there has been
some recent work in formalizing the meta theory of the $\pi$-calculus
in this framework \cite{bengtson07fossacs}.
Hirschkoff \cite{hirschkoff97tphol} also used Coq but
employed deBruijn numbers \cite{debruijn72} instead of explicit names.
In the papers that address bisimulation, formalizing names and their
scopes, occurrences, freshness, and substitution is considerable work.
In our approach, much of this same work is required, of course, but it
is available in rather old technology, particularly, via Church's
Simple Theory of Types (where bindings in terms and formulas were put
on a firm foundation via $\lambda$-terms), Gentzen's sequent calculus,
Huet's unification procedure for $\lambda$-terms \cite{huet75tcs},
etc.  More modern work on proof search in higher-order logics is also
available to make our task easier and more declarative.

The encoding of transitions for the $\pi$-calculus into logics and type
systems have been studied in a number of previous
works~\cite{honsell98,despeyroux00ifiptcs,honsell01tcs,roeckl01fossacs,bengtson07fossacs}.
Our encoding, presented as a definition in Figure~\ref{late pi def},
has appeared in \cite{miller99surveys,miller03lics}.  The material on
proof automation in Section~\ref{sec:auto} clearly seems related to
{\em symbolic bisimulation} (for example, see
\cite{hennessy95tcs,boreale96ic}) and on using unification and logic
programming techniques to compute symbolic bisimulations (for example,
see \cite{basu01iclp,boreale01icalp}).  Since the technologies used to
describe these other approaches are rather different than what is
described here, a detailed comparison is left for future work.

It is, of course, interesting to consider the general $\pi$-calculus
where infinite behaviors are allowed (by including $!$ or recursive
definitions).  In such cases, one might be able to still do many
proofs involving bisimulation if the proof system included induction
and co-induction inference rules. We have illustrated with a simple
example in Section~\ref{sec:pi rep} how such a proof might be done.
Inference rules for induction and co-induction appropriate for the 
sequent calculus have been presented 
in \cite{momigliano03types} and a version of these rules that also
involves the $\nabla$ quantifier has been presented in the first
author's PhD thesis \cite{tiu04phd}.  Open bisimulation, however, has not
been studied in this setting. We plan to investigate further how these
stronger proof systems can be used to establish properties about
$\pi$-calculus expressions with infinite behaviors.

Specifications of operational semantics using a logic should make
it possible to formally prove properties concerning that operational
semantics.  This was the case, for example, with specifications of the
evaluation and typing of simple functional and imperative programming
languages: a number of common theorems (determinacy of evaluation,
subject-reduction, etc) can be naturally inferred using logical
specifications \cite{mcdowell02tocl}.  We plan to investigate using
our logic (also incorporating rules for induction and
co-induction) for formally proving parts of the theory of the
$\pi$-calculus.  It seems, for example, rather transparent to prove
that open bisimilarity is a congruence in our setting (see
\cite{ziegler05sos} for a more general class of congruence relations).

\section{Conclusion}
\label{sec:conc}

In this paper we presented a meta-logic that allows for declarative specifications of
judgments related to the $\pi$-calculus.
These specifications are done entirely within the logic and without any
additional side conditions.  The management of name bindings in the
specification of one-step transition, bisimulation, and modal logic is
handled completely by the logic's three levels of binding,
namely, $\lambda$-bindings within terms, the formula-level
binders (quantifiers) $\forall$, $\exists$, and $\nabla$, and the
proof-level bindings for eigenvariables and local (generic) contexts.

This paper can be seen as part of a tradition of treating
syntax more abstractly.   The early, formal treatments of syntax by, for
example, Church and G\"odel, formalized terms and formulas as strings.
Eventually, that treatment of syntax was replaced by more abstract
objects such as parse trees: it is on parse trees that most
syntactic descriptions of the $\lambda$-calculus and $\pi$-calculus
are now given.  Unfortunately, parse trees do not come equipped with
primitive notions of bindings.  To fix that problem, for example,
Prawitz introduced ``discharge functions'' \cite{prawitz65} and de
Bruijn introduced ``nameless dummies'' \cite{debruijn72}.  The move
from parse trees to $\lambda$-trees, along with the use of a logic
able to deal intimately with syntactic abstractions, is another way to
fix this problem.

A significant part of this paper deals with establishing adequacy
results that show a formal connection between the ``standard''
definitions of judgments concerning the $\pi$-calculus and the
definitions given in logic (see the appendices for the details).
These adequacy results are all 
rather tedious and shallow but seem necessary to ensure that we have
not invented our own problems for which we provide good solutions.  It
would seem, however, that the tediousness nature of the adequacy
results can be
attributed to the large gap between our proof-theory approach and
the ``standard'' approach used to encode the
$\pi$-calculus: now that some of these basic adequacy results have
been written down, the adequacy results for any additional logical
specifications using $\lambda$-tree syntax should follow more
immediately.

We note that our effort in developing a proof theoretic setting for
the $\pi$-calculus has led us to find new description for, in
particular, the underlying assumptions on names in open and late
bisimulatons.  This examination has led us to characterize the
differences between open and late bisimulations in a simple and
logical fashion: in particular, as the difference in name
quantification and in the assumption about decidability of name
equality.

\smallskip
\noindent
{\em Acknowledgments.}  We are grateful to the reviewers of earlier
drafts of this paper for their detailed and useful comments.  We also
benefited from support from INRIA through the ``Equipes
Associ{\'e}es'' Slimmer, and from the Australian Research Council
through the Discovery Project ``Proof Theoretical Methods for 
Reasoning about Process Equivalence.''
\par % needed this to fix spacing problems in the above paragraph when
     % the reference section is added next.

\newpage
\appendix
 
\def\onef{\hbox{\sl one}_f}
\def\oneb{\hbox{\sl one}_b}
\def\pidef{{\bf D}_\pi}
\def\defrule{\mbox{\it def}}

\section{Properties of one-step transitions}

To prove the adequacy results for the encodings of bisimulation and 
modal logics, we shall consider 
some derived rules which allow us to enumerate all possible next states
from a given process. 
In the following, we use the notation $\alpha^n \ra \beta$ to denote the type
$\underbrace{\alpha \ra \cdots \ra \alpha}_{n} \ra \beta$, and we write $\alpha^* \ra \beta$
to denote $\alpha^n \ra \beta$ for some $n \geq 0.$ Due to space limits, some results
in this section are stated without proofs, but they can be found
in the electronic appendix of the paper.

\begin{definition}
The judgments $\Judg{\sigma}{\one{P}{A}{Q}}$ and
$\Judg{\sigma}{\onep{P}{A}{Q}}$ are 
{\em higher-order patterned judgments}, or patterned judgments for short, if
\begin{enumerate}
\item every occurrence of the free variables in the judgment is applied
to distinct names, which are either in $\sigma$ or bound by $\lambda$-abstractions, \ie,
$M~a_1 \cdots a_n$, where $a_i \in \sigma$ or it is bound by some
$\lambda$-abstraction, and $a_1, \ldots, a_n$ are pairwise distinct,
\item the only occurrences of free variables in $P$ are those of type
$\name^n \ra \name$ where $n \geq 0$, and the only occurrences of 
free variables in $A$ are those of type 
$\name^n \ra \name$ or $\name^n \ra \action$,
\item and $Q$ is of the form $(M ~ \vec \sigma)$ for some variable $M.$ 
\end{enumerate}
The process term $P$ in the transition predicate $\one{P}{A}{Q}$ and 
$\onep{P}{A}{Q}$ is called a {\em primary} process term. 
The notion of patterned judgments extends to non-atomic judgments,
which are defined inductively as follows: 
\begin{itemize}
\item $\Judg{\sigma}{\top}$ is a patterned judgment,
\item if $\Judg{\sigma}{B}$ and $\Judg{\sigma}{C}$ are patterned judgments
such that both judgments have no free variables in common which are
of type $\name^* \ra \proc,$  then $\Judg{\sigma}{B \land C}$ is a patterned judgment, 
\item if $\Judg{\sigma x}{B}$ is a patterned judgment, then 
$\Judg{\sigma}{\nabla x.B}$ is a patterned judgment, 
\item and if $\Judg{\sigma}{B[h\,\vec \sigma/y]}$ is a patterned judgment
then $\Judg{\sigma}{\exists y.B}$ is a patterned judgment, provided that 
$h$ is of type $\name^n \ra \action$ or $\name^n \ra \proc$, and
$h$ is not free in $\exists y.B$. 
\end{itemize}
Two patterned judgments $\Ascr$ and $\Bscr$ are {\em $\proc$-compatible}
if they do not have variables in common which are of type $\name^* \ra \proc.$
\end{definition}
The restrictions on the occurences of free variables in 
patterned judgments are similar to the restrictions used in higher-order
pattern unification. This is to ensure that proof search for patterned judgments 
involves only higher-order pattern unification.

Let $\rho$ be a substitution and let $\Sigma$ be a signature.
We write $\Sigma \vdash \rho$ if for every $x \in \dom{\rho}$ of type $\tau$,
we have $\Sigma \vdash \rho(x): \tau.$ 
Two signatures $\Sigma$ and $\Sigma'$ are said to be compatible
if whenever $x : \tau_1 \in \Sigma$ and $y :\tau_2 \in \Sigma'$, 
$x = y$ implies $\tau_1 = \tau_2.$
Given two signature-and-substitution pairs 
$(\Sigma_1, \rho_1)$ and $(\Sigma_2, \rho_2)$ such that $\Sigma_1$ and $\Sigma_2$
are compatible, and $\Sigma_1 \vdash \rho_1$ and $\Sigma_2 \vdash \rho_2$, 
we write $(\Sigma_1, \rho_1) \circ (\Sigma_2, \rho_2)$ to denote the pair
$
(\Sigma_1\rho_2 \cup \Sigma_2, \rho_1 \circ \rho_2).
$ 
This definition of composition extends straightforwardly
to composition between a pair and a set or a list of pairs. 

Let us call a  signature-substitution pair $(\Sigma,\rho)$ a {\em solution}
for a patterned judgment $\Cscr$ if $\Sigma \vdash \rho$ and 
$\NSeq{\Sigma}{.}{\Cscr\rho}$ is provable. 
In proving the adequacy of the encoding of bisimulation and
modal logics for the $\pi$ calculus, we often want to find
all possible solutions to a given transition relation, which
corresponds to enumerating all possible continuations of
a given process. 
For this purpose, we define a construction of ``open'' derivation 
trees for a given list of patterned judgments $\Delta$. Open derivation trees 
are trees made of nodes which are instances of certain inference rules. 
This construction gives us a set of derivation trees for the 
sequent $\Delta \vdash \bot$, following a certain order of rule applications. 
As we shall see, the construction of the trees basically amounts to 
application of left-introduction rules to $\Delta$. We are interested 
in collecting all the substitutions generated by the $\defL$ rule
in these trees, which we will show to correspond 
to the solutions for the patterned judgments in $\Delta.$

\begin{definition}
\label{def:patterned-judgment}
Let $\Delta$ be a list of patterned judgments such that its elements are
pairwise $\proc$-compatible, and let $(\Sigma,\theta)$ be a pair 
such that $\Sigma \vdash \theta$, and that the free variables of $\Delta$ are in $\Sigma.$
An {\em open inference rule} is an inference on triples of the form
$(\Sigma', \Delta', \theta')$ where $\Sigma'$ is a signature, $\Delta'$ is a list of patterned
judgments and $\theta'$ is a substitution such that $\Sigma' \vdash \theta'.$
We will use the notation $(\Sigma',\theta') \vdash \Delta'$ to denote such a triple.
{\em Open derivation trees} are derivations constructed using 
the following open inference rules:
$$
\infer[open]
{(\Sigma, \theta) \vdash [\,] }
{}
\qquad
\infer[\top]
{(\Sigma, \theta) \vdash \Judg{\bar n}{\top}, \Delta'}
{(\Sigma, \theta) \vdash \Delta'}
$$
$$
\infer[\land]
{(\Sigma,\theta) \vdash \Judg{\bar n} {A \land B}, \Delta'}
{(\Sigma,\theta) \vdash \Judg{\bar n} A, \Judg{\bar n} B, \Delta'}
\qquad
\infer[\exists]
{(\Sigma, \theta) \vdash \Judg{\bar n} {\exists x.B\,x}, \Delta' }
{(\Sigma \cup \{h\}, \theta) \vdash \Judg{\bar n} {B\,(h\,\bar n)}, \Delta'}
$$
$$
\infer[\defrule]
{(\Sigma, \theta) \vdash \Ascr, \Delta}
{
\{(\Sigma\rho, \theta \circ \rho) \vdash \Bscr \rho, \Delta\rho \mid \rho \in CSU(\Ascr,\Hscr), ~ \Hscr \defeq \Bscr \}
}
$$
In the $\exists$-rule, the eigenvariable $h$ is new, \ie, it is not in $\Sigma.$
In the $\defrule$-rule, we require that for every $\rho \in CSU(\Ascr,\Hscr)$,
the judgments $\Bscr \rho, \Delta \rho$ are patterned judgments. 
That is, we restrict the CSU's to those that preserves the pattern
restrictions on judgments.
The instances of the $open$-rule in an open derivation are 
called {\em open leaves} of the derivation. 
Given an open derivation $\Pi$, we denote with $\Lscr(\Pi)$ the set of 
signature-substitution pairs in the open leaves of $\Pi.$ 
\end{definition}

\begin{definition}
The measure of a patterned judgment $\Judg{\sigma}{B}$,
written $|\Judg{\sigma}{B}|$, is the
number of process constructors occuring in the primary terms in $B.$
The measure of a list of judgments $\Delta$ is the multiset of
measures of the judgments in $\Delta.$
\end{definition}

\begin{lemma}
\label{lm:open-drv-exists}
Let $\Delta$ be a list of patterned judgments such that its elements are
pairwise $\proc$-compatible, and whose variables
are in a given signature $\Sigma$. Let $\theta$ be a substitution such that
$\Sigma \vdash \theta$. Then there exists an open derivation $\Pi$ of $(\Sigma, \theta) \vdash \Delta.$ 
\end{lemma}

\begin{lemma}
\label{lm:solution-comp}
Let $\Sigma_1$, $\Sigma_2$, $\theta_1$ and $\theta_2$ be signatures and substitutions
such that $\Sigma_1 \vdash \theta_1$ and $\Sigma_2 \vdash \theta_2$. Let $\Delta$ be a list of
pairwise $\proc$-compatible patterned jugdments such that all its free variables are in $\Sigma_2.$
If there exists an open derivation $\Pi_1$ of 
$(\Sigma_1 \theta_2 \cup \Sigma_2, \theta_1 \circ \theta_2) \vdash \Delta$, then there exists an open 
derivation $\Pi_2$ of $(\Sigma_2, \theta_2) \vdash \Delta$ of the same height such that 
$\Lscr(\Pi_1) = (\Sigma_1, \theta_1) \circ \Lscr(\Pi_2)$
and vice versa.
\end{lemma}

The following lemma states that the open leaves in an open derivation are solutions
of the patterned judgments on the root of the derivation tree. This can be proved
by induction on the height of derivation and case analysis on the definition clauses
of one-step transitions.

\begin{lemma}
\label{lm:open-drv-sound}
Let $\Delta$ be a list of patterned judgments such that its elements are pairwise
$\proc$-compatible and whose variables are in a given signature $\Sigma$. 
Let $\Pi$ be an open derivation of $(\Sigma, \epsilon) \vdash \Delta.$ 
Then for every element $\Cscr \in \Delta$ and every pair
$(\Sigma', \theta) \in \Lscr(\Pi)$, the sequent
$\NSeq{\Sigma'}{.}{\Cscr \theta}$ is provable.
\end{lemma}

We are now ready to define the following derived rules.
The rule $\onef$ enumerates all possible free-actions that a process can 
perform. Given a patterned judgment  $\Judg{\bar n}{\one P A Q}$
and an open derivation $\Pi$ of $(\Sigma, \epsilon) \vdash \Judg{\bar n}{\one P A Q}$,
the $\onef$ rule, applied to this judgment, is as follows:
$$
\infer[\onef]
{\NSeq{\Sigma}{\Judg{\bar n}{\one P A Q}, \Gamma}{\Cscr}}
{
\{
\NSeq {\Sigma'} {\Gamma\theta} {\Cscr\theta}
\mid (\Sigma', \theta) \in \Lscr(\Pi)
\}
}
$$
The corresponding rule for bound input or bound output transition is defined 
analogously, \ie,
$$
\infer[\oneb.]
{\NSeq{\Sigma}{\Judg{\bar n}{\onep P X M}, \Gamma} \Cscr}
{
\{
\NSeq {\Sigma'} {\Gamma\theta} {\Cscr\theta}
\mid (\Sigma', \theta) \in \Lscr(\Pi)
\}
}
$$
where $\Pi$ is an open derivation of $(\Sigma, \epsilon) \vdash \Judg{\bar n}{\onep P X M}.$
Since open inference rules are essentially invertible left-rules of $\FOLDNb$,
these derived rules are sound and invertible.
\begin{lemma}
The rules $\onef$ and $\oneb$ are invertible and derivable in $\FOLDNb$.
\end{lemma}

We can now prove Proposition~\ref{prop:neg one step}.

\begin{proof}
Suppose that $\one{\Ppi}{\alpha}{\Qpi}$ does not hold in the $\pi$-calculus.
We show that the sequent $\neg \nabla \bar n. \trans{\one \Ppi \alpha \Qpi}$ is 
derivable in $\FOLDNb.$ 
This is equivalent to proving the sequent
$
\NSeq{}{\Judg{ \vec n}{ \trans{\one \Ppi \alpha \Qpi}}}{\bot}.
$
We apply either $\onef$ or $\oneb$ to the sequent (bottom-up), 
depending on whether $\alpha$ is a free or a bound action. 
In both cases, if the premise of the $\onef$ or $\oneb$ is 
empty, then we are done. Otherwise, there exists a substitution
$\theta$ such that $(\nabla \vec n. \trans{\one \Ppi \alpha \Qpi})\theta$
is derivable in $\FOLDNb$. Since the transition judgment is ground, this would
mean that $\nabla \vec n. \trans{\one \Ppi \alpha \Qpi}$ is derivable, and
by Proposition~\ref{prop:one step}, the transition $\one \Ppi \alpha \Qpi$ holds
in the $\pi$-calculus, contradicting our assumption.

Conversely, suppose that $\neg \nabla \bar n. \trans{\one \Ppi \alpha \Qpi}$
is derivable in $\FOLDNb.$ Then $\one \Ppi \alpha \Qpi$ cannot be a transition
in the $\pi$-calculus, for otherwise, we would have $\vdash \nabla \bar n. \trans{\one \Ppi \alpha \Qpi}$
by Proposition~\ref{prop:one step}, and by cut, we would have a proof of $\bot$, which
is impossible. 
\end{proof}

\section{Adequacy of the specifications of bisimulations}

We need some auxiliary lemmas that concern the structures of cut free proofs.
The next three lemmas can be proved by simple permutations of inference rules. 

\begin{lemma}
\label{lm:right-first}
Let $\Pi$ be a cut-free derivation of $\NSeq{\cdot}{\Gamma}{\Cscr}$,  where $\Cscr$ 
contains a non-equality atomic formula and 
every judgment in $\Gamma$ is in one of the following forms:
$$
\Judg{\bar n}{\forall x\forall y. x = y \lor x \not = y}
\qquad
\Judg{\bar n}{\forall y. a = y \lor a \not = y}
\qquad
\Judg{\bar n}{a = b \lor a \not = b}
$$
$$
\Judg{\bar n}{a = a \lor a \not = a}
\qquad
\Judg{\bar n}{a = a}
\qquad
\Judg{\bar n}{a \not = b}
$$
for some $\bar n$ and distinct names $a,b$ in $\bar n$.
Then there exists a derivation of the sequent which ends with a right-introduction rule on $\Cscr.$
\end{lemma}

\begin{lemma}
\label{lm:lbisim-inv}
The $\defR$ rule, applied to $\lbisim P Q$, for any $P$ and $Q$, is
invertible. 
\end{lemma}

\begin{lemma}
\label{lm:ebisim-inv}
The $\defR$ rule, applied to $\ebisim P Q$, for any $P$ and $Q$, is
invertible. 
\end{lemma}

\subsection{Adequacy of the specification of late bisimulation}

In the following, we use the notation
$x_1 \not = x_2 \not = \cdots \not = x_{n-1}\not  = x_n$ to abbreviate the conjunction
$$
\bigwedge \{ x_i \not = x_j \mid i, j\in \{1,\ldots,n\},  i \not = j \}.
$$
With a slight abuse of notation, we shall write $\Xscr \oimp B$,
where $\Xscr$ is a finite set of formula $\{B_1, \ldots, B_n \}$ , to mean 
$B_1 \land \cdots \land B_n \oimp B$, and we shall write $\nabla y. \Xscr$ to mean the formula
$\nabla y.B_1 \land \cdots \land \nabla y.B_2.$

\begin{lemma}
\label{lm:lbisim-complete}
Let $\Ppi$ and $\Qpi$ be two late-bisimilar finite $\pi$-processes and let
$n_1, \ldots, n_k$ be the free names in $\Ppi$ and $\Qpi.$
Then for some finite set $\Xscr \subset \Escr$, we have
\begin{equation}
\vdash \forall n_1 \cdots \forall n_k. (\Xscr \land n_1 \not = n_2 \not = \cdots \not = n_k \oimp \lbisim \Ppi \Qpi).
\label{eq:lbisim-completeness}
\end{equation}
\end{lemma}
\begin{proof}
We construct a proof of formula (\ref{eq:lbisim-completeness}) by induction on the size of
$\Ppi$ and $\Qpi,$ \ie, the number of action prefixes in $\Ppi$ and $\Qpi.$ 
It can be easily shown that the number of prefixes in a process is reduced
by transitions, for finite processes. 
By applying the introduction rules
for $\forall$, $\oimp$ and unfolding the definition of $lbisim$ (bottom up) 
to the formula (\ref{eq:lbisim-completeness}), we get the following three sequents:
\begin{enumerate}
\item $\NSeq{n_1, \cdots, n_k, A, P'}{\Xscr, n_1 \not = \cdots \not = n_k, \one{\Ppi}{A}{P'}}
           {\exists Q'. \one {\Qpi}{A}{Q'} \land \lbisim{P'}{Q'}}$
\item $\NSeq{n_1, \cdots, n_k, X, P'}{\Xscr, n_1 \not = \cdots \not = n_k, \onep{\Ppi}{\inact X}{P'}}
           {\exists Q'. 
             \begin{array}{l}
              \onep {\Qpi}{\inact X}{Q'} ~ \land \\
               \forall w. \lbisim{(P'w)}{(Q'w)}
             \end{array}
           }$
\item $\NSeq{n_1, \cdots, n_k, X, P'}{\Xscr, n_1 \not = \cdots \not = n_k, \onep{\Ppi}{\outact X}{P'}}
           {\exists Q'. 
             \begin{array}{l}
              \onep {\Qpi}{\outact X}{Q'} ~ \land  \\
              \nabla w. \lbisim{(P'w)}{(Q'w)}
             \end{array}
           }$
\end{enumerate}
and their symmetric counterparts (obtained by exchanging the role of $\Ppi$ and $\Qpi$).
The set $\Xscr$ is left unspecified above, since it will be constructed by induction
hypothesis (in the base case, where both $\Ppi$ and $\Qpi$ are deadlocked processes, 
define $\Xscr$ to be the empty set). 
We show here how to construct proofs for these three sequents; their symmetric counterparts can be
proved similarly. In all these three cases, we apply either the $one_f$ rule (for sequent 1) 
or the $one_b$ rule (for sequent 2 and 3). If this application of $one_f$ (or $one_b$)
results in two distinct name-variables, say $n_1$ and $n_2$, to be identified, then
the sequent is proved by using the assumption $n_1 \not = n_2.$ Therefore the only 
interesting cases are when the name-variables $n_1, \cdots, n_k$ are instantiated to 
distinct name-variables, say, $m_1, \cdots, m_k$. In the following we assume
that the substitution in the premises of $one_f$ or $one_b$ are non-trivial,
meaning that they do not violate the assumption on name-distinction above.
\begin{description}
\item[Sequent 1] In this case, after applying the $one_f$ rule bottom up
and discharging the trivial premises, we need to prove, 
for each $\theta$ associated with the rule, the sequent
$$
\NSeq{m_1, \cdots, m_k, \Sigma}{\Xscr, m_1 \not = \cdots \not = m_k}{\exists Q'. \one{\Qpi\theta}{A\theta}{Q'}
\land \lbisim{(P'\theta)}{Q'}}
$$
for some signature $\Sigma.$ We give a top-down construction of a derivation of this sequent as follows. 
By Lemma~\ref{lm:open-drv-sound}, we know that 
$$\vdash \NSeq{m_1,\cdots, m_k, \Sigma}{.}{\one{\Ppi\theta}{A\theta}{P'\theta}}.$$
Since $m_1, \ldots, m_k$ are the only free names in $\Ppi\theta$, we can show by
induction on proofs that $\Sigma$ in the sequent is redundant and can be removed,
thus 
$$\vdash \NSeq{m_1,\cdots, m_k}{.}{\one{\Ppi\theta}{A\theta}{P'\theta}}.$$
By the adequacy of one-step transition (Proposition~\ref{prop:one step}),
we have
$\one{\Ppi\theta}{A\theta}{P'\theta}.$ Notice that $\Ppi$ is a renaming of $\Ppi\theta$, since
$m_1, \ldots, m_k$ are pairwise distinct. We recall that both one-step
transitions and (late) bisimulation are closed under injective renaming 
(see, \eg, \cite{milner92icII}). Therefore, there exist $\alpha$ and $\Rpi$ such that
$\one{\Ppi}{\alpha}{\Rpi}$, where $\alpha$ and $\Rpi$ are obtained from $A\theta$ and $P'\theta$, respectively,
under the same injective renaming. Since $\Ppi$ and $\Qpi$ are bisimilar, there exists
$\Tpi$ such that $\one{\Qpi}{\alpha}{\Tpi}$, hence, by injective renaming and the adequacy
result for one-step transitions, the sequent 
$\NSeq{m_1,\cdots, m_k}{.}{\one{\Qpi\theta}{A\theta}{\Tpi\theta}}$ is provable. It remains to show
that 
$$
\vdash \NSeq{m_1,\cdots, m_k}{\Xscr, m_1 \not = \cdots \not = m_k}{\lbisim {(P'\theta)}{(\Tpi\theta)}}
$$
By induction hypothesis (note that the size of $(\Rpi,\Tpi)$ is smaller
than $(\Ppi,\Qpi)$), we have 
$$
\vdash \forall x_1 \cdots \forall x_j. \Xscr' \land x_1 \not = \cdots \not = x_j \oimp \lbisim \Rpi \Tpi
$$
where $\{x_1,\ldots,x_j\}$ is a subset of $\{n_1,\ldots, n_k\}.$
We can weaken the formula with extra variables and assumptions to get
$$
\vdash \forall n_1 \cdots \forall n_k. \Xscr' \land n_1 \not = \cdots \not = n_k \oimp \lbisim \Rpi \Tpi.
$$
Now since the $\forall R$ and $\oimp R$ rules are invertible, this means 
$$
\vdash \NSeq{n_1, \ldots, n_k}{\Xscr', n_1\not = \cdots \not = n_k}{\lbisim \Rpi \Tpi}.
$$
Now define $\Xscr$ to be $\Xscr'$ and apply a renaming
substitution which maps each $n_i$ to $m_i$, we get a derivation of
$$
\NSeq{m_1, \ldots, m_k}{\Xscr, m_1\not = \cdots \not = m_k}{\lbisim {(P'\theta)} {(\Tpi\theta)}}.
$$
Since provability is closed under weakening of signature, we have
$$
\vdash \NSeq{m_1, \ldots, m_k, \Sigma }{\Xscr, m_1\not = \cdots \not = m_k}{\lbisim {(P'\theta)} {(\Tpi\theta)}},
$$
and together with provability of 
$\NSeq{m_1,\ldots,m_k}{.}{\one{\Qpi\theta}{A\theta}{\Tpi\theta}}$, we get 
$$
\vdash \NSeq{m_1, \cdots, m_k, \Sigma}{\Xscr, m_1 \not = \cdots \not = m_k}{\one{\Qpi\theta}{A\theta}{\Tpi\theta}
\land \lbisim{(P'\theta)}{\Tpi\theta}}.
$$
Finally, applying an $\exists R$ to this sequent, we get
$$
\vdash \NSeq{m_1, \cdots, m_k, \Sigma}{\Xscr, m_1 \not = \cdots \not = m_k}{\exists Q'. \one{\Qpi\theta}{A\theta}{Q'}
\land \lbisim{(P'\theta)}{Q'}}. 
$$

\item[Sequent 2.]
In this case, we need to prove the sequent
$$
(*) ~ \NSeq{m_1, \cdots, m_k, \Sigma}{\Xscr, m_1 \not = \cdots \not = m_k}{\exists Q'. \onep{\Qpi\theta}{\inact {X\theta}}{Q'}
\land \forall w. \lbisim{((P'\theta) w)}{(Q' w)}}
$$
for each non-trivial $\theta$ in the premises of $one_b$ rule. 
By the same reasoning as in the previous case, we obtain, for every transition
$
\one{P\theta}{x(w)}{\Rpi},
$
where $\Rpi = (P'\theta)\,w,$ another transition
$
\one{Q\theta}{x(w)}{\Tpi}
$
such that for all name $z$
$
\Rpi[z/w] \sim_l  \Tpi[z/w].
$
It is enough to consider $k$+1 cases for $z$, \ie, those in which
$z$ is one of $m_1, \ldots, m_k$ and another where $z$ is a new name,
say $m_{k+1}.$
By induction hypothesis, we have, for each $i \in \{1,\ldots, k\}$, a provable formula
$F_i$
$$
\forall m_1 \cdots \forall m_{k}. \Xscr_1 \land m_1 \not = \cdots \not = m_{k} \oimp
\lbisim{(\Rpi[m_i/w])}{(\Tpi[m_i/w])}
$$
and a provable formula $F_{k+1}$:
$$
\forall m_1 \cdots \forall m_{k+1}. \Xscr_{k+1} \land m_1 \not = \cdots \not = m_{k+1} \oimp
\lbisim{(\Rpi[m_{k+1}/w])}{(\Tpi[m_{k+1}/w])}.
$$
Let $\Xscr$ be the set
$
\{\forall x \forall y. x = y \lor x \not = y \}  \cup 
\{\Xscr_i \mid i \in \{1,\ldots,k+1\}  \}. 
$
Then the sequent $(*)$ is proved, in a bottom-up fashion, by instantiating $Q'$ to $\lambda w. \Tpi$,
followed by an $\landR$-rule, resulting in the sequents:
$$
\NSeq{m_1, \ldots, m_k, \Sigma}{\Xscr, m_1 \not = \cdots \not = m_k}{\one{\Qpi \theta}{A\theta}{\lambda w.\Tpi}} 
\hbox{\qquad and}
$$
$$
\NSeq{m_1, \ldots, m_k, \Sigma}{\Xscr, m_1 \not = \cdots \not = m_k}{\forall w. \lbisim{\Rpi}{\Tpi}} 
$$
The first sequent is provable following the adequacy of one-step transition.
For the second sequent, we apply the $\forall R$-rule to get the sequent
$$
\NSeq{m_1, \ldots, m_k, m_{k+1}, \Sigma}{\Xscr, m_1 \not = \cdots \not = m_k}
   {\lbisim{(\Rpi[m_{k+1}/w])}{(\Tpi[m_{k+1}/w])}}. 
$$
We then do a case analysis on the name $m_{k+1}$, using the assumption
$\forall x\forall y. x = y \lor x \not = y$ in $\Xscr.$ 
Let $\Rpi_{k+1} = \Rpi[m_{k+1}/w]$ and let $\Tpi_{k+1} = \Tpi[m_{k+1}/w].$
We consider $k$ instantiations, each instantiation compares $m_{k+1}$
with $m_i$, for $i \in \{1,\dots, k\}.$ We thus get the following sequents:
$$
\begin{array}{ll}
(S_1) & \NSeq{\Sigma'}{\Delta, m_1 = m_{k+1}}
   {\lbisim{\Rpi_{k+1}}{\Tpi_{k+1}}} \\
(S_2) & \NSeq{\Sigma'}{\Delta, m_1 \not = m_{k+1}, m_2 = m_{k+1}} {\lbisim{\Rpi_{k+1}}{\Tpi_{k+1}}} \\
 & \vdots \\
(S_k) & \NSeq{\Sigma'}{\Delta, m_1 \not = m_{k+1}, \ldots, m_{k-1} \not = m_{k+1}, m_k = m_{k+1}}
   {\lbisim{\Rpi_{k+1}}{\Tpi_{k+1}}} \\
(S_{k+1}) & \NSeq{\Sigma', m_{k+1}}{\Delta,
m_1 \not = m_{2}, \cdots, m_{k-1} \not = m_{k}, m_k \not = m_{k+1}}
   {\lbisim{\Rpi_{k+1}}{\Tpi_{k+1}}} \\
\end{array}
$$
Here $\Sigma'$ denotes the set $\{m_1, \ldots, m_{k+1}\} \cup \Sigma$ and
$\Delta$ denotes the set $\{\Xscr, m_1 \not = \cdots \not = m_k \}.$
Provability of these sequents follow from provability of $F_1, \ldots, F_{k+1}.$

\item[Sequent 3] In this case, we need to prove the sequent
$$
(**) \qquad \NSeq{m_1, \cdots, m_k, \Sigma}{\Xscr, m_1 \not = \cdots \not = m_k}
{
\begin{array}{l}
\exists Q'. \one{\Qpi\theta}{A\theta}{Q'} ~ \land \\
\qquad \nabla w. \lbisim{((P'\theta) w)}{(Q' w)}
\end{array}
}
$$
for each non-trivial $\theta$ in the premises of $one_b$ rule. As in the previous case, we obtain
$\Rpi$ and $\Tpi$ such that
$
\one{\Ppi \theta}{\bar x(w)}{\Rpi}
$
and
$
\one{\Qpi \theta}{\bar x(w)}{\Tpi}
$
where $\lambda w.\Rpi = P'\theta.$ We assume, without loss of generality, that $w$ is fresh.
By the induction hypothesis, $\Rpi \sim_l \Tpi$ and 
$$
\vdash\forall m_1 \cdots \forall m_k \forall w. \Xscr' \land m_1 \not = \cdots \not = m_k \not = w \oimp 
\lbisim{\Rpi}{\Tpi}.
$$
Now apply Proposition~\ref{prop:forall nabla} to replace $\forall w$ with $\nabla w$,
$$
\vdash\forall m_1 \cdots \forall m_k \nabla w. \Xscr' \land m_1 \not = \cdots \not = m_k \not = w \oimp 
\lbisim{\Rpi}{\Tpi}.
$$
And since $\nabla$ distributes over all propositional connectives, we also have
$$
\vdash\forall m_1 \cdots \forall m_k.  (\nabla w \Xscr') \land \nabla w. (m_1 \not = \cdots \not = m_k) \land 
\nabla w (\bar m \not = w) \oimp \nabla w. \lbisim{\Rpi}{\Tpi}.
$$
Let $\Xscr = \nabla w \Xscr'$.
Now, since the right-introduction rules for $\forall$, $\nabla$ and $\oimp$ are all invertible,
we have that the sequent
$$
(i) \qquad \NSeq{m_1,\ldots,m_k}{\Xscr, \nabla w. (m_1 \not = \cdots \not = m_k), 
\nabla w (\bar m \not = w)}{\nabla w. \lbisim{\Rpi}{\Tpi}}
$$
is provable. It can be easily checked that the following sequents are provable:
$$
\begin{array}{l}
\nabla w. m_i \not = m_j \seqsym m_i \not = m_j, \hbox{ for any $i$ and $j$.} \\
\seqsym \nabla w. m_i \not = w, \hbox{ for any $i$ (since $w$ is in the scope of $m_i$).}
\end{array}
$$
By applying the cut rules to these sequents and sequent $(i)$ above, we
obtain
$$
(ii) \qquad \NSeq{m_1,\ldots,m_k}{\Xscr, m_1 \not = \cdots \not = m_k}{\nabla w. \lbisim{\Rpi}{\Tpi}},
$$
Provability of sequent $(**)$ then follows from provability of sequent $(ii)$ above
and the adequacy of the one-step transition (\ie, by instantiating $Q'$
with $\lambda w.\Tpi$).\qed
\end{description}
\end{proof}

The following lemma shows that {\em lbisim} is symmetric. Its proof is
straightforward by induction on derivations. 

\begin{lemma}
\label{lm:lbisim-sym}
Let $\Ppi$ and $\Qpi$ be two $\pi$-processes and let $\bar n$ be the
list of all free names in $\Ppi$ and $\Qpi$.
If $\vdash \Xscr \oimp \nabla \bar n. \lbisim \Ppi \Qpi$, for
some $\Xscr \subset \Escr$, then 
$\vdash \Xscr \oimp \nabla \bar n. \lbisim \Qpi \Ppi.$
\end{lemma}

\subsection{Proof for Theorem~\ref{thm:lbisim} (adequacy of late bisimulation specification)}

{\em Soundness.} 
We define a set $\Sscr$ as
$$
\Sscr = \{ (\Ppi, \Qpi) \mid \vdash \Xscr \oimp \nabla \bar{n}\ \lbisim{\Ppi}{\Qpi} \mbox{, where 
$\fn{\Ppi, \Qpi} \subseteq \{\bar{n}\}$ and $\Xscr \subseteq_f \Escr$} \}
$$
and show that $\Sscr$ is a bisimulation, \ie, it is symmetric and 
closed with respect to the conditions 1, 2 and 3 in Definition~\ref{def:lbisim}.
The symmetry of $\Sscr$ follows from Lemma~\ref{lm:lbisim-sym}.

Suppose that $(\Ppi, \Qpi) \in \Sscr$, that is, 
$\vdash \Xscr \oimp \nabla \bar{n}\ \lbisim{\Ppi}{\Qpi}$ for some $\Xscr$. 
Since $\defR$ on $lbisim$ is invertible (Lemma~\ref{lm:lbisim-inv}), and since
$\landR$, $\oimpR$, $\nablaR$ and $\forallR$ are also invertible, there is a proof of the formula that ends 
with applications of these invertible rules. From this and the definition of $lbisim$, 
we can infer that provability of $\Xscr \oimp \nabla \bar n \ \lbisim \Ppi \Qpi$
implies provability of six other sequents, three of which are given 
in the following (the other three are symmetric counterparts of these):
$$
\begin{array}{cl}
(a) & \NSeq{P',A}{\Xscr, \Judg{\bar{n}}{\one{\Ppi}{A\bar{n}}{(P'\bar{n})}}}
                {\Judg{\bar{n}}{\exists Q'.\one{\Qpi}{A\bar{n}}{Q'}~\land~\lbisim{(P'\bar{n})}{Q'}}} \\
(b) & \NSeq{M,X}{\Xscr, \Judg{\bar{n}}{\onep{\Ppi}{\inact (X\bar{n})}{(M\bar{n})}}}
                {\Judg{\bar{n}}{\exists N.
                    \onep{\Qpi}{\inact (X\bar{n})}{N}~\land~
                    \forall y. \lbisim{(M\bar{n}y)}{(Ny)} 
                }} \\   
(c) & \NSeq{M,X}{\Xscr,\Judg{\bar{n}}{\onep{\Ppi}{\outact (X\bar{n})}{(M\bar{n})}}}
                {\Judg{\bar{n}}{\exists N.\onep{\Qpi}{\outact (X\bar{n})}{N}~\land~
                 \nabla y.\lbisim{(M \bar{n}y)}{(Ny)}               }
                }                       
\end{array}
$$
By examing the structure of proofs of these three sequents, we show 
that $\Sscr$ is closed under all possible transitions from $\Ppi$ and $\Qpi.$
We examine the three cases in Definition~\ref{def:lbisim}:

\noindent (1) Suppose $\one{\Ppi}{\alpha}{\Ppi'}$ for some free action $\alpha.$ 
Since $\one{\Ppi}{\alpha}{\Ppi'}$, by the adequacy result for one-step transitions,
we have that $\Judg{\bar n}{\one{\Ppi}{\alpha}{\Ppi'}}$ is derivable.
Let $\rho = [\lambda \bar n. \alpha / A, \lambda \bar n.\Ppi' / P']$. Applying $\rho$ to the derivation
of sequent $(a)$, we get 
$$
\vdash \NSeq{\cdot}{\Xscr, \Judg{\bar{n}}{\one{\Ppi}{\alpha}{\Ppi'}}}
                {\Judg{\bar{n}}{\exists Q'.\one{\Qpi}{\alpha}{Q'}~\land~\lbisim{\Ppi'}{Q'}}}. 
$$
By a cut between $\Judg{\bar n}{\one{\Ppi}{\alpha}{\Ppi'}}$ and this sequent,
we obtain a derivation of 
$$
\NSeq{\cdot}{\Xscr}{\Judg{\bar{n}}{\exists Q'.\one{\Qpi}{\alpha}{Q'}~\land~\lbisim{\Ppi'}{Q'}}}. 
$$
By Lemma~\ref{lm:right-first}, we know that there exists a derivation of this sequent
which ends with a right-rule, hence, there exists a process $\Qpi'$ such that
$\vdash \NSeq{\cdot}{\Xscr}{\Judg{\bar n}{\one{\Qpi}{\alpha}{\Qpi'}}}$
and
$\vdash \NSeq{\cdot}{\Xscr}{\Judg{\bar n}{\lbisim{\Ppi'}{\Qpi'}}.}$
It is easy to show that $\Xscr$ plays no part in the proof of the first sequent,
so it can be removed from the sequent. Hence by the adequacy of one-step transitions,
we have $\one{\Qpi}{\alpha}{\Qpi'}$. Provability of the second sequent implies
that $(\Ppi', \Qpi')$ is in the set $\Sscr$.  Thus
$\Sscr$ is indeed closed under the $\alpha$-transition.

\noindent(2) Suppose that $\one{\Ppi}{a(y)}{\Ppi'}.$
Applying a similar argument as in the previous case to sequent $(b)$ with
substitution $\rho = [\lambda \bar n.a/X, \lambda \bar n\lambda y.\Ppi'/M]$,
we obtain a provable sequent 
$$
\NSeq{\cdot}{\Xscr}{\Judg{\bar{n}}{\exists N.
                    \onep{\Qpi}{\inact a}{N}~\land~
                    \forall y. \lbisim{\Ppi'}{(Ny)}}}.
$$
Again, as in the previous case, using Lemma~\ref{lm:right-first}, we can show that
$\one{\Qpi}{a(y)}{\Qpi'}$ for some process $\Qpi'$ such that
$
\vdash \NSeq{\cdot}{\Xscr}{\Judg{\bar n}{\forall y. \lbisim {\Ppi'} {\Qpi'}}}.
$
This implies that 
$$
\begin{array}{cl}
(i) & \Xscr \oimp \nabla \bar n \forall y. \lbisim {\Ppi'}{\Qpi'},\\
(ii) & \Xscr \oimp \nabla \bar n. \lbisim {(\Ppi'[w/y])}{(\Qpi'[w/y])} \hbox {, where $w \in \{\bar n\},$} \\
(iii) & (\nabla y. \Xscr) \oimp \nabla y\nabla \bar n. \lbisim {\Ppi'} {\Qpi'}
\end{array}
$$
are all provable.
The formula $(ii)$ is obtained from $(i)$ by instantiating $y$ with one of $\bar n.$
The formula $(iii)$ is obtained from $(i)$ as follows:
Since 
$$
\nabla x \forall y. P\,x\,y \oimp \forall y\nabla x.P\,x\,y
\quad
\hbox{ and }
\quad
(A \oimp \forall y.B) \oimp (\forall y.(A \oimp B)),
$$
where $y$ is not free in $A$, are theorems of $\FOLDNb$, 
we can enlarge the scope of $y$ in $(i)$ to the outermost level:
hence, we have that $\forall y(\Xscr \oimp \nabla \bar n.\lbisim {\Ppi'} {\Qpi'})$
is provable. Now apply Proposition~\ref{prop:forall nabla} to turn $\forall y$ into $\nabla y$, then distribute
the $\nabla y$ over the implication $\oimp$ and conjunction $\land$, and we have $(iii).$

It remains to show that for every name $w$, the pair
$(\Ppi'[w/y], \Qpi'[w/y])$ is in $\Sscr.$ There are two cases to consider:
The case where $w$ is among $\bar n$ follows straightforwardly from $(ii)$,
the other case, where $w$ is a new name, follows from $(iii).$

\noindent(3) Suppose $\one{\Ppi}{\bar a(y)}{\Ppi'}.$ Using the same argument as in the previous case,
we can show that there exists a process $\Qpi'$ such that $\one{\Qpi}{\bar a(y)}{\Qpi'}$
and such that
$$
\vdash \NSeq{\cdot}{\Xscr}{\Judg{\bar n}{\nabla y. \lbisim {\Ppi'} {\Qpi'}}}.
$$
The latter entails that $(\Ppi', \Qpi') \in \Sscr$, as required. \qed

\subsection{Completeness} We are given $\Ppi \sim_l \Qpi$ and
we need to show that $\vdash \Xscr \oimp \nabla\bar{n}.\lbisim \Ppi \Qpi,$ where 
$\Xscr \subseteq_f \Escr$ and $\bar n = \{n_1, \ldots, n_k \}$ includes all the free names in
$\Ppi$ and $\Qpi.$ From Lemma~\ref{lm:lbisim-complete} we have that 
$$
\vdash \forall n_1 \cdots \forall n_k (\Xscr' \land n_1 \not = \cdots \not = n_k \oimp \lbisim \Ppi \Qpi)
$$
for some $\Xscr' \subseteq_f \Escr$. By Proposition~\ref{prop:forall nabla},
we can turn all the $\forall$ into $\nabla$, hence
$$
\vdash \nabla n_1 \cdots \nabla n_k (\Xscr' \land n_1 \not = \cdots \not = n_k \oimp \lbisim \Ppi \Qpi).
$$
Since $\nabla$ distributes over all propositional connectives, we have
$$
\vdash (\nabla \bar n. \Xscr') \land \nabla \bar n. (n_1 \not = \cdots \not = n_k) \oimp \nabla \bar n. \lbisim \Ppi \Qpi.
$$
Now, $\nabla \bar n. n_1 \not = \cdots \not = n_k$ is a theorem of $\FOLDNb$ (since any
two distinct $\nabla$-quantified names are not equal), therefore by modus ponens
we have
$$
\vdash \nabla \bar n. \Xscr' \oimp \nabla \bar n. \lbisim \Ppi \Qpi.
$$
Let $\Xscr = \nabla \bar n.\Xscr'$, then we have $\Xscr \oimp \nabla \bar n.\lbisim \Ppi \Qpi$
as required.
\qed

\subsection{Adequacy of the specification of early bisimulation}

The proof for the adequacy of the specification of early bisimulation follows a 
similar outline as that of late bisimulation. The proof is rather tedious and is not
enlightening. We therefore omit the proof and refer interested readers to
the electronic appendix of the paper for more details.

\subsection{Adequacy of the specification of open bisimulation}

{\em Proof of Lemma~\ref{lm:prefix}: }
The proof proceeds by induction on the length of the quantifier prefix $\Qscr \bar x.$
At each stage of the induction, we construct a quantifier prefix $\Qscr \bar y$
such that $\Qscr \bar x.P \oimp \Qscr \bar y.P\theta$ and $D\theta$ corresponds to
the $\Qscr \bar y$-distinction.
In the base case, where the quantifier prefix $\Qscr \bar x$ is empty, the quantifier
$\Qscr \bar y$ is also the empty prefix. In this case we have
$P\theta = P$, therefore $P \oimp P\theta$ holds trivially.
There are the following two inductive cases.

\noindent (1) Suppose $\Qscr \bar x. P = \Qscr' \bar u \nabla z. P.$
Let $D'$ be the distinction that corresponds to $\Qscr' \bar u.$
Note that by definition, we have
$D = D' \cup \{(z, v), (x, v) \mid v \in D' \}$.
Let $\theta'$ be the substitution $\theta$ with domain restricted to $\{ \bar u\}.$
Since $\theta$ respects $D$, obviously $\theta'$ respects $D'$ and 
$\theta(z) \not = \theta(v)$ for all $v \in D'.$ By induction hypothesis, we have
a proof of the formula
$\Qscr \bar u (\nabla z. P) \oimp \Qscr \bar m (\nabla z.P)\theta'$ 
for some quantifier prefix $\Qscr \bar m$ such that
$D'\theta'$ is the $\Qscr \bar m$-distinction.
Note that since $z$ is not in the domain of $\theta'$, 
we have $(\nabla z.P)\theta' = \nabla z.(P\theta').$
Let $w = \theta(z).$ Since $w$ is distinct from all other free names in
$D'\theta'$, we can rename $z$ with $w$, thus, 
$$
\vdash \Qscr \bar u \nabla z.P \oimp \Qscr \bar m \nabla w.P(\theta' \circ [w/z])
$$
But $\theta' \circ [w/z]$ is exactly $\theta$. Let $\Qscr \bar y$ be the
prefix $\Qscr \bar m\nabla w.$ It then follows that 
$$
\vdash \Qscr \bar x.P \oimp \Qscr \bar y.P\theta.
$$
Moreover, $D\theta$ can be easily shown to be the $\Qscr \bar y$-distinction.

\noindent(2) Suppose $\Qscr \bar x = \Qscr' \bar u \forall z.P.$
Note that in this case, the $\Qscr \bar x$-distinction and
$\Qscr' \bar u$-distinction co-incide, \ie, both are the same
distinction $D.$ Moreover, $z \not \in \fv{D}.$
Let $\theta'$ be the substitution $\theta$ restricted to the domain
$\{\bar u \}.$
By induction hypothesis, we have that 
$\vdash \Qscr \bar u (\forall z.P) \oimp \Qscr \bar m (\forall z.P)\theta'$,
for some quantifier prefix $\Qscr \bar m$ 
such that $\Qscr \bar m$ corresponds to $D\theta'.$
Note that $D\theta' = D\theta$, because $z \not \in \fv D.$
There are two cases to consider when constructing $\Qscr \bar y.$
The first case is when $z$ is identified, by $\theta$, with
some name in $\{\bar u \}.$ In this case, by the property of 
universal quantification, we have that 
$\vdash \Qscr \bar u \forall z.P \oimp \Qscr \bar m.P \theta$.
In this case, we let $\Qscr \bar y = \Qscr \bar m.$
Note that $D\theta'$ is the same as $D\theta$ in this case.
Therefore $D\theta$ is the $\Qscr \bar y$-distinction. 
For the second case, we have that $z$ is instantiated by $\theta$
to a new name, say $w$. 
Then following the same argument as the case with $\nabla$, we have that
$\vdash \Qscr \bar u \forall z.P \oimp \Qscr \bar m \forall w.P\theta$.
In this case, we let $\Qscr \bar y = \Qscr \bar m \forall w.$
Note that in this case the $\Qscr \bar y$-distinction also coincides
with $\Qscr \bar m$-distinction, \ie, both are the same set
$D\theta.$ \qed

In the proof of soundness of open bisimulation to follow, we make use of
a property of the structure of proofs of certain sequents. 
The following three lemmas state some meta-level properties of $\FOLDNb$.
Their proofs are easy and are omited.

\begin{lemma}
\label{lm:distinction}
Suppose the sequent $\NSeq{\Sigma}{\Delta}{\Cscr}$ is provable, where $C$ is an existential
judgment and $\Delta$ is a set of inequality between distinct terms, \ie, every element of
$\Delta$ is of the form $\Judg{\bar n}{s \not = t}$, for some $\bar n$, $s$ and $t.$
Then there exists a proof of the sequent ending with $\existsR$ applied to $C.$
\end{lemma}

\begin{lemma}
\label{lm:context}
For any positive formula context $C[]$, $\vdash C[\forall x.B] \oimp C[B[t/x]].$ 
\end{lemma}

\begin{lemma}
\label{lm:modus ponens}
Let $\Qscr \bar x$ be a quantifier prefix. 
If $\Qscr \bar x. P$ and 
$\Qscr \bar x. P \oimp Q$ are provable 
then $\Qscr \bar x. Q$ is provable.
\end{lemma}

\begin{lemma}
\label{lm:nabla distinction}
Let $D$ be a conjunction of inequalities between terms.
If $\vdash \Qscr \bar x.D \oimp \nabla y. P$, where $y$ is not free in $D$, 
then  $\vdash \Qscr \bar x \nabla y. D \oimp P$.
\end{lemma}

The following lemma is a simple corollary of Proposition~\ref{prop:one step}
and Proposition~\ref{prop:nabla forall}.
\begin{lemma}
\label{lm:mixed prefix one step}
$\one{\Ppi}{\alpha}{\Qpi}$ if and only if $\Qscr \bar n.\trans{\one{\Ppi}{\alpha}{\Qpi}}$
is provable, where $\Qscr \bar n$ is a quantifier prefix and $\bar n$ are the free names of $\Ppi.$
\end{lemma}

To prove soundness of open bisimulation specification, we define 
a family of sets $\Sscr$ in the following, and show that it is indeed an open
bisimulation.
$$
\Sscr_D = \{ (\Ppi, \Qpi) \mid  
\begin{array}[t]{l}
\mbox{$\vdash \Qscr \bar{n}. [D'] \oimp \lbisim \Ppi \Qpi$ and
      $\fn{\Ppi,\Qpi, D'} = \{\bar{n}\}$ and}\\
\mbox{$D = D' \cup D''$, where $D''$ is the $\Qscr\bar{n}$-distinction. $\}$} 
\end{array}
$$
Suppose $(\Ppi, \Qpi) \in \Sscr_D.$ That is, 
$
\vdash \Qscr \bar n. [D'] \oimp \lbisim \Ppi \Qpi.
$
Let $D''$ be the distinction that corresponds to the prefix $\Qscr \bar n.$
We have to show that for every name substitution $\theta$ which respects $D$,
the set $\Sscr$ is closed under conditions 1, 2, and 3 in Definition~\ref{def:obisim}.
Since $\theta$ respects $D$, it also respects $D''$ (since $D''$ is a subset of $D$).
Therefore, it follows from Lemma~\ref{lm:prefix} that there exists a prefix $\Qscr \bar x$
such that $D''\theta$ is the $\Qscr \bar x$-distinction, and 
$\vdash \Qscr \bar x. [D'\theta] \oimp \lbisim {(\Ppi\theta)} {(\Qpi\theta)}$.
By the invertibility of $\defR$ on $lbisim$ and the right-introduction rules 
for $\forall$, $\nabla$, $\oimp$ and $\land$, we can infer that provability of the
above formula implies provability of six other formulas, three of which
are given in the following (the other three are symmetric variants of these formulas):
$$
\begin{array}{ll}
(a) & \Qscr \bar x. [D'\theta] \oimp \forall P'\forall A. \one{\Ppi\theta}{A}{P'} 
\oimp \exists Q'. \one{\Qpi\theta}{A}{Q'} \land \lbisim {P'} {Q'} \\
(b) & \Qscr \bar x. [D'\theta] \oimp \forall M\forall X. \onep{\Ppi\theta}{\inact X}{M} 
\oimp \exists N. \onep{\Qpi\theta}{\inact X}{N} \land \forall w. \lbisim {(M w)} {(N w)} \\
(c) & \Qscr \bar x. [D'\theta] \oimp \forall M\forall X. \onep{\Ppi\theta}{\outact X}{M} 
\oimp \exists N. \onep{\Qpi\theta}{\outact X}{N} \land \nabla w. \lbisim {(M w)} {(N w)} \\
\end{array}
$$
Using provability of these formulas, we show that $\Sscr$ is closed under
free actions, bound input actions and bound output actions.

\begin{itemize}
\item Suppose $\one{\Ppi\theta}{\alpha}{\Rpi}$ where $\alpha$ is a free action. 
By Lemma~\ref{lm:mixed prefix one step}, we have that 
\begin{equation}
\label{eq:obisim1}
\vdash \Qscr \bar x. \one{\Ppi\theta}{\alpha}{\Rpi}.
\end{equation}
From formula $(a)$ and Lemma~\ref{lm:context}, we have that
\begin{equation}
\label{eq:obisim2}
\vdash \Qscr \bar x. [D'\theta] \oimp \one{\Ppi\theta}{\alpha}{\Rpi} 
\oimp \exists Q'. \one{\Qpi\theta}{\alpha}{Q'} \land \lbisim {\Rpi} {Q'}.
\end{equation}
Applying Lemma~\ref{lm:modus ponens} to formula (\ref{eq:obisim1}) and (\ref{eq:obisim2}) above,
we have that 
$$
\vdash \Qscr \bar x. [D'\theta] \oimp \exists Q'. \one{\Qpi\theta}{\alpha}{Q'} \land \lbisim {\Rpi} {Q'}.
$$
The latter implies, by the invertibility of the right rules for $\nabla$ and $\forall$,
provability of the sequent
$$
\NSeq{\Sigma}{D_1}{\Judg{\bar m}{\exists Q'. \one{\Qpi'}{\alpha'}{Q'} \land \lbisim {\Rpi'} {Q'}}}
$$
where $\Sigma$ are the eigenvariables corresponding to the universally quantified variables in $\Qscr \bar x$
(with appropriate raising) and $\bar m$ corresponds to the $\nabla$-quantified variables
in the same prefix. The terms $\Qpi'$, $\Rpi'$, $D_1$ and $\alpha'$ are obtained from, respectively,  
$\Qpi\theta$, $\Rpi$, $[D'\theta]$ and $\alpha$ by replacing their free names with their raised counterparts.
Note that since $\theta$ respects $D'$, the inequality in $D_1$ are those that relate
distinct terms, hence, by Lemma~\ref{lm:distinction}, provability of the above sequent implies
the existence of a term $T$ such that 
$\vdash \NSeq{\Sigma}{D_1}{\Judg{\bar m}{\one{\Qpi'}{\alpha'}{T'}}}$
and
\begin{equation}
\label{eq:obisim4}
\vdash \NSeq{\Sigma}{D_1}{\Judg{\bar m}{\lbisim {\Rpi'} {T'}}}.
\end{equation}
It can be shown by induction on the height of derivations that $D_1$ in the first sequent can
be removed, hence we have that
$$
\vdash \NSeq{\Sigma}{.}{\Judg{\bar m}{\one{\Qpi'}{\alpha'}{T'}}}.
$$
Applying the appropriate introduction rules to this sequent (top down), we 
``unraise'' the variables in $\Sigma$ and obtain 
$
\vdash \Qscr \bar x. \one{\Qpi\theta}{\alpha}{\Tpi}, 
$
where $\Tpi$ corresponds to $T'.$ By Lemma~\ref{lm:mixed prefix one step}, this means that 
$\one{\Qpi\theta}{\alpha}{\Tpi}.$
It remains to show that $(\Rpi,\Qpi) \in \Sscr.$ This is obtained from the sequent
(\ref{eq:obisim4}) above as follows. 
We apply the introduction rules for quantifiers and implication (top down) to sequent (\ref{eq:obisim4}), 
hence unraising the variables in $\Sigma$ and obtain the provable formula
$
\Qscr \bar x. [D'\theta] \oimp \lbisim {\Rpi}{\Tpi},
$
from which it follows that $(\Rpi, \Tpi) \in \Sscr_{D\theta }.$

\item Suppose $\one{\Ppi\theta }{a(y)}{\Rpi}.$
As in the previous case, using Lemma~\ref{lm:mixed prefix one step}, Lemma~\ref{lm:context}, 
Lemma~\ref{lm:modus ponens} and formula $(b)$
we can show that 
$$
\vdash \Qscr \bar x. [D'\theta] \oimp \exists Q'. \onep{\Qpi\theta}{\inact a}{N} \land \forall w.\lbisim {(\Rpi[w/y])} {(N\,y)}.
$$
From this formula, we can show that there exists $\Tpi$ such that $\Qscr \bar x. \onep{\Qpi\theta}{\inact a}{\lambda z.\Tpi}$,
therefore $\one{\Qpi\theta}{a(z)}{\Tpi}$, and that 
\begin{equation}
\label{eq:obisim5}
\vdash \Qscr \bar x. [D'\theta] \oimp \forall w.\lbisim {(\Rpi[w/y])}{(\Tpi[w/z])}.
\end{equation}
We need to show that for a fresh name $w$, $(\Rpi[w/y], \Tpi[w/z]) \in \Sscr_{D\theta}.$ 
From provability of formula (\ref{eq:obisim5}), and the fact that
$(A \oimp \forall x.B) \oimp \forall x(A \oimp B),$ we obtain
$$
\vdash \Qscr \bar x \forall w. [D'\theta] \oimp \lbisim {(\Rpi[w/y])}{(\Tpi[w/z])}.
$$
Since the $\Qscr\bar x\forall w$-distinction is the same as $\Qscr \bar x$-distinction,
the overal distinction encoded in the above formula is $D\theta$, therefore, by definition
of $\Sscr$, we have 
$(\Rpi[w/y], \Tpi[w/z]) \in \Sscr_{D\theta}.$ 

\item Suppose $\one{\Ppi\theta}{\bar a(y)}{\Rpi}.$ This case is similar to the bound input
case. Applying the same arguments shows that there exists a process $\Tpi$
such that $\one{\Qpi\theta}{a(z)}{\Tpi}$ and 
\begin{equation}
\label{eq:obisim6}
\vdash \Qscr \bar x. [D'\theta] \oimp \nabla w. \lbisim {(\Rpi[w/y])}{(\Tpi[w/z])}.
\end{equation}
We have to show that, for a fresh $w$, 
$(\Rpi[w/y], \Tpi[w/z]) \in \Sscr_{D_2}$ where $D_2 = D\theta \cup \{w \} \times \fn{D\theta , \Ppi\theta, \Qpi\theta }.$
Note that the free names of $D\theta$, $\Ppi\theta$ and $\Qpi\theta$ are all in $\bar x$ by definition.
From formula (\ref{eq:obisim6}) and Lemma~\ref{lm:nabla distinction}, we have that
$$
\vdash \Qscr \bar x \nabla w. [D'\theta] \oimp \lbisim {(\Rpi[w/y])}{(\Tpi[w/z])}.
$$
Notice that the $\Qscr \bar x \nabla w$-distinction is 
$D''\theta \cup \{w \} \times \{\bar x \}$, and since $\bar x$ is the free names of $D\theta$, $\Ppi\theta$
and $\Qpi\theta$, the overall distinction encoded by the above formula is exactly
$D_2,$ hence $(\Rpi[w/y], \Tpi[w/z]) \in \Sscr_{D_2}$ as required. \qed
\end{itemize}

The proof of Theorem~\ref{thm:open bisim complete}
is analogous to the completeness proof for Theorem~\ref{thm:lbisim}.
Suppose $\Ppi$ and $\Qpi$ are open $D$-bisimilar. We construct a derivation of
the formula 
\begin{equation}
\label{eq:obisim-completeness}
\forall n_1 \cdots \forall n_k ([D] \oimp \lbisim \Ppi \Qpi)
\end{equation}
by induction on the number of action prefixes in $\Ppi$ and $\Qpi.$ 
By applying the introduction rules
for $\forall$, $\oimp$ and unfolding the definition of $lbisim$ (bottom up) 
to the formula (\ref{eq:obisim-completeness}), we get the following sequents:
\begin{enumerate}
\item $\NSeq{n_1, \cdots, n_k, A, P'}{[D], \one{\Ppi}{A}{P'}}
           {\exists Q'. \one {\Qpi}{A}{Q'} \land \lbisim{P'}{Q'}}$
\item $\NSeq{n_1, \cdots, n_k, X, P'}{[D], \onep{\Ppi}{\inact X}{P'}}
           {\exists Q'. \onep {\Qpi}{\inact X}{Q'} \land \forall w. \lbisim{(P'w)}{(Q'w)}}$
\item $\NSeq{n_1, \cdots, n_k, X, P'}{[D], \onep{\Ppi}{\outact X}{P'}}
           {\exists Q'. \onep {\Qpi}{\outact X}{Q'} \land \nabla w. \lbisim{(P'w)}{(Q'w)}}$
\end{enumerate}
and their symmetric counterparts. 
We show here how to construct proofs for these three sequents; the rest can be
proved similarly. In all these three cases, we apply either the $one_f$ rule (for sequent 1) 
or the $one_b$ rule (for sequent 2 and 3). If this application of $one_f$ (or $one_b$)
results in two distinct name-variables, say $n_1$ and $n_2$, in $D$ to be identified, then
the sequent is proved by using the assumption $n_1 \not = n_2$ in $D$. Therefore the only 
interesting cases are when the instantiations of name-variables $n_1, \cdots, n_k$ respect the
distinction $D$. 
In the following we assume the names $n_1, \ldots, n_k$ are instantiated to 
$m_1, \ldots, m_l$ and the distinction $D$ is respected. 
Note that $l$ may be smaller than $k$, depending on $D$, \ie, it may allow
some names to be identified. 

\begin{description}
\item[Sequent 1] In this case, after applying the $one_f$ rule bottom up
and discharging the trivial premises (\ie, those that violates the distinction $D$), 
we need to prove, for each $\theta$ associated with the rule, the sequent
\begin{equation}
\label{eq: open-comp1}
\NSeq{m_1, \cdots, m_l, \Sigma}{[D\theta]}{\exists Q'. \one{\Qpi\theta}{A\theta}{Q'}
\land \lbisim{(P'\theta)}{Q'}}
\end{equation}
for some signature $\Sigma.$  
By Lemma~\ref{lm:open-drv-sound}, we know that 
$\NSeq{m_1,\cdots, m_l, \Sigma}{.}{\one{\Ppi\theta}{A\theta}{P'\theta}}$ is provable.
Since $m_1, \ldots, m_l$ are the only free names in $\Ppi\theta$, we can show by
induction on proofs that $\Sigma$ in the sequent is redundant and can be removed,
thus the sequent
$\NSeq{m_1,\cdots, m_l}{.}{\one{\Ppi\theta}{A\theta}{P'\theta}}$ is also provable.
By the adequacy of one-step transition (Proposition~\ref{prop:one step})
and Proposition~\ref{prop:nabla forall}, we have
$\one{\Ppi\theta}{\alpha}{\Rpi}$ for some free action $\alpha$ and $\Rpi$
where $\alpha = A\theta$ and $P'\theta = \Rpi.$
Let $\theta'$ be $\theta$ with domain restricted to $\{n_1, \ldots, n_k \}.$
Obviously, $\theta'$ respects $D$ and $D\theta' = D\theta.$
Since $\Ppi$ and $\Qpi$ are open $D$-bisimilar, we have that
there exists $\Tpi$ such that $\one{\Qpi\theta}{\alpha}{\Tpi}$
and $\Rpi \sim_o^{D\theta'} \Tpi,$ hence by induction hypothesis, we have that
\begin{equation}
\label{eq: open-comp2}
\vdash \forall m_1 \cdots \forall m_l. [D\theta] \oimp \lbisim {P'\theta}{\Tpi}.
\end{equation}
Provability of sequent (\ref{eq: open-comp1}) follows from these facts,
by instantiating $Q'$ with $\Tpi.$

\item[Sequent 2.]
In this case, we need to prove the sequent
\begin{equation}
\label{eq: open-comp3}
\NSeq{m_1, \cdots, m_l, \Sigma}{[D\theta]}{\exists Q'. \onep{\Qpi\theta}{\inact {X\theta}}{Q'}
\land \forall w. \lbisim{((P'\theta) w)}{(Q' w)}}
\end{equation}
for each non-trivial $\theta$ in the premises of $one_b$ rule. 
By the same reasoning as in the previous case, we obtain, for every transition
$
\one{\Ppi\theta}{x(w)}{\Rpi},
$
where $\Rpi = (P'\theta)\,w,$ another transition
$
\one{\Qpi\theta}{x(w)}{\Tpi}
$
such that (we assume w.l.o.g. that $w$ is fresh)
$
\Rpi \sim_o^{D\theta}  \Tpi.
$
The former implies that $\one{\Qpi\theta}{\inact x}{\lambda w. \Tpi}$ is derivable, and the latter
implies, by induction hypothesis, that 
$$
\forall m_1 \cdots \forall m_l \forall w. [D\theta] \oimp \lbisim {\Rpi}{\Tpi}
$$
is derivable. 
As in the previous case, from these two facts, we can prove the sequent (\ref{eq: open-comp3})
by instantiating $Q'$ with $\lambda w.\Tpi$.

\item[Sequent 3] In this case, we need to prove the sequent
\begin{equation}
\label{eq: open-comp4}
\NSeq{m_1, \cdots, m_l, \Sigma}{[D\theta]}{\exists Q'. \one{\Qpi\theta}{A\theta}{Q'}
\land \nabla w. \lbisim{((P'\theta) w)}{(Q' w)}}
\end{equation}
for each non-trivial $\theta$ in the premises of $one_b$ rule. As in the previous case, we obtain
$\Rpi$ and $\Tpi$ such that
$\one{\Ppi \theta}{\bar x(w)}{\Rpi}$ and $\one{\Qpi \theta}{\bar x(w)}{\Tpi}$
where $\lambda w.\Rpi = P'\theta.$ We assume, without loss of generality, that $w$ is fresh, therefore
since $\Ppi \sim_o^{D} \Qpi$, by definition we have that $\Rpi \sim_o^{D'} \Tpi,$
where $D' = D\theta \cup \{x\}  \times \fn{D\theta, \Ppi\theta, \Qpi\theta}.$
Note that the free names of $D\theta$, $\Ppi\theta$ and $\Qpi\theta$ are exactly
$m_1, \ldots, m_l$, so $D' = D\theta \cup \{x\} \times \{m_1, \ldots m_l\}.$
Thus by induction hypothesis, the formula
$$
\forall m_1 \cdots \forall m_l \forall w. [D'] \oimp \lbisim{\Rpi}{\Tpi}.
$$
Now apply Proposition~\ref{prop:forall nabla} to replace $\forall w$ with $\nabla w$,
$$
\forall m_1 \cdots \forall m_k \nabla w. [D'] \oimp \lbisim{\Rpi}{\Tpi}.
$$
And since $\nabla$ distributes over all propositional connectives, we also have
$$
\forall m_1 \cdots \forall m_k.  (\nabla w.[D']) \oimp \nabla w. \lbisim{\Rpi}{\Tpi}.
$$
It can be shown that $\NSeq{m_1,\ldots, m_l}{.}{\nabla w.[D'] \oimp [D\theta]}$ is provable, 
since the inequalities between $w$ and $m_1, \ldots, m_k$ trivially true. 
Therefore we have that
\begin{equation}
\label{eq: open-comp5}
\vdash \forall m_1 \cdots \forall m_k.  [D\theta] \oimp \nabla w. \lbisim{\Rpi}{\Tpi}.
\end{equation}
Now in order to prove sequent (\ref{eq: open-comp4}), we instantiate
$Q'$ with $\lambda w.\Tpi$, and the rest of the proof proceeds as in the previous case,
\ie, with the help of formula (\ref{eq: open-comp5}).\qed
\end{description}

\subsection{ ``Early'' open bisimulation}

The proof of Theorem~\ref{thm:open-early-bisim} is by induction on the
number of input prefixes in $\Ppi$ and $\Qpi.$ 
We prove a more general result: $\vdash \Qscr \bar n. \lbisim \Ppi \Qpi$
if and only if $\vdash \Qscr \bar n. \ebisim \Ppi \Qpi$, for any 
quantifier prefix $\Qscr \bar n.$
By Lemma~\ref{lm:lbisim-inv} and Lemma~\ref{lm:ebisim-inv}, and the invertibility of $\nablaR$ and $\forallR$
rules, we know that if $\vdash \Qscr \bar n.\lbisim \Ppi \Qpi$
and $\vdash \Qscr \bar n. \ebisim \Ppi \Qpi$, then their unfolded instances
are also provable. We show that one can construct a derivation for one instance
from the other. The non-trivial case is when the bound input transition is 
involved. That is, given a derivation of
$$
\Qscr \bar n. [\forall X \forall P'. \onep{\Ppi}{\inact X}{P'}
\oimp \forall w \exists Q'. \onep{\Qpi}{\inact X}{Q'} \land \ebisim {(P'w)}{Q'w)}]
$$
we can construct a derivation of
$$
\Qscr \bar n. [\forall X \forall P'. \onep{\Ppi}{\inact X}{P'}
\oimp \exists Q'. \onep{\Qpi}{\inact X}{Q'} \land \forall w. \ebisim {(P'w)}{(Q'w)}]
$$
and vice versa. Note that we cannot do any analysis on the universally quantified
name $w$ in both formulas, since we do not have any assumptions on names
(\eg, the excluded middle on names as in the adequacy theorem for late bisimulation).
It is then easy to check that the choice of $Q'$ in both cases is independent of
the name $w$, and their correspondence follows straightforwardly from the induction
hypothesis.
\qed

\section{Adequacy of the specifications of modal logics}

The completeness proof of the modal logics specification shares similar structures
with the completeness proofs for specifications of bisimulation. 
In particular, we use an analog of Lemma~\ref{lm:lbisim-complete}, given in the following. 
\begin{lemma}
\label{lm:modal-complete}
Let $\Ppi$ be a process and $\Api$ an assertion such that $\Ppi \models \Api.$
Then 
$$
\vdash \forall n_1 \cdots \forall n_k. \Xscr \land n_1 \not = \cdots \not = n_k \oimp \trans {\Ppi \models \Api}
$$
for some $\Xscr \subseteq_f \Escr$ and some names $n_1, \ldots , n_k$
such that $\fn{\Ppi, \Api} \subseteq \{n_1,\ldots, n_k \}.$
\end{lemma}
The proof of lemma proceeds by induction on the size of $\Api$. The crucial
step is when its interpretation in $\FOLDNb$ contains universal quantification
over names, \eg, when $\Api = \mBox{a(y)}\Bpi$. In this case, we again
use the same technique as in the proof of Lemma~\ref{lm:lbisim-complete}, \ie,
using the excluded middle assumptions on names to enumerate all possible
instances of the judgments. A more detailed proof can be found in the electronic
appendix of this paper.

\subsection{Proof of Theorem~\ref{thm:modal adequacy} (Adequacy of the
modal logic encoding)}

First consider proving the soundness part of this theorem.
Suppose we have a derivation $\Pi$ of $\NSeq{\cdot}{\Xscr}{\nabla \bar n. \trans{\stf \Ppi \Api}}.$
We want to show that $\stf \Ppi \Api$. 
This is proved by induction on the size of $\Api$. The proof also uses the property
of invertible rules and the fact that applications of the excluded middles in $\Xscr$
in deriving the sequent can be permuted up over all the right introduction rules.
The latter is a consequence of Lemma~\ref{lm:right-first}.
We look at a couple of interesting cases involving bound input and bound output.

\begin{description}

\item[out:] Suppose $\Api$ is $\mBox{\bar x(y)}\Bpi$. We need to show that for every
$\Ppi'$ such that $\one \Ppi {\bar x (y)} \Ppi'$, we have $\stf {\Ppi'} \Bpi.$ 
(By $\alpha$-conversion we can assume without loss of generality that 
$y$ is not free in $\Ppi$ and $\Api$.) Note that here the occurrence of $y$ in $\Ppi'$ is bound 
in the transition judgment $\one \Ppi {\bar x(y)} {\Ppi'}$.
By Lemma~\ref{lm:right-first} and the invertibility of certain inference rules, we can
show that provability of $\NSeq{\cdot}{\Xscr}{\nabla \bar n. \trans{\stf \Ppi \Api}}$
implies the existence of a derivation $\Pi'$ of 
$$
  \NSeq{M}{\Xscr, \Judg{\bar n}{\onep{\trans \Ppi}{\outact x}{M\bar n}}}
   {\Judg{\bar n}{\nabla y.\stf{M\bar n y}{\trans \Bpi}}}
$$
for some eigenvariable $M$. 
By the adequacy of one-step transitions, we have that 
$\vdash \nabla \bar n. \onep{\trans \Ppi}{\outact x}{\lambda y.\trans {\Ppi'}}$.
Let $\theta$ be the substitution $[(\lambda \bar n\lambda y.\trans {\Ppi'})/M].$
Applying $\theta$ to $\Pi'$ we get the derivation $\Pi'\theta$ of
$
\NSeq{\cdot}{\Judg{\bar n}{\onep{\trans \Ppi}{\outact x}{\lambda y.\trans {\Ppi'}}}}
{\Judg{\bar n}{\nabla y.\stf{\Ppi'}{\trans \Bpi}}}.
$
By cutting this derivation with the one-step transition judgment above, we obtain
a derivation of 
$
\NSeq{\cdot}{.}{\Judg{\bar n}{\nabla y.\stf{\Ppi'}{\trans \Bpi}}}.
$
Hence by induction hypothesis, we have that $\stf {\Ppi'} \Bpi$.

\item[in:] Suppose $\Api$ is $\mBox{x(y)}^L \Bpi$. We show that there exists
a process $\Ppi'$ such that $\one \Ppi {x(y)} \Ppi'$ and
for all name $w$, $\stf {\Ppi'[w/y]} {\Bpi[w/y]}$. It is enough to consider
the case where $w$ is a name in $\fn{\Ppi,\Api}$ and the case where $w$
is a new name not in $\fn{\Ppi,\Api}$. By Lemma~\ref{lm:right-first} and the invertibility
of some inference rules, we can show that provability of 
$
\NSeq{\cdot}{\Xscr}{\Judg{\bar n}{\stf{\trans \Ppi}{\trans{\mBox{x(y)}^L \Bpi}}} }
$
implies the existence of two derivations $\Pi_1$ and $\Pi_2$, of the sequents
$\NSeq{\cdot}{\Xscr}{\Judg{\bar n}{\onep \Ppi {\inact x} {N}}}$
and 
$\NSeq{\cdot}{\Xscr}{\Judg{\bar n}{\forall y. \stf {N y} {\trans{\Bpi}}}}$,
respectively, for some closed term $N$.

By the adequacy result in Proposition~\ref{prop:one step},
there exists a process $\Ppi'$ such that $\trans{\Ppi'} = N y$ and $\one \Ppi {x(y)} {\Ppi'}.$
By Proposition~\ref{prop:subst}, we can instantiate $y$ with any of the free names occurring
in $\Ppi$ or $\Api$ (since they are all in the list $\bar n$), and hence for any name $w \in \fn{\Ppi,\Api}$
by induction hypothesis we get $\stf {\Ppi'[w/y]} {\Bpi[w/z]}$. The case where $w$ is a new name
is dealt with as follows. Without loss of generality we assume that $y = w$ (since we can always
choose $y$ to be sufficiently fresh).  
From $\Pi_2$ it follows that $\vdash \Xscr \oimp \nabla \bar n. \forall y. \stf {\trans{\Ppi'}} {\trans{\Bpi}}.$
Using the $\FOLDNb$ theorems
$$
(\nabla x \forall y. P) \oimp \forall y\nabla x.P
\quad
\hbox{ and }
\quad
(P \oimp \forall z.Q) \oimp \forall z(P \oimp Q)
$$
where $z$ is not free in $P$, we can move the $\forall y$ quantification in 
$\Xscr \oimp \nabla \bar n. \forall y. \stf {\trans{\Ppi'}} {\trans{\Bpi}}$ to the outermost level and get
the provable formula 
$\forall y(\Xscr \oimp \nabla \bar n. \stf {\trans{\Ppi'}} {\trans{\Bpi}})$.
We then apply Proposition~\ref{prop:forall nabla}, to turn $\forall y$ into $\nabla y$, thus obtaining
a derivation of 
$
\nabla y(\Xscr \oimp \nabla \bar n. \stf {\trans{\Ppi'}} {\trans{\Bpi}}),
$
and by distributing $\nabla$ over $\oimp$, we get
$(\nabla y.\Xscr) \oimp \nabla y\nabla \bar n.\stf {\trans{\Ppi'}} {\trans{\Bpi}}.$
We can now apply the induction hypothesis to get $\Ppi' \models \Bpi.$
\end{description}

Next we consider proving the completeness part of Theorem~\ref{thm:modal adequacy}.
Given $\Ppi \models \Api$, we would like to show that 
$\NSeq{\cdot}{\Xscr}{\nabla \bar n. \trans {\Ppi \models \Api}}$ is provable.
By Lemma~\ref{lm:modal-complete}, there are $m_1,\ldots, m_k$ and $\Xscr'$ such that
$$
\vdash \forall m_1 \cdots \forall m_k. \Xscr' \land m_1 \not = m_2 \cdots \not = m_k \oimp \trans{\Ppi \models \Api}.
$$ 
Let $\bar n = m_1, \ldots, m_k$ and let $\Xscr = \nabla \bar n.\Xscr'.$
By Proposition~\ref{prop:forall nabla}, we have a derivation of 
$$
\nabla m_1 \cdots \nabla m_k. \Xscr' \land m_1 \not = m_2 \cdots \not = m_k \oimp \trans{\Ppi \models \Api}.
$$
By distributing the $\nabla$'s over implication and conjunction we obtain
$$
\Xscr \land (\nabla \bar n. m_1 \not = m_2 \cdots \not = m_k) \oimp \nabla \bar n. \trans{\Ppi \models \Api}.
$$
But since $\nabla \bar n. m_1 \not = m_2 \cdots \not = m_k$ is provable, by cut we obtain
a derivation of 
$$
\NSeq{}{\Xscr}{\nabla \bar n. \trans{\Ppi \models \Api}.\qquad\qquad\qed}
$$

\section{Characterisation of open bisimulation}

\begin{lemma}
\label{lm:ch-open-sound}
Let $\Ppi$ and $\Qpi$ be two processes. 
If for all $\Api \in \Lscr\Mscr$, 
$\vdash (\Qscr \bar n. \Ppi \models \Api)$ if and only if $\vdash (\Qscr \bar n. 
\Qpi \models \Api)$, where $\fn{\Ppi,\Qpi,\Api} \subseteq \{ \bar n\}$, 
then $\Ppi \sim_o^{D} \Qpi,$ where $D$ is the $\Qscr \bar n$-distinction.
\end{lemma}
\begin{proof}
Let $\Sscr$ be the following family of relations 
$$
\begin{array}{ll}
\Sscr_D = \{ (\Ppi, \Qpi) \mid & \hbox{for all $\Api$, $\vdash (\Qscr \bar n. \Ppi \models \Api)$ iff
$\vdash (\Qscr \bar n. \Qpi \models \Api)$,} \\
 & \hbox{where $\fn{\Ppi,\Qpi,\Api} \subseteq \{\bar n\}$ and $D$ is the $\Qscr \bar n$-distinction}  \} 
\end{array}
$$
We then show that $\Sscr$ is an open bisimulation. $\Sscr$ is obviously symmetric, so it remains
to show that it is closed under one-step transitions. 
We show here a case involving bound output; 
the rest are treated analogously.

Suppose $(\Ppi, \Qpi) \in \Sscr_D$. Then we have that for all $\Api$, 
$\vdash \Qscr \bar n. \Ppi \models \Api$ iff $\vdash \Qscr \bar n. \Qpi \models \Api$,
for some prefix $\Qscr \bar n.$
Let $\theta$ be a substitution that respects $D$. Suppose 
$\one{\Ppi\theta }{\bar x(y)}{\Ppi'}.$ We need to show that there exists a $\Qpi'$
such that $\one{\Qpi\theta}{\bar x(y)}{\Qpi'}$ and $\Ppi' \sim_o^{D'} \Qpi'$
where $D' = D\theta \cup \{ y  \} \times \fn{\Ppi,\Qpi,D}.$ (Here we
assume w.l.o.g.\ that 
$y$ is chosen to be sufficiently fresh.)
Suppose $\theta$ identifies the following pairs of names in $\Ppi$ and $\Qpi$:
$(x_1,y_1), \ldots, (x_k, y_k)$, and suppose that $\theta(z) = x.$
Then by the definition of $\Sscr_D$:
$$
\vdash \Qscr \bar n. \Ppi \models [x_1 = y_2][x_2 = y_2] \cdots [x_{k} = y_k] \langle \bar z(y)\rangle \Bpi 
$$
if and only if for all $\Bpi$,
$$
\vdash \Qscr \bar n. \Qpi \models [x_1 = y_2][x_2 = y_2] \cdots [x_{k} = y_k] \langle \bar z(y)\rangle \Bpi.
$$

Note that the statement cannot hold vacuously, since for at least
one instance of $\Bpi$, \ie, $\Bpi = \mTrue$, both judgments must be true.
By analysis on the (supposed) cut-free proofs of both judgments, for any $\Bpi$, 
the above statement reduces to
$$
\vdash \Qscr \bar m. \Ppi\theta \models \langle \bar x(y) \rangle  \Bpi\theta 
\qquad \hbox{ iff } \qquad
\vdash \Qscr \bar m. \Qpi\theta \models \langle \bar x(y)\rangle \Bpi\theta, 
$$
for some prefix $\Qscr \bar m$ such that $\Qscr \bar m$-distinction
is the result of applying $\theta$ to the $\Qscr \bar n$-distinction.

Now let $\{ \Qpi_i \}_{i \in I}$ be the set of all $\Qpi'$ such that
$\one{\Qpi\theta}{\bar x(y)}{\Qpi'},$ and suppose that for all $i \in I$,
$\Ppi' \not \sim_o^{D'} \Qpi_i.$ That means that there exists an 
$\Api_i$, for each $i \in I$, that separates $\Ppi'$ and $\Qpi_i,$
\ie, $\vdash (\Qscr \bar m   \nabla y . \Ppi' \models \Api_i)$ but
$
\not \vdash (\Qscr \bar m \nabla y. \Qpi' \models \Api_i).
$
Note that we can assume w.l.o.g. that $\bar m$ include all the free names
of $\Api_i$ (recall that $\bar n$ is really a schematic list of names,
dependent on the choice of $\Api$ in the first place). 
Let $\Bpi\theta$ be $\bigwedge_{i\in I} \Api_i.$
Then, by analysis of cut-free proofs, we can show that 
$
\vdash (\Qscr \bar m. \Ppi\theta \models \langle \bar x(y) \rangle \Bpi\theta)
$
but
$
\not \vdash (\Qscr \bar m. \Qpi \models \langle \bar x(y) \rangle \Bpi\theta),
$
which contradicts our initial assumption.
Therefore, there must be one $\Qpi'$ such that
$\one{\Qpi}{\bar x(y)}{\Qpi'}$ and $\Ppi' \sim_o^{D'} \Qpi'.$
\qed
\end{proof}

\begin{lemma}
\label{lm:ch-open-complete}
Let $\Ppi$ and $\Qpi$ be two processes such that $\Ppi \sim_o^D \Qpi$ for some distinction $D$.
Then for all $\Api \in \Lscr \Mscr$ and for all prefix $\Qscr \bar n$ such that
$D$ corresponds to the $\Qscr \bar n$-distinction and $\fn{\Ppi, \Qpi, D} \subseteq \{ \bar n\}$,
$\vdash \Qscr \bar n. \Ppi \models \Api$ if and only if 
$\vdash \Qscr \bar n. \Qpi \models \Api.$
\end{lemma}
\begin{proof}
Suppose that $\Ppi \sim_o^D \Qpi$ and $\vdash \Qscr \bar n. \Ppi \models \Api.$ 
We show, by induction on the size of $\Api,$
that  $\vdash \Qscr \bar n. \Qpi \models \Api.$ The other direction
is proved symmetrically, since open bisimulation is symmetric. 
We look at the interesting cases.
\begin{itemize}

\item Suppose $\Api = \langle \bar x(y) \rangle \Bpi$ for some $\Bpi.$
By analysis on the cut free derivations of $\Qscr \bar n. \Ppi \models \Api$, 
it can be shown that
$$
\vdash \Qscr \bar n. \exists M. \onep{\Ppi}{\outact x}{M} \land \nabla y. (M\,y) \models \Bpi.
$$
This entails that there exists a process $\Ppi'$ such that
$$
\vdash \Qscr \bar n. \onep{\Ppi}{\outact x}{\lambda y.\Ppi'} \land \nabla y. \Ppi' \models \Bpi.
$$
And by the invertibility of the right-introduction rules for $\forall$, $\nabla$ and $\land $, this in turn entails that
$
\vdash \Qscr \bar n. \onep{\Ppi}{\outact x}{\lambda y.\Ppi'}
$
and
$
\vdash \Qscr \bar n \nabla y. \Ppi' \models \Bpi.
$ The former implies, by the adequacy of one-step transition, that
$\one{\Ppi}{\bar x(y)}{\Ppi'}.$ Since $\Ppi \sim_o^D \Qpi$, this means
that there exists $\Qpi'$ such that $\one{\Qpi}{\bar x(y)}{\Qpi'}$ and
$\Ppi' \sim_o^{D'} \Qpi'$, where $D' = D \cup \{
y \} \times \fn{\Ppi,\Qpi,D}.$ At this point we are almost ready to
apply the induction hypothesis to $\Qscr \bar n\nabla
y.\Ppi' \models \Bpi$, except that $D'$ may not corresponds to the
$\Qscr \bar n \nabla y$-distinction, since the latter may contain more
inequal pairs than $D'$.  However, since open bisimulation is closed
under extensions of distinctions (see Lemma
6.3. in \cite{sangiorgi96acta}), we can assume without loss of
generality that $D'$ is indeed the $\Qscr \bar n\nabla
y$-distinction. Therefore by the adequacy of one-step transition and
induction hypothesis, we conclude that $
\vdash \Qscr \bar n. \onep{\Qpi}{\outact x}{\lambda x.\Qpi'}
$
and
$
\vdash \Qscr \bar n \nabla y. \Qpi' \models \Bpi,
$
and from these, it follows that $\Qscr \bar n. \Ppi \models \Api$ is also provable.

\item Suppose $\Api = \langle x(y) \rangle \Bpi.$ This case is analogous to the previous case. The only
difference is that the bound input is universally quantified, instead of $\nabla$-quantified.
So we apply the induction hypothesis to $\Qscr \bar n \forall y. \Ppi' \models \Bpi$, which can be done
without resorting to extensions of the distinction $D$, since in this case the $\Qscr \bar n \forall y$-distinction
is exactly $D.$

\item For the cases where $\Api$ is prefixed by either $[x(y)]^L$ or $[\bar x(y)]$, the proof follows
a similar argument as in the completeness proof of open bisimulation (Theorem~\ref{thm:open bisim complete}).
For instance, for the case where $\Api = [x(y)]^L \Bpi$, from the fact that 
$\vdash \Qscr \bar n. \Ppi \models \Api$, it follows that 
$$
\vdash \Qscr \bar n. \forall M(\onep{\Ppi}{\inact x}{M} \oimp \exists y. (M\,y) \models \Bpi).
$$
As in the proof of Theorem~\ref{thm:open bisim complete}, we can further show that there is 
a derivation of this formula that ends with $\oneb$-rule, such that every $\theta$ in this
premise is a $D$-respecting substitution. 
Since $\Ppi \sim_o^D \Qpi$, we can show that every bound input 
action of $\Ppi\theta$, for any $D$-respecting $\theta$, can be imitated by $\Qpi\theta$ and vice versa. 
From this and induction hypothesis, we can therefore obtain a derivation of 
$$
\Qscr \bar n. \forall N(\onep{\Qpi}{\inact x}{N} \oimp \exists y. (N\,y) \models \Bpi),
$$
hence $\vdash \Qscr \bar n. \Qpi \models \Api.$  \qed
\end{itemize}
\end{proof}

Finally, the proof of Theorem~\ref{thm:ch-open} now follows 
immediately from Lemma~\ref{lm:ch-open-sound} and
Lemma~\ref{lm:ch-open-complete}. 
\qed

 \begin{received}
 Received May 2008;
 revised December 2008;
 accepted February 2009
 \end{received}

 \end{document}